\documentclass[article,11pt,apl,amsmath,amssymb,showpacs]{revtex4-2}
\usepackage{epsf}      
\usepackage{graphicx}
\usepackage{booktabs}
\usepackage{color}
\usepackage{soul}
\usepackage{gensymb}
\usepackage{sidecap}
\usepackage{amsmath}
\usepackage{mathtools}
\usepackage[hidelinks,colorlinks=true,linkcolor=blue,citecolor=blue]{hyperref}
\usepackage[table,xcdraw]{xcolor} 
\usepackage{colortbl}
\usepackage{longtable}
\usepackage{mhchem}

\begin{document}
\title{Synergistic Role of Transition Metals and Polyanionic Frameworks in Phosphate-Based Cathode  Materials for Sodium-Ion Batteries}
\author{Madhav Sharma}
\affiliation{Department of Physics, Indian Institute of Technology Delhi, Hauz Khas, New Delhi-110016, India}
\author{Riya Gulati}
\affiliation{Department of Physics, Indian Institute of Technology Delhi, Hauz Khas, New Delhi-110016, India}
\author{Rajendra S. Dhaka}
\email{rsdhaka@physics.iitd.ac.in}
\affiliation{Department of Physics, Indian Institute of Technology Delhi, Hauz Khas, New Delhi-110016, India}

\date{\today}      

\begin{abstract}
Ongoing research in the area of advanced cathode materials for sodium-ion batteries (SIBs) is expected to reduce reliance on lithium-ion batteries (LIBs), providing more affordable and sustainable energy storage solutions. Polyanionic compounds have emerged as promising options due to their stable structure and ability to withstand high-voltage conditions as well as fast charging capabilities. This review offers a thorough discussion of phosphate-based polyanionic cathodes for SIBs, exploring their structure, electrochemical performance with various transition metals, and existing challenges. We discuss different polyanionic frameworks, such as ortho-phosphates, fluoro-phosphates, pyro-phosphates, mix pyro-phosphates, and NASICON-based phosphates, highlighting their unique structural characteristics and ability to perform well across a wide potential range. Further, we delve into the mechanisms governing sodium storage and tunability of redox potentials in polyanionic materials, providing insights into the factors that affect their electrochemical performance. Finally, we outline future research directions and potential avenues for the practical applications of polyanionic high-voltage cathodes in sodium-ion battery technologies.
\newline

\noindent\textbf{Broader context}

The need and considerations for post-LIBs have presented a significant scope for SIBs to grow and deliver the goal of sustainability. However, there is still a significant gap in their energy densities, demanding a breakthrough in SIBs. Many potential cathodes face challenges in achieving high capacities, but the energy density can also be boosted by elevating the working voltage of SIBs. This higher voltage can be achieved through the use of the highly electronegative PO$_4$ polyanionic unit. Nevertheless, the arrangement with which these units are attached to the transition metal causes a significant change in the working voltage. To substantiate the correlation between the local structure and its influence on the electrochemical activity, a compilation of the crystallographic data of recent cathodes is presented and discussed, and an attempt is made to offer a discussion relating to the local structure and its influence on the electrochemical performance. Moreover, the discussion of the structural polymorphs with the same formula unit but differing electrochemical behaviors is also documented to help the readers gain a better understanding. This correlation is not limited to SIBs but can also be extended to a wide range of intercalation-based cathode materials, thereby offering valuable insights for the development of improved batteries in the future. 

\end{abstract}

\maketitle
\section{\noindent~Introduction}

\begin{figure*} 
\includegraphics[width=6.5in]{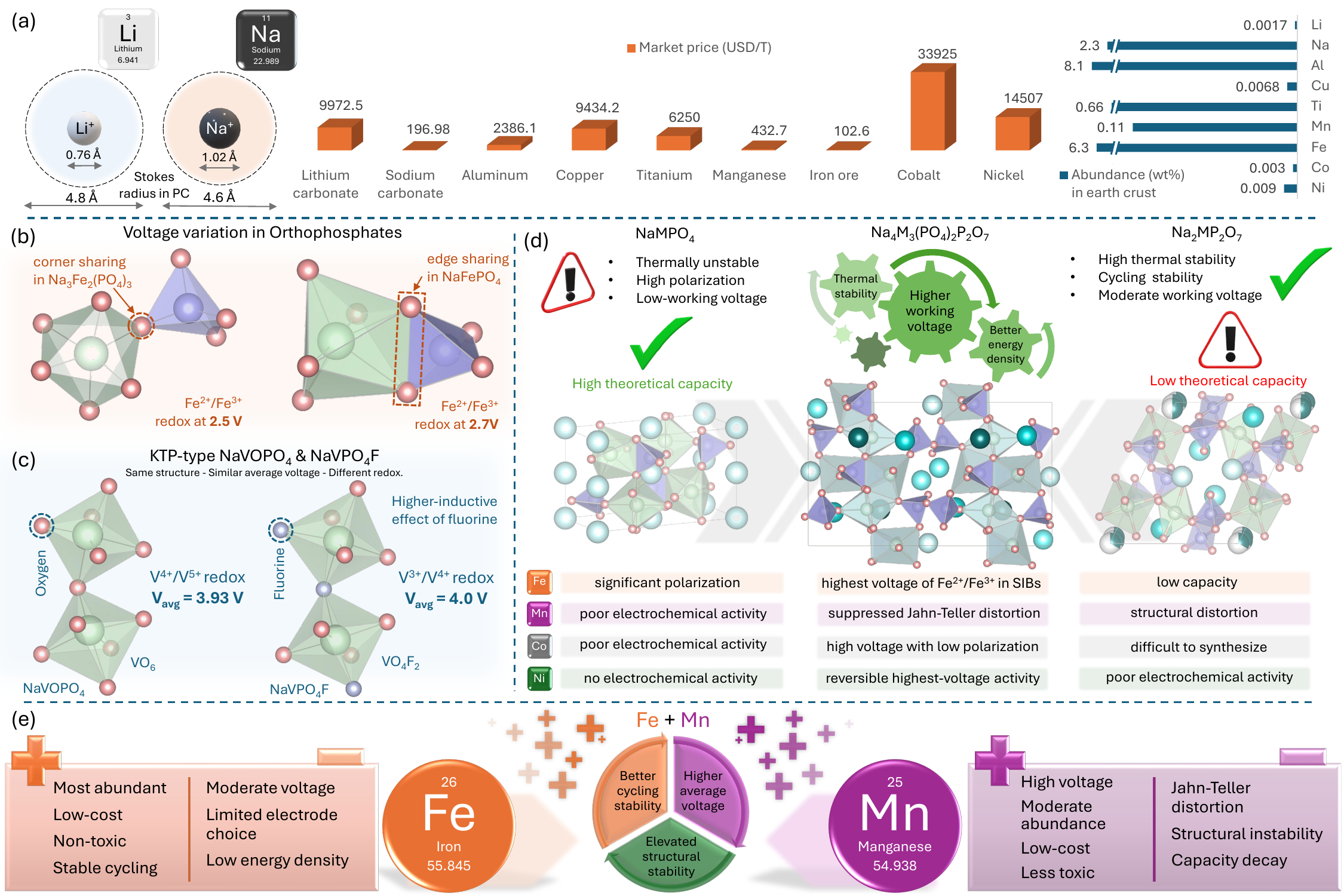} 
\caption {(a) A comparison between Li$^+$ and Na$^+$ as charge carriers, including global market prices (cost conversion 07-Apr-2025, 1 USD = 7.28 CNY) \cite{TE} of key battery elements and their corresponding elemental abundance (wt\%) in the Earth's crust \cite{PT}. (b) The schematic illustration of  the variation in redox voltage with changes in MO$_6$--PO$_4$ connectivity. (c) The influence of different ligands on redox voltage. (d) The strategic advantages of mixed-phosphate polyanionic cathodes compared to ortho- and pyro-phosphate frameworks. (e) An overview of the individual advantages and limitations of Fe and Mn, and how their appropriate combination can overcome respective challenges in sodium-ion batteries.}
\label{Figure0}
\end{figure*}

The success of the LiCoO$_2$ (LCO) and its derivatives as cathode have revolutionized portable devices, and the current utilization of lithium-ion batteries (LIBs) in electric vehicles, which have a rigorous impact on the demand and supply \cite{Manthiram_NatE_21, Nishi_JPS_01, EV_Report_24}. However, the high cost and scarcity of Li and Co resources [Fig. \ref{Figure0}(a)] makes the dream of sustainable energy economically challenging. To deal with Co, scientists are opting for Co-free alternatives, utilizing Ni, Mn, Fe, etc. \cite{Li_NE_20}; for example, the polyanionic structured LiFePO$_4$ (LFP) came into the market as a low-cost alternative to LCO, utilizing abundant resources and giving a decent energy density of around 160 Wh/kg in a pouch/cylindrical cell configuration compared to the average 210--220 Wh/kg of LiNi$_x$Mn$_y$Co$_{1-x-y}$O$_2$ (NMC) based LIBs \cite{Padhi_JES_97, Link_B_23}. Regardless of having lower energy densities than layered cathodes, the LFP offers numerous benefits in terms of safety, longer cycle life, eco-friendliness, and many more \cite{Kim_JPS_22, Park_EES_11}. The LFP batteries are expected to deliver comparatively better performance during their second life after retirement and offer easier recyclability at the end of their lifespan \cite{Nallah_JES_23, Dobo_ER_23, Lei_ESM_25}. These advantages position the polyanionic cathode as a key factor driving the exponential growth of the LFP battery market, both in terms of size and market share \cite{LFP_FBI_24}. However, the challenge of limited and non-uniform availability of lithium demand replacing lithium ions (-3.05 V vs. SHE), i.e., exploring new and innovative chemistries as post-LIB alternatives for energy storage applications \cite{Manish_CEJ_23, Sharma_IJPAP_24, Sapra_JMCA_25}. 

In this context, sodium, potassium, magnesium, calcium, aluminum, and zinc ion batteries are being considered due to their abundance, low cost, and comparable energy storage capabilities \cite{Kundu_ACIE_15, Wang_AFM_24, Li_AEM_23, Gummow_AM_18, Lin_NAT_15, Ting_EC_24, Palacin_JPE_24}. Out of these options, sodium (Na) and potassium (K) offer an alternative low-cost energy storage ecosystem, as most of the Na and K-based chemistries derive their motivation from the vast and established Li-based chemistries. The choice between Na and K is difficult as K offers more negative reduction potential (-2.93 V vs. SHE) than Na (-2.71 V vs. SHE). In contrast, Na has lower atomic weight, smaller ionic radii, and natural abundance \cite{Pramudita_AEM_17, Chen_ASS_18}. However, compared to lithium, the Na$^+$ exhibits a smaller Stokes radius of 4.6 \AA~in propylene carbonate (PC) versus 4.8 \AA~for Li$^+$ [see Fig.~\ref{Figure0}(a)], offering an advantage in ionic mobility within liquid electrolytes \cite{Wu_JPS_21}. Nevertheless, the larger ionic radius and higher atomic weight of sodium pose significant challenges in developing high-performance electrodes with desirable electrochemical performance in sodium-ion batteries (SIBs)\cite{Chayambuka_AEM_20}. 

The research community is therefore focusing on developing high-voltage and high-capacity cathode materials to achieve energy density and cycle life comparable to the commercial LIBs \cite{Pati_JMCA_22}. Till then, SIBs can be viewed as complementary to LIBs rather than competitive, considering LIBs have expanded the market of portable devices, which will definitely open up opportunities for SIBs \cite{Slater_AFM_13, Tarascon_J_20, Delmas_AEM_18, Rudola_NE_23}. On the other hand, SIBs offer an advantage in applications where cost is a critical factor and energy density is less of a concern, providing a valuable place for their use \cite{Dhaka_CW_24, Hu_ACSEL_21}. Further, the lack of alloy formation between aluminum and sodium allows the use of aluminum as current collectors, which significantly reduces the weight and price of SIBs. With their cost-effectiveness, scalability, and alignment with global sustainability goals, SIBs have captured the attention of leading industries like CATL \cite{CATL}, Faradion \cite{Faradion}, Natron Energy \cite{Natron_energy}, and HiNa \cite{HiNa}, to revolutionize the energy landscape. The first-generation commercialized SIBs have already achieved an impressive energy density of 160 Wh/kg, with an ambitious target of 200 Wh/kg set for future generation, promising even greater performance and efficiency \cite{CATL, Rudola_JMCA_21}.

\begin{figure*} 
\includegraphics[width=6.5in]{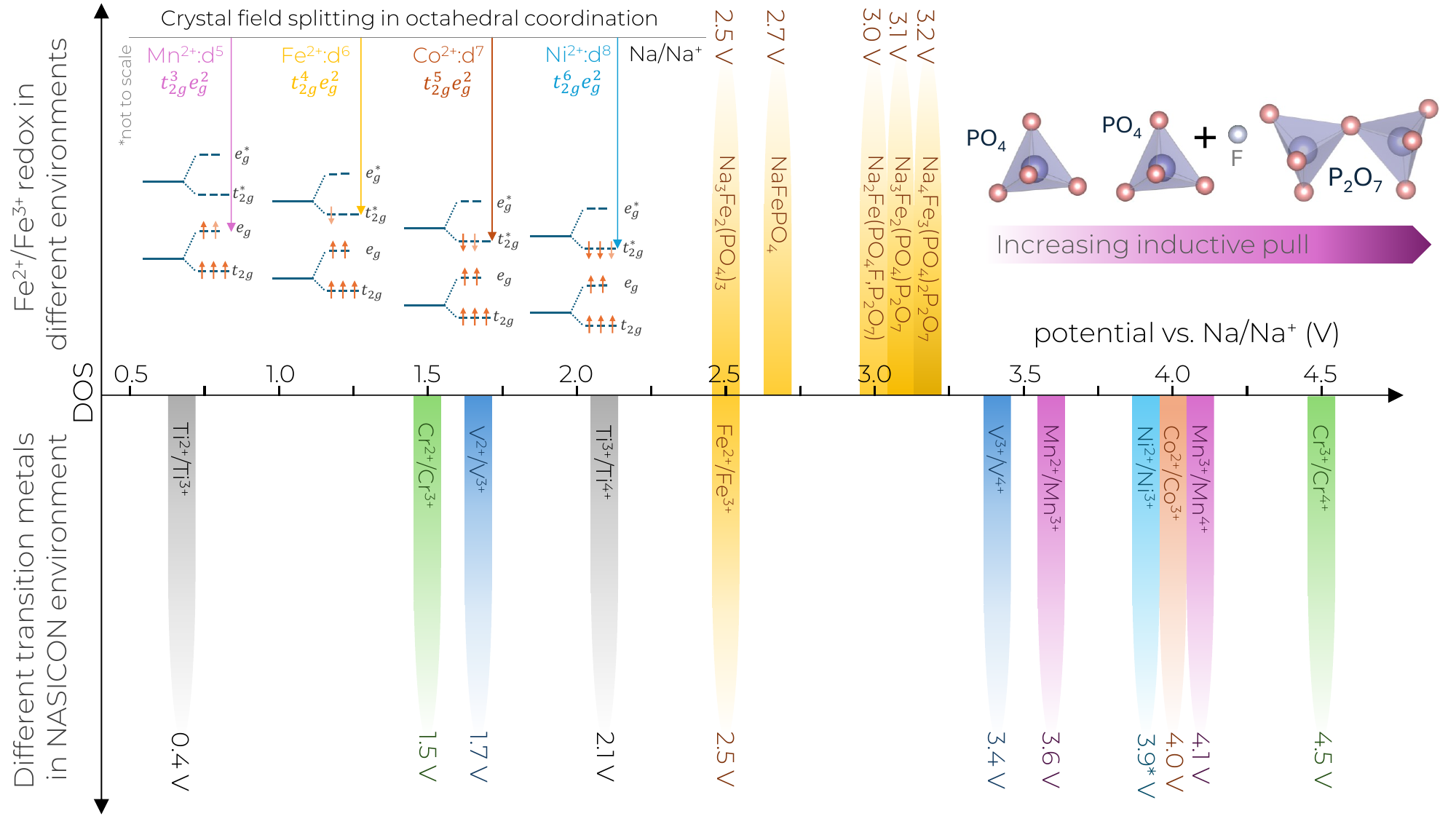} 
\caption {A schematic comparison of experimental operating voltages of Fe$^{2+}$/Fe$^{3+}$ redox in various phosphate-based anionic environments and of different transition metals in NASICON (Na$_x$MM$^\prime$(PO$_4$)$_3$) structured materials. The depiction of crystal field splitting for M$^{2+}$ (M=Mn, Fe, Co, Ni) cations in an octahedral coordination environment is also presented.}
\label{Abstract}
\end{figure*}

If we look from another perspective, the voltage of the electrodes is a property of the host structure and the active transition metal [see Figs.~ \ref{Figure0}(b, c)]. Moreover, the mass of Li and Na is only a fraction of the weight of the total mass of active materials, which may not alter the capacities significantly. However, finding a good host structure is a quest for researchers to elevate the voltage of positive electrodes, provided the host structure retains its structural integrity and cycling stability during high-voltage operation. Additionally, a balanced combination of intrinsic electronic and ionic conductivity is essential for the host structure to maintain overall charge neutrality during the redox activity and transport of alkali metal ions. The main aim of this article is to discuss advancement and capabilities towards the quest for high-voltage cathodes for SIBs, and the local structural factors that influence the voltage. Therefore, understanding the localized arrangement of the host structure becomes important when engineering novel cathode materials, which could help mitigate the potential concerns in SIBs. In this direction, polyanion-type cathodes have drawn great attention after the success of LFP, as these polyanionic combinations offer structural diversity and multi-electron redox activity, resulting in excellent electrochemical performance. Here, different polyanion combinations (XO$_4$, X=S, P, Si) have different degrees of inductive effect and structural variety, which can be used to tune the redox voltage \cite{Sapra_WEE_21, Masquelier_CR_13}. Additionally, polyanionic cathode materials exhibit exceptional thermal, air, and moisture stability, which are the figures of merit for the long-term reliability and safety of SIBs \cite{Jin_CSR_20}. Unlike layered oxide materials, where cation migration and oxygen evolution occur at high voltages \cite{Gent_NC_17}, the inherent rigid structure of polyanionic materials offers stability against high-voltage redox reactions, thus enhancing the overall safety and durability of SIB systems \cite{Li_AFM_20}. These unique combinations of properties position polyanionic cathode materials as key contenders in the development.

Note that the polyanionic units of sulfates offer the highest inductive pull, which raises the redox voltages and enhances the energy density \cite{Barpanda_NC_14}. Nevertheless, SO$_4$-based systems are hygroscopic, making their synthesis and storage complicated. Also, silicates (SiO$_4$) are promising candidates for low-cost cathode materials owing to silicon's abundance in the Earth's crust. However, the poor rate capability, drastic capacity fades, and complex synthesis procedures limit their potential \cite{Wei_ER_20}. On the other hand, the air and water stability of the PO$_4$-based cathode materials becomes the first choice for further exploration to utilize it to its full potential \cite{Hao_AM_24}. The work on the mixed polyanionic systems is also fascinating, in which the unique quality of each polyanionic unit is included in material engineering \cite{Lu_CM_17, Yahia_JPS_18}. For example, introducing PO$_4$ in SO$_4$ based cathode materials made the system immune to moisture instability \cite{Liu_NE_24, Pati_JPS_24}, while small addition of SiO$_4$ at PO$_4$ sites improves the sodium ion conduction \cite{Pal_ACSAEM_20, Hou_S_23}. Overall, there is a significant influence of unique structural arrangements on the working voltage of polyanionic cathode materials [see Fig.~\ref{Figure0}(d)]. For example, notable changes are observed in the electrochemical properties of polymorphs that share identical formula units but exhibit different arrangements of components within the crystal structure. Therefore, a systematic review is vital to get a better understanding of the correlation between structure and electrochemical behavior as well as the influencing factors. To ensure a fair comparison, we focus our discussion on different phosphate-based polyanionic materials, including ortho-phosphates, fluoro-phosphates, pyro-phosphates, mix pyro-phosphates, and NASICON-based (sodium superionic conductors) phosphates. Each of them holds its structural distinctiveness and offers a unique working voltage (see Fig.~\ref{Abstract}) and sodium storage mechanism, which are critical to understand how the PO$_4$ influences the local environment \cite{Senthilkumar_SM_19}. This article offers a comprehensive review linking the electrochemical performance to the crystal structures, i.e., the dependence of the working voltage on the local environment of the transition metal induced by the host crystal structure and the anionic unit attached \cite{Senthilkumar_SM_19, Fan_EMA_24}. The crystal structure symmetries are expressed using Hermann-Mauguin notation where the crystallographic space groups describe the symmetry of a crystal structure by combining translational and point group symmetries. They provide essential information about the lattice type and symmetry elements, aiding in the understanding of atomic arrangements and structural behavior. A systematic and detailed compilation of these polyanionic materials, including their theoretical parameters and experimentally measured electrochemical values, are presented in Tables: \ref{tab:Summary1}, \ref{tab:Summary2}, and \ref{tab:Summary3}. 

\section{\noindent~Structure Environment and Voltage of a battery} 

\begin{figure*} 
\includegraphics[width=6.5in]{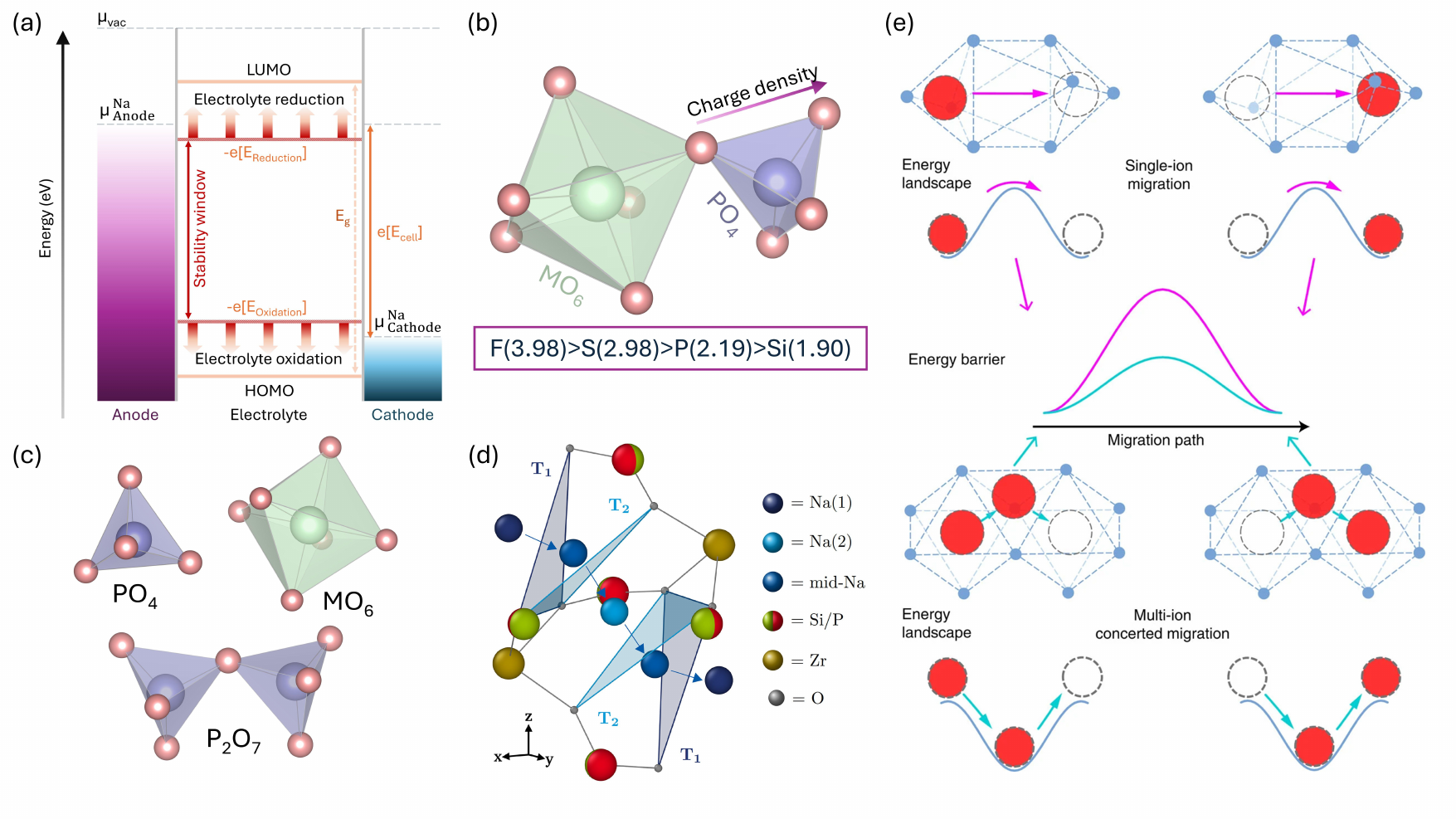}
\caption {(a) The appropriate representation for the minimum and maximum potential thresholds concerning electrolyte stability, as well as the energy levels associated with the highest occupied molecular orbital (HOMO) and lowest unoccupied molecular orbital (LUMO), adapted from \cite{Peljo_EES_18}. (b) The attraction of electrons towards PO$_4$ tetrahedra with high electronegativity and the electronegativity trend of elements F, S, P, and Si, as per the Pauling scale. (c) The visual representation of PO$_4$ tetrahedra, MO$_6$ octahedra, and P$_2$O$_7$ unit. (d) The illustration of the primary pathway for sodium diffusion in NASICON structure (Na$_3$Zr$_2$Si$_2$PO$_{12}$). Here, blue spheres represent the sodium ions' positions within the crystal structure, and blue triangles indicate key structural points labeled T$_1$ and T$_2$, which act as bottlenecks for sodium movement, specifically from Na(1) to the midpoint and from the midpoint to Na(2), respectively \cite{Haarmann_SSI_21}. (e) A schematic representation of the difference between single-ion migration and multi-ion concerted migration and their related energy barrier \cite{He_NatCom_17}.}
\label{Figure1}
\end{figure*}

The output voltage of a battery depends on the difference between the chemical potential of positive and negative electrodes. The crystal structure of host materials that render higher voltage as well as capacity is needed as a positive electrode (cathode), which is crucial in limiting the energy output of SIBs \cite{You_AEM_18}. The cathode materials generally consist of 3$d$ transition metals in order to achieve high specific capacities owing to their low molecular weight. The transition metal acts as the deciding factor for the voltage of the cathode provided the local environment remains invariant. In an octahedron environment where the electric field created by surrounding ligands, the five 3$d$ orbitals of transition metal split into two non-degenerate higher energy $e_g$ set (consisting of $d_{z^2}$ and $d_{x^2-y^2}$) and the lower-energy $t_{2g}$ set (consisting of $d_{xy}$, $d_{xz}$ and $d_{yz}$) \cite{Zuckerman_JCE_65}. The observed splitting in orbitals is caused due to the repulsion between the ligand orbitals and $d-$orbitals, resulting in orbitals closer to the ligands having higher energy. The extent of this splitting is further influenced by the oxidation state of the transition metal, the nature of the ligand, and the ligand's geometrical arrangement around the transition metal. The $e_g$ and $t_{2g}$ will have bonding and antibonding molecular orbitals denoted as $e_g^*$ and $t_{2g}^*$ which are even higher in energy as this requires additional pairing energy for insertion of an electron in these antibonding orbitals. This combined effect of crystal field theory and molecular orbital theory lays the foundation for more advanced ligand field theory, which considers the interaction between metal $d-$orbitals and surrounding ligands through both electrostatic and covalent effects, helping to explain the splitting of $d-$orbitals in various geometries (e.g., octahedral, tetrahedral) and the resulting electronic configurations \cite{Liu_MT_16}. With an increase in the atomic number of the transition metal, the redox voltage is expected to be elevated under the same local structure of metal ion and the same oxidation states participating in redox, with the exception of the case between Mn and Fe. The answer lies in the electronic configuration of the Fe$^{2+}$ and Mn$^{2+}$ ions having $d^6$ and $d^5$ configuration, respectively. The sixth electron in Fe will be occupying the high energy $t_{2g}^*$ orbitals, which requires electron pairing energy, therefore reducing the gap between the Fermi level of sodium and the electronic state involved in the redox reaction [see Fig.~\ref{Abstract}]. It is easier for the Fe$^{2+}$/Fe$^{3+}$ redox to reach a half-filled stable $d^5$ configuration. While Mn$^{2+}$ does not have to pay any energy cost and is already in stable $d^5$ configuration, which will require more energy to reach a higher oxidation state \cite{Muraliganth_JPCC_10}.

Other important factors like the ionic state, size, electronegativity, and the local surroundings of the cations in cathode materials affect the electrochemical potential because they influence the types of bonds formed between the ligands and transition metals. The electrochemically stable window of the electrolytes is also a crucial factor in battery research, which enables electrodes to undergo their redox activities safely. The redox potential of the anode and cathode must reside within the electrolyte stability window to prevent significant electrolyte degradation. The electrolyte stability window [see Fig.~\ref{Figure1}(a)] refers to the potential range between electrolyte reduction at negative potentials and electrolyte oxidation at positive potentials, which is different from the potential range between the highest occupied molecular orbital (HOMO) and the lowest unoccupied molecular orbital (LUMO) \cite{Goodenough_CM_10, Peljo_EES_18}. If the redox activity of the electrode falls beyond this window, the side reactions from the interfacial oxidation/reduction of electrolyte by electrode catalytic reaction result in electrode-electrolyte interphase, also known as solid-electrolyte interface (SEI). The SEI is electrically insulating and ionically conducting, isolating electrodes and electrolytes from electronic contact and further prohibiting side reactions. The transfer of solvated sodium ions from the electrolyte to the bulk electrode occurs across this interface and involves an associated energy barrier. A lower barrier enhances sodium-ion kinetics, thereby supporting faster charging of sodium-ion batteries. Note that a stable SEI is also a critical aspect in achieving better Coulombic efficiency, electrolytic stability, and long cycle-life \cite{Li_ACSEL_21}.

In polyanionic materials, the strong X--O covalent bonds offer a large inductive effect, which pushes electron density away from the transition metal (M) center in the M--O--X structure [see Fig.~\ref{Figure1}(b)]. This redistribution of electron density weakens the covalent nature of the M--O bonds and contributes to the higher potential observed in polyanionic cathodes \cite{Gutierrez_CM_13}. The local coordination of transition metal ions and their connection with polyanionic units (sharing corner/edge/face and M--O--X bond angle) plays a key role in determining redox energies. When the MO$_6$ octahedra and XO$_4$ tetrahedra share edges or faces instead of just corners, the distance between them decreases, resulting in reduced covalency (higher voltage compared to Na$^+$/Na). Additionally, alkali metal ions located near the redox center of transition metal ions induce a secondary inductive effect, thereby altering the covalency of the bonds \cite{Gutierrez_CM_13, Melot_ACSAMI_14}. For cathode materials containing multiple transition metal ions, the changes in the local electronic structure further affect the redox potential by influencing the covalency \cite{Huang_JACS_24}. So, there will be an overall effect of each factor (covalency, local coordination, and corner/edge/face sharing) on the redox potential that will assess the M--O covalency along with the ligand field theory considerations.

\section{\noindent~Phosphate-based Cathodes}

Note that the stable and strong bond between phosphorus and oxygen contributes to both chemical and structural stability, along with the open framework giving the polyanionic compounds the ability to reversibly accommodate sodium ions within their crystal structures, also resulting in minimal volumetric changes. However, these properties come at the cost of reduced theoretical capacity because of the substantial framework of phosphate cathodes \cite{Hautier_CM_11}. The basic structure of polyanionic cathode materials builds up on the PO$_4$ tetrahedral units and MO$_6$/MO$_4$F$_2$ octahedra units [see Fig.~\ref{Figure1}(c)]. Their respective arrangement creates unique crystallographic sites for sodium, which are further classified by their position and oxygen coordination. The polyanionic electrode materials are intercalation-based active hosts for sodium ions to store the energy in the form of an electrochemical reaction. This puts a requirement on the polyanionic cathodes to have intrinsic electronic as well as ionic conductivities, which should be in sync to have smooth redox activity during the charge storage process. In polyanionic electrode materials, electronic transport primarily occurs via polaron hopping between transition-metal sites rather than through band-like conduction, resulting in inherently low electronic conductivity. The ease of small-hole polaron formation at the transition-metal site significantly affects its mobility, which is strongly influenced by the nature of the transition metal involved. Another key factor is the connectivity between adjacent MO$_6$/MO$_4$F$_2$ sites. If these sites are directly linked across the crystal lattice, polaron hopping is more efficient. In contrast, when resistive units like PO$_4$ or P$_2$O$_7$ groups interrupt the path between metal sites, the polaron encounters higher resistance, ultimately reducing the material’s overall electrochemical performance \cite{Johannes_PRB_12, Luong_JSAMD_22}. To tackle this and to enhance the surface electronic conduction, carbon coatings, nanostructuring, and surface modifications of polyanionic materials seem to become a common practice and effective ways \cite{Tiwari_CEJ_23, Chung_NM_02}. The ionic conduction of sodium ions between the unique sites follows the classical diffusion model, where the sodium ion conducts by hopping from one position in the lattice to another via interconnected diffusion pathways within the crystal's structural framework \cite{He_NatCom_17}. Here, the connectivity of the polyhedra and the octahedra and their relative arrangement in the crystal structure constructs these pathways for the sodium ion migration, and the adjacent sodium sites are connected through the structural faces formed by the neighboring lattice oxygen [visually depicted in Fig.~\ref{Figure1}(d)]. The area of these structural faces (also called the bottleneck) and the separation length of these sites determine the ease of Na ion hopping, which in turn relates to the activation energy for the ionic conduction \cite{Haarmann_SSI_21}. Additionally, the ionic conduction in polyanionic cathodes follows the concerted migration of multiple ions, which offers lower migration energy relative to single-ion migration, as represented in Fig.~\ref{Figure1}(e) \cite{He_NatCom_17}. The ionic conductivity of the sodium ions within the bulk electrode is a crucial parameter in designing future cathode materials that can follow up with the high current rates \cite{Chen_S_25}. 

Also, during the de/intercalation process, a redox reaction occurs, which results in a change in the oxidation state of the transition metal. The charge transfer associated with this redox process is often the rate-limiting step, as it involves both sodium-ion migration and electronic conduction \cite{Johannes_PRB_12, Chen_S_25}, and if either of these processes is compromized, the electrochemical reaction becomes inefficient at high current rates, leading to significant polarization. Moreover, the change in the oxidation state is accompanied by a variation in the metal's ionic radius, causing dynamic volumetric changes in the crystal structure. These structural shifts impact the size and connectivity of the ionic channels, which in turn affects ion transport. For stable electrochemical performance, these small volumetric changes must remain reversible during the repeated insertion and extraction of sodium ions. The transport of alkali-metal ions plays a key role in the electrochemical behavior of cathode materials. This process is collectively influenced by factors such as migration energy, dimensional characteristics, and potential lattice defects within the crystal structure. For efficient battery operation, the migration energy required for ion hopping should be minimal. If migration energy is high, it increases resistance to ion movement, leading to higher polarization during battery cycling and potentially compromising the electrochemical activity \cite{Liu_MT_16}. Further, reducing grain size to sub-micron or nanometer scales is often suggested as a means to shorten ion migration pathways and enhance electrochemical performance \cite{Bruce_ACIE_08}. However, this reduction in grain size requires advanced synthesis techniques, which may complicate the production process. Now, we discuss the development of cathode materials having different phosphate groups and the effect of their structural environment on the energy density of SIBs. 

\subsection{~PO$_4$}

\begin{figure*} 
\includegraphics[width=6in]{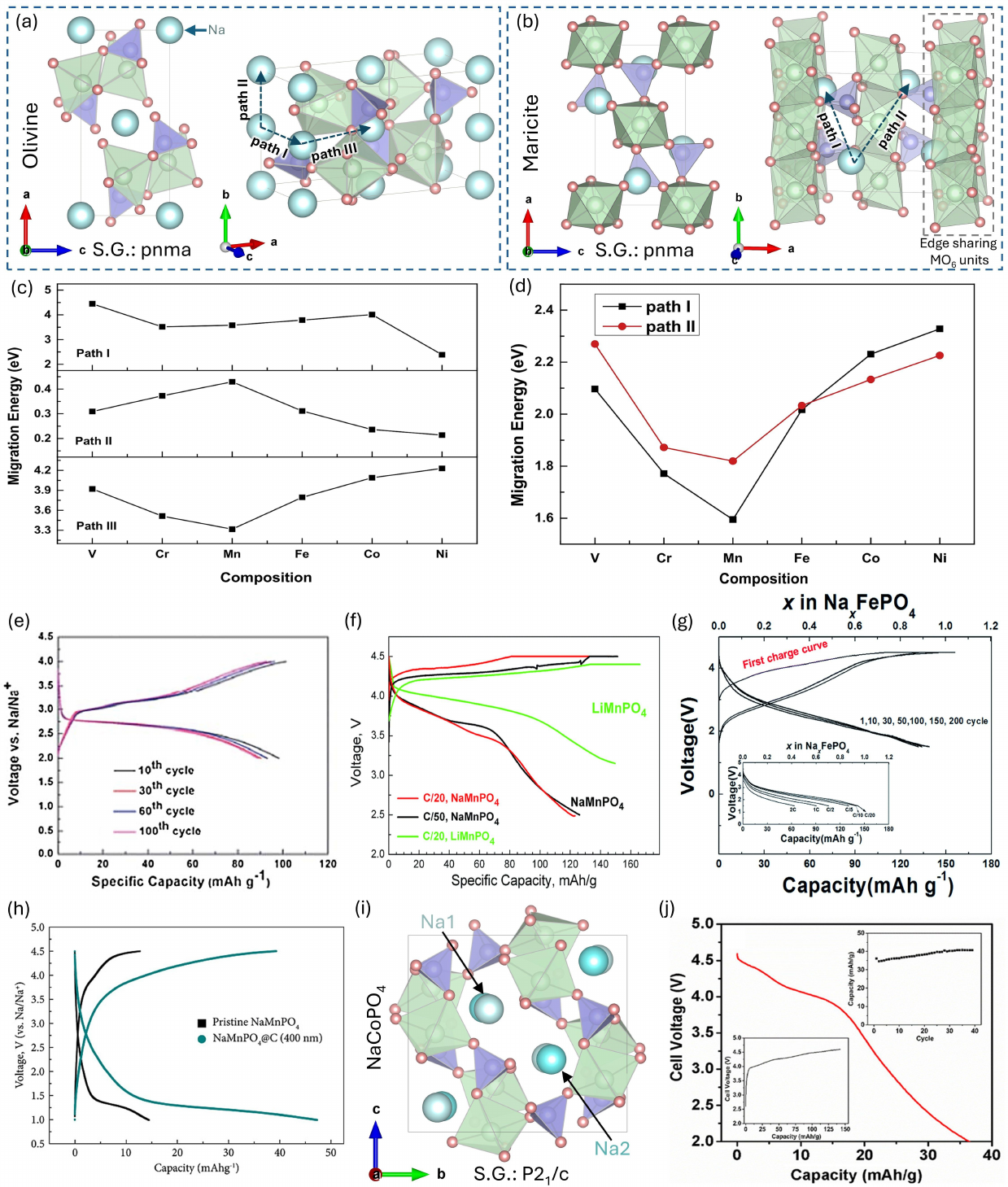}
\caption {The crystal structure of the NaFePO$_4$ along the $b$ direction and arrows indicating the plausible sodium ion diffusion pathways in (a) Olivine, and (b) Maricite phase, respectively, adapted from \cite{Zhu_JPS_19}. The illustration of migration energy of sodium ions along different paths in (c) Olivine and (d) Maricite phase, respectively \cite{Zhu_JPS_19}. The GCD profiles of (e) Olivine NaFePO$_4$ after 10$^{th}$, 30$^{th}$, 60$^{th}$, and 100$^{th}$ cycle \cite{Zhu_NS_13}, (f) Olivine NaMnPO$_4$ (against lithium) \cite{Boyadzhieva_RSCA_15}, (g) Maricite NaFePO$_4$ \cite{Kim_EES_15_Fe}, and (h) Maricite NaMnPO$_4$  pristine and milled-coated \cite{Mohsin_IJE_23}. (i) The crystal structure of the red phase NaCoPO$_4$, and (j) the corresponding GCD discharging profile along with the charging profile and capacity retention shown in the inset \cite{Gutierrez_ACSAMI_17}. The CIF file to construct (a) is taken from (mp-746030), (b) from (mp-19226), (i) from (mp-562436) \cite{Jain_APLM_13}.}
\label{NaMPO4}
\end{figure*}

The sodium-based NaMPO$_4$ cathode materials draw inspiration from the widely used LiFePO$_4$ (LFP) cathode in the lithium-ion battery industry. With a theoretical capacity of 170 mAh/g and a working voltage of 3.3 V, the LFP serves as a benchmark for SIBs to match or surpass to emerge as competitive alternatives of LIBs \cite{Padhi_JES_97}. The NaMPO$_4$ materials typically crystallize in the orthorhombic Pnma space group, characterized by a three-dimensional framework of interconnected NaO$_6$ octahedra, MO$_6$ octahedra, and PO$_4$ tetrahedra linked by shared corners and edges. The NaMPO$_4$ cathode offers one sodium ion per formula unit for reversible storage corresponding to M$^{2+}$/M$^{3+}$ (M= Fe, Mn, Co, and Ni) redox offering approx $\sim$154 mAh/g theoretical capacity based on the reversible redox process outlined below: 
\begin{equation}
\small
NaM^{(II)}PO_4 \rightleftharpoons Na^+ + M^{(III)}PO_4 + e^-
\end{equation} 
Two prominent polymorphs, Olivine (O) and Maricite (M), share the same space group but differ in the spatial arrangement of sodium ions and transition metals, as shown in Figs.~\ref{NaMPO4}(a, b). Here, we use crystallographic data from Materials Studio and parent article \cite{Jain_APLM_13} for structural information, and VESTA software was employed for visualization \cite{Momma_JAC_11}. In the Olivine phase, sodium ions occupy the 4$a$ sites, forming NaO$_6$ octahedra that create edge-sharing chains along the $b-$axis. This structural arrangement influences both the physical and chemical properties of the crystal, distinguishing it from the Maricite phase, where sodium ions and transition metals swap places. The Olivine phase is less thermally stable and easily transforms into the high-temperature stable Maricite phase above 480\degree C \cite{Liao_ESM_24}. The positioning of these ions within the lattice is crucial in determining the polymorph and its properties, making NaMPO$_4$ materials particularly interesting for SIB technology \cite{Ong_EES_11}.

Ion diffusion studies reveal distinct pathways in these phases; for example, the Olivine phase offers three potential routes for sodium migration, as shown in Fig.~\ref{NaMPO4}(a), with path II along the $b-$axis providing efficient conduction consisting of a one-dimensional sodium migration pathway. However, paths I and III present significant energy barriers, limiting ionic movement along these directions. Conversely, the Maricite phase, depicted in Fig.~\ref{NaMPO4}(b), features two pathways, but both exhibit high migration energies, rendering the phase electrochemically inactive as the Maricite phase lacks interconnected sodium-ion sites, effectively preventing sodium hopping. The ionic radii and the nature of the transition metal also influence the pathways, which in turn changes the migration energies, and the theoretically calculated values are illustrated in Fig.~\ref{NaMPO4}(c, d). There are also high possibilities of anti-site disorder between Fe and Na in the Olivine structure, which also could negatively impact ionic diffusion. Thus, the differences in ionic arrangements and diffusion pathways between Olivine and Maricite phases highlight the critical impact of crystal structure on the electrochemical performance of NaMPO$_4$ materials \cite{Zhu_JPS_19}. The expected activity of Olivine-NaFePO$_4$ vs. Na$^+$/Na should be 0.3 V lower (at 3.0 V) than the redox of LiFePO$_4$ vs. Li$^+$/Li (3.3 V). However, the galvanostatic charge-discharge (GCD) profile shown in Fig.~\ref{NaMPO4}(e) depicted that the average potential is around 2.7 V with a capacity of 100 mAh/g at 0.1C and can maintain 90\% of this capacity after 100 cycles \cite{Zhu_NS_13} where it is noticeable that the size of the sodium causes significant polarization in the GCD profile. 

The Olivine-NaMnPO$_4$ is expected to have electrochemical activity of Mn$^{2+}$/Mn$^{3+}$ redox at 3.59 V vs. Na$^+$/Na. However, due to the bigger size of Mn$^{2+}$ ion, the structure offers higher sodium ion migration energy [see Fig.~\ref{NaMPO4}(c)] compared to other transition metals \cite{Zhu_JPS_19}. When the O-NaMnPO$_4$ cathode material was tested against lithium, the GCD profiles were observed to be almost similar to O-LiMnPO$_4$. Still, a higher polarization was observed for the O-NaMnPO$_4$ and with lower initial discharge capacity (nearly 120 mAh/g), as shown in Fig.~\ref{NaMPO4}(f) \cite{Boyadzhieva_RSCA_15}. Interestingly, despite being regarded as electrochemically inactive, Kim {\it et al.} reported reversible intercalation of sodium ions in the stable Maricite structure of NaFePO$_4$ by nano-sizing the active material, in turn lowering the diffusion pathways and achieving 142 mAh/g capacity at 0.05 C, equivalent to 92\% of its theoretical capacity. However, their experimental analysis revealed that initial desodiation results in the formation of amorphous FePO$_4$ having sodium diffusion pathways of lower activation energies ($\sim$0.73 eV). Further, de/sodiation retains the amorphous-FePO$_4$ structure, which is evident from the GCD profiles depicted in Fig.~\ref{NaMPO4}(g), showing stable operation giving 95\% capacity retention after 200 cycles \cite{Kim_EES_15_Fe}. When the Maricite phase of NaMnPO$_4$ was tested, a significant overpotential of approximately 2.5 V was observed in the GCD curves. The poor electrochemical performance of M-NaMnPO$_4$ is largely attributed to the high energy barrier for sodium-ion migration. As shown in Fig.~\ref{NaMPO4}(d), the calculated migration energy for the Maricite structure is lowest for manganese, but it remains too high for efficient storage operations \cite{Zhu_JPS_19}. This results in substantial polarization in the GCD profile, as illustrated in Fig. \ref{NaMPO4}(h). Although the milled-coated cathode material shows a better discharge capacity of 47 mAh/g compared to pristine material, it suffers from extremely low energy efficiency \cite{Mohsin_IJE_23}. 

Alike, the NaFePO$_4$ and NaMnPO$_4$, the NaCoPO$_4$ crystallizes in three polymorphs namely $\alpha$ (M), $\beta$, and red NaCoPO$_4$ having space groups Pnma, $P6_5$, and $P2_1/c$, respectively. The $\alpha$-NaCoPO$_4$ is isostructural with Maricite NaFePO$_4$ and NaMnPO$_4$, but no electrochemical study is reported. At the same time, the red NaCoPO$_4$ polymorph is completely different [see Fig.~\ref{NaMPO4}(i)], consisting of two Co sites in 5-fold symmetry, labeled as Co1 and Co2, connected with corner and edge-sharing with PO$_4$, respectively \cite{Gutierrez_ACSAMI_17}. Two distinguishable sodium sites have been identified; both sites are in corner and edge-sharing with PO$_4$ and CoO$_5$, respectively. Notably, one sodium site exhibits edge sharing with a PO$_4$ unit, setting it apart from the other. This arrangement yields an open structure having expanded channels aligned parallel to the $a-$axis, which facilitates efficient sodium ion diffusion. The GCD discharging curve of red NaCoPO$_4$ is shown in Fig.~\ref{NaMPO4}(j), and the charging curve is shown in the inset. The first charge reaches a capacity of 135 mAh/g (theoretical capacity 154 mAh/g), with two distinctive plateaus at a high potential of 4.1 and 4.4 V. However, during discharge, only 35 mAh/g capacity was achieved because of the partial activation of the Co$^{2+}$/Co$^{3+}$ redox, confirmed using the XANES measurement, where only a small shift of $\sim$0.5 eV was observed in the Co K-edge, but is found to be reversible. This reversibility can also be seen in the capacity retention with the number of cycles, shown in the inset of Fig.~\ref{NaMPO4}(j). The NaNiPO$_4$ is also found to be stable in both Olivine and Maricite phases at calcination temperatures of 400$\degree$C and 550$\degree$C, respectively. However, only the capacitance study is performed, and no other electrochemical study is found in the literature. Analogously, the Olivine phase is found to give a higher capacitance of 85 F/g and retention of nearly 100\% when compared to the Maricite phase, giving 70 F/g capacitance and retention of only 40\% after 50 cycles \cite{Minakshi_NS_16}. Overall, the NaMPO$_4$ category offers high theoretical capacity among phosphate-based cathodes, with a TM content of approximately 32.13 wt.\%, which is comparable to that of LiFePO$_4$ (~35.4 wt.\%) \cite{Liu_MF_23}. Among the NaMPO$_4$ series, the NaFePO$_4$ stands out as the only candidate exhibiting decent electrochemical activity. However, its energy density remains significantly lower than that of LFP, primarily due to its low working voltage and pronounced polarization. To address the sluggish kinetics associated with heavier sodium ions, a more suitable host framework needs to be explored that can better accommodate and facilitate Na$^+$ transport. 

\begin{figure*} 
\includegraphics[width=5.8in]{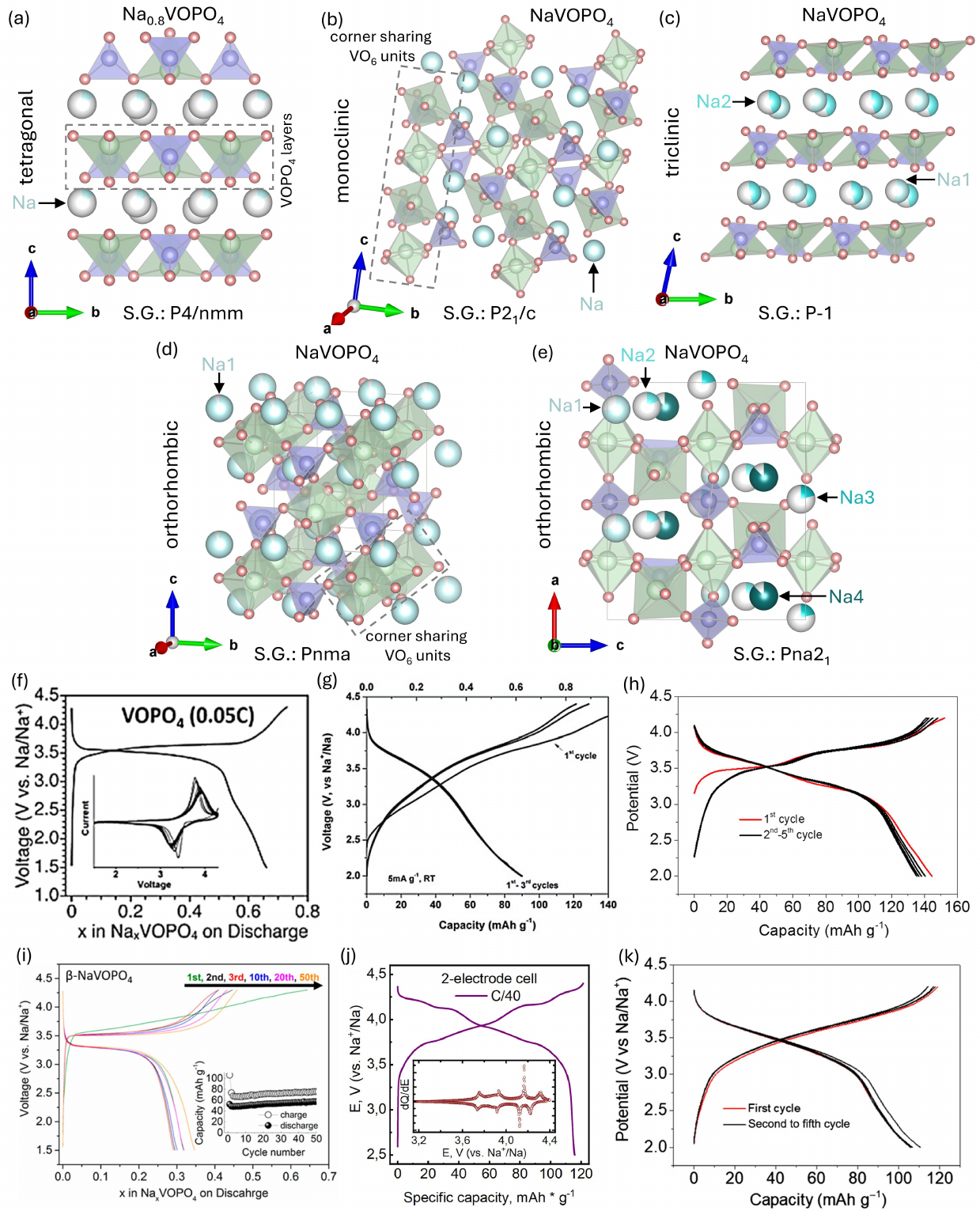}
\caption {The crystal structure of NaVOPO$_4$ in (a) tetragonal (Na$_{0.8}$VOPO$_4$), (b) monoclinic, (c) triclinic phase, (d) $\beta$-orthorhombic (space group $Pnma$), (e) orthorhombic ($Pna2_1$) and the corresponding GCD profile of (f) tetragonal phase (cyclic voltametery is in inset) \cite{He_CM_16_T}, (g) the monoclinic phase for 1$^{st}$--3$^{rd}$ cycles \cite{Song_CC_13}, (h) triclinic phase for five cycles \cite{Fang_C_18}, (i) $\beta$-orthorhombic ($Pnma$) for 1$^{st}$-50$^{th}$ cycles and cycling stability in inset \cite{He_CM_16_O}, (j) orthorhombic ($Pna2_1$) at C/40 current rate (dQ/dE shown in inset) \cite{Shraer_ESM_24}, and (k) amorphous phase for first five cycles \cite{Fang_CCSC_21}. The CIF file to construct (a) is taken from \cite{He_CM_16_T}, (b) from \cite{Song_CC_13}, (c) from \cite{Fang_C_18}, (d) from \cite{He_CM_16_O}, and (e) from \cite{Shraer_ESM_24}.}
\label{NaVOPO4}
\end{figure*}

Furthermore, the oxy-orthophosphate NaVOPO$_4$ cathode material exists in various polymorphs namely: $\alpha_1$-tetragonal ($P4/nmm$, sodiated VOPO$_4$), $\alpha$-monoclinic ($P2_1/c$), triclinic ($P\bar{1}$), $\beta$-orthorhombic ($Pnma$), and KTiOPO$_4$ (KTP)-type orthorhombic ($Pna2_1$) \cite{He_CM_16_T, Song_CC_13, Fang_C_18, He_CM_16_O, Shraer_ESM_24}. These crystalline polymorphs are reported to exhibit electrochemical activity, while the amorphous phase was also reversibly accommodating sodium ions \cite{Fang_CCSC_21}. The VOPO$_4$ in its pristine form offers a theoretical capacity of 165 mAh/g and delivers 150 mAh/g at 0.05 C with an average voltage of 3.5 V, corresponding to V$^{4+}$/V$^{5+}$ redox. Structurally, the compound adopts a layered form consisting of VOPO$_4$ layers with tetragonal symmetry and space group $P4/n$, where the VO$_5$ complex has a square pyramidal configuration. In this arrangement, ligand field splitting causes the $d_{x^2-y^2}$ orbital to be the highest energy $d-$orbital, which lies in the plane of the square base. The other $d-$orbitals ($d_{xy}$, $d_{xz}$, $d_{yz}$, and $d_{z^2}$) are lower in energy, with the $d_{z^2}$ orbital typically being slightly higher than $d_{xy}$, $d_{xz}$, and $d_{yz}$ due to its interaction with the axial oxygen ligand \cite{Jurca_JACS_11}. However, in the tetragonal and triclinic phases, we find in refs.~\cite{He_CM_16_T, Fang_C_18} that the V ion is slightly above the plane of the square base, causing the $d_{xy}$ orbital to have the lowest energy, while the energy levels of $d_{xz}$ and $d_{yz}$ lie between the $d_{xy}$ and $d_{z^2}$ orbitals \cite{Jurca_JACS_11}. The V ion retains the same square pyramidal configuration in the chemically sodiated Na$_{0.8}$VOPO$_4$ and adopts the $P4/nmm$ space group under the same tetragonal symmetry, with sodium ions residing between the VOPO$_4$ layers, see  Fig.~\ref{NaVOPO4}(a) \cite{He_CM_16_T}. 

The sodium-rich NaVOPO$_4$ has a theoretical capacity of 145 mAh/g corresponding to one sodium ion and the V$^{4+}$/V$^{5+}$ redox couple following the reversible reaction: 
\begin{equation}
\small
NaV^{(IV)}OPO_4 \rightleftharpoons Na^+ + V^{(V)}OPO_4 + e^-
\end{equation} 
The monoclinic phase of as-synthesized sodium-rich NaVOPO$_4$ stabilizes in $P2_1/c$  space group, and the respective crystal structure is depicted in Fig.~\ref{NaVOPO4}(b) \cite{Song_CC_13}. Additionally, Fang {\it et al.} stabilized the triclinic phase of NaVOPO$_4$ by solvothermal reducing VOPO$_4$·2H$_2$O with NaI, resulting in an interlayer distance of 5.698 \AA~and the $P\bar{1}$ space group \cite{Fang_C_18}. The tetragonal and triclinic phases are layered structures consisting of VOPO$_4$ layers made up of distorted VO$_5$ pentahedra and PO$_4$ tetrahedra via corner-sharing, and the Na ions can freely migrate in between the $ab$ planes, facilitating the 2D sodium ion migration [see Fig.~\ref{NaVOPO4}(a, c)]. The VO$_5$ pentahedra features four identical bonds and one characteristic short vanadyl bond where the distance between the oxygen atom opposite the vanadyl bond is large enough to be considered non-bonding with VO$_5$ pentahedra from adjacent layers \cite{He_CM_16_T, Fang_C_18}. However, in the triclinic phase, the VOPO$_4$ layers are glided in such a way that the oxygen atom of the short vanadyl bond is not directly below the VO$_5$ pentahedra of the next layers, which is not in the case of the tetragonal phase where the oxygen atom is right below VO$_5$ pentahedra of the next layers. The VO$_6$ is also distorted in the monoclinic phase, featuring a short vanadyl bond of 1.681 \AA and a longer bond of 2.091 \AA, along with four equatorial V--O bonds. This structure comprises chains of corner-sharing VO$_6$ octahedra aligned along the $c-$axis, interconnected by PO$_4$ tetrahedra, which facilitate the three-dimensional movement of sodium ions \cite{Song_CC_13}. However, it is important to emphasize that the open 3D structure does not necessarily result in rapid ion diffusion within the monoclinic phase. In fact, the ionic conductivity of Na ions in the 2D and 1D channels of the tetragonal and $\beta$-orthorhombic structures, respectively, is superior to that in the monoclinic structure \cite{He_CM_16_O}. This is despite the fact that the $\beta$-orthorhombic phase features distorted VO$_6$ infinite chains [as shown in Fig.~\ref{NaVOPO4}(d)], which are similar to those in the monoclinic phase. These chains are aligned along $a-$direction and connected via corner-sharing PO$_4$ units along the $bc$ plane, forming tunnels along the [010] direction that facilitate preferential sodium ion diffusion \cite{He_CM_16_O} The novel KTP-type-orthorhombic phase, illustrated in Fig.~\ref{NaVOPO4}(e), is constructed from helical chains of two types of VO$_6$ octahedra sharing corners, which are interconnected by corner sharing PO$_4$ tetrahedra. Like the monoclinic and $\beta$-orthorhombic phase, the KTP-orthorhombic structure also features infinite VO$_6$ chains in a helical structure formed through vanadyl bonds. The arrangement of oxygen atoms shared between octahedral corners distinguishes two types of VO$_6$ octahedra. Here, there are four distinct partially occupied Na sites interconnected via 3D spacious channels of this robust framework \cite{Shraer_ESM_24}. 

Interestingly, the tetragonal VOPO$_4$ and triclinic NaVOPO$_4$ exhibit nearly identical discharge/charge profiles, both with an average voltage of 3.5 V. At a rate of 0.05 C, the GCD profile presented in Fig.~\ref{NaVOPO4}(f) shows that the tetragonal VOPO$_4$ achieves a capacity of 150 mAh/g \cite{He_CM_16_T}. Whereas the monoclinic phase only reaches a capacity of 90 mAh/g at 1/15 C [see Fig.~\ref{NaVOPO4}(g)], with an average potential near 3.6 V. The sloppy GCD profile could relate to the poor intrinsic ionic conductivity, as they have also reported that the non-ball-milled sample delivers only a capacity of 20 mAh/g compared to 90 mAh/g of ball-milled sample \cite{Song_CC_13}. In contrast, the GCD curve of triclinic NaVOPO$_4$ delivers 144 mAh/g at 0.05 C, presented in Fig.~\ref{NaVOPO4}(h) \cite{Fang_C_18}. The high reversible capacity of the triclinic phase results from the broad interlayer spacing, which measures 5.698 \AA~ compared to the 5.119 \AA~ of tetragonal Na$_{0.8}$VOPO$_4$. The $\beta$-orthorhombic structure showed a large initial charge capacity, but during discharge, only $\sim$0.3 Na ions per f.u. could be reversibly inserted, as presented in Fig.~\ref{NaVOPO4}(i). The GCD profile shows a flat plateau at 3.3 V, but the obtained capacities are not satisfactory \cite{He_CM_16_O}. The KTP-type orthorhombic phase displayed a distinct multi-step feature in the GCD profile, see Fig.~\ref{NaVOPO4}(j). This characteristic is associated with the sequential redox reactions of two different types of V ion sites, resulting in a high average voltage of 3.93 V and a discharge capacity of 110 mAh/g at a rate of 0.1 C \cite{Shraer_ESM_24}. The GCD of the amorphous phase is featured in Fig.~\ref{NaVOPO4}(k), indicating a working voltage of 3.5 V with a specific capacity of 110 mAh/g at 0.05 C. It shows a sloppy GCD profile, evidence of its amorphous nature, while having some similarities with the tetragonal and triclinic phases. As the amorphous NaVOPO$_4$ was obtained from the crystalline tetragonal VOPO$_4$·2H$_2$O, which may account for its resemblance with the electrochemical performance of the layered structure of NaVOPO$_4$ \cite{Fang_CCSC_21}. This discussion highlights the superior performance of NaVOPO$_4$ in a layered structure, which offers enhanced 2D ionic conduction pathways and features elevated working voltage of V$^{4+}$/V$^{5+}$ redox with competitive specific capacities.

\begin{table*}[]
\centering
\caption{A summary of different parameters of phosphate-based polyanionic cathode materials utilizing PO$_4$ \& PO$_4$F groups.}
\label{tab:Summary1}
\resizebox{\textwidth}{!}{%
\begin{tabular}{lcccccccc}
\hline
Formula &
  \begin{tabular}[c]{@{}c@{}}Space\\ Group\end{tabular} &
  \begin{tabular}[c]{@{}c@{}}Active\\ Redox\end{tabular} &
  \begin{tabular}[c]{@{}c@{}}Working\\ Voltage\\ (V vs. Na/Na$^+$)\end{tabular} &
  \begin{tabular}[c]{@{}c@{}}Q$_{th}$(mAh/g)\\ (n Na/f.u.)\end{tabular} &
  \begin{tabular}[c]{@{}c@{}}Q$_{dis}$\\ (mAh/g)\end{tabular} &
  Electrolyte &
  \begin{tabular}[c]{@{}c@{}}Voltage\\ Window\\ (V vs. Na/Na$^+$)\end{tabular} &
  Ref. \\ \hline
\multicolumn{8}{c}{PO$_4$} &
   \\ \hline
O-NaFePO$_4$ &
  $Pnma$ &
  Fe$^{2+}$/Fe$^{3+}$ &
  2.7 &
  154 (1 Na) &
  120 (0.05 C) &
  \begin{tabular}[c]{@{}c@{}}1M NaClO$_4$ in EC/PC\\ (v/v 1:1) with 10\% FEC\end{tabular} &
  2.0-4.0 &
  \cite{Zhu_NS_13} \\ \hline
O-NaMnPO$_4$ &
  $Pnma$ &
  Mn$^{2+}$/Mn$^{3+}$ &
  \begin{tabular}[c]{@{}c@{}}$\sim$3.5\\ vs. Li$^+$/Li\end{tabular} &
  155 (1 Na) &
  85 (C/20) &
  \begin{tabular}[c]{@{}c@{}}1 M NaPF$_6$ in PC/FEC\\ (95:5 v/v ratio)\end{tabular} &
  2.5-4.5 &
  \cite{Boyadzhieva_RSCA_15} \\ \hline
M-NaFePO$_4$ &
  $Pnma$ &
  Fe$^{2+}$/Fe$^{3+}$ &
  $\sim$2.7 &
  154 (1 Na) &
  142 (C/20) &
  1 M NaPF$_6$ (1:1 EC/PC) &
  1.5-4.5 &
  \cite{Kim_EES_15_Fe} \\ \hline
M-NaMnPO$_4$ &
  $Pnma$ &
  Mn$^{2+}$/Mn$^{3+}$ &
  - &
  155 (1 Na) &
  47 (0.1 C) &
  \begin{tabular}[c]{@{}c@{}}1 M NaClO$_4$ in EC:DMC \\ 1:1 vol\% + 5 vol\% FEC\end{tabular} &
  1.0-4.5 &
  \cite{Mohsin_IJE_23} \\ \hline
NaCoPO$_4$ &
   $P2_1/c$ &
  Co$^{2+}$/Co$^{3+}$ &
  $\sim$4.1 &
  154 (1 Na) &
  35 (C/50) &
  \begin{tabular}[c]{@{}c@{}}1 M NaPF$_6$ in 3:7 (by wt.) \\ EC/EMC 2 wt \% FEC\end{tabular} &
  2.0-4.6 &
  \cite{Gutierrez_ACSAMI_17} \\ \hline
VOPO$_4$ &
   $P4/nmm$ &
  V$^{4+}$/V$^{5+}$ &
  3.4 &
  165 (1 Na) &
  150 (0.05 C) &
  \begin{tabular}[c]{@{}c@{}}1 M NaClO$_4$ in EC/PC\\ (1:1 v/v)\end{tabular} &
  1.5-4.3 &
  \cite{He_CM_16_T} \\ \hline
NaVOPO$_4$ &
   $P2_1/c$ &
  V$^{4+}$/V$^{5+}$ &
  3.6 &
  145 (1 Na) &
  90 (1/15 C) &
  1 M NaClO$_4$ in PC &
  2.0-4.4 &
  \cite{Song_CC_13} \\ \hline
NaVOPO$_4$ &
   $P\bar{1}$ &
  V$^{4+}$/V$^{5+}$ &
  3.5 &
  145 (1 Na) &
  144 (0.05 C) &
  \begin{tabular}[c]{@{}c@{}}1 M NaClO$_4$ in EC/DEC\\ (1:1 v/v)\end{tabular} &
  2.0-4.2 &
  \cite{Fang_C_18} \\ \hline
NaVOPO$_4$ &
   $Pnma$ &
  V$^{4+}$/V$^{5+}$ &
  3.3 &
  145 (1 Na) &
  $\sim$43 (0.05 C) &
  \begin{tabular}[c]{@{}c@{}}1 M NaClO$_4$ in EC/PC\\ (1:1 v/v)\end{tabular} &
  1.5-4.3 &
  \cite{He_CM_16_O} \\ \hline
NaVOPO$_4$ &
   $Pna2_1$ &
  V$^{4+}$/V$^{5+}$ &
  3.93 &
  145 (1 Na) &
  110 (0.1 C) &
  \begin{tabular}[c]{@{}c@{}}1 M NaPF$_6$ in EC:PC:FEC\\ (47.5:47.5:5 v/v)\end{tabular} &
  2.5-4.4 &
  \cite{Shraer_ESM_24} \\ \hline
NaVOPO$_4$ &
   amorphous &
  V$^{4+}$/V$^{5+}$ &
  3.5 &
  145 (1 Na) &
  110 (0.05 C) &
  \begin{tabular}[c]{@{}c@{}}1 M NaClO$_4$ in EC/DEC\\ (1:1 v/v)\end{tabular} &
  2.0-4.2 &
  \cite{Fang_CCSC_21} \\ \hline
\multicolumn{8}{c}{PO$_4$F} &
   \\ \hline
Na$_2$FePO$_4$F &
   $Pbcn$ &
  Fe$^{2+}$/Fe$^{3+}$ &
  3 &
  124 (1 Na) &
  110 (0.05 C) &
  \begin{tabular}[c]{@{}c@{}}1 M NaClO$_4$ in PC\\ with FEC additive\end{tabular} &
  2.0-3.8 &
  \cite{Kawabe_EC_11} \\ \hline
Na$_2$MnPO$_4$F &
   $P2_1/n$ &
  Mn$^{2+}$/Mn$^{3+}$ &
  3.6 &
  125 (1 Na) &
  102.3 (0.05 C) &
  \begin{tabular}[c]{@{}c@{}}1 M NaClO$_4$ in PC\\ with 5\% FEC\end{tabular} &
  1.5-4.5 &
  \cite{Wu_JPS_18} \\ \hline
Na$_2$CoPO$_4$F &
   $Pbcn$ &
  Co$^{2+}$/Co$^{3+}$ &
  4.3 &
  122 (1 Na) &
  107 (0.5 C) &
  \begin{tabular}[c]{@{}c@{}}1M NaPF$_6$ in EC:DMC:FEC\\ (49:49:2 vol\%)\end{tabular} &
  2.0-5.0 &
  \cite{Zou_ECSEL_15} \\ \hline
NaVPO$_4$F &
   $C2/c$ &
  V$^{3+}$/V$^{4+}$ &
  3.33 &
  143 (1 Na) &
  133 (0.1 C) &
  \begin{tabular}[c]{@{}c@{}}1M NaClO$_4$ in EC/PC\\ (v/v 1:1)\end{tabular} &
  2.0-4.2 &
  \cite{Law_ESM_18} \\ \hline
NaVPO$_4$F &
   $I4/mmm$ &
  V$^{3+}$/V$^{4+}$ &
  3.95 &
  143 (1 Na) &
  120.9 (0.05 C) &
  1 M NaClO$_4$ in PC &
  1.5-4.3 &
  \cite{Ruan_EA_15} \\ \hline
NaVPO$_4$F &
   $Pna2_1$ &
  V$^{3+}$/V$^{4+}$ &
  4 &
  143 (1 Na) &
  136 (14.3 mA/g) &
  \begin{tabular}[c]{@{}c@{}}1M NaPF$_6$ in EC:PC:FEC\\ (47.5:47.5:5 vol.)\end{tabular} &
  2.0-4.5 &
  \cite{Shraer_NatCom_22} \\ \hline
\end{tabular}%
}
\end{table*}

\subsection{~PO$_4$F}

\begin{figure*} 
\includegraphics[width=6.4in]{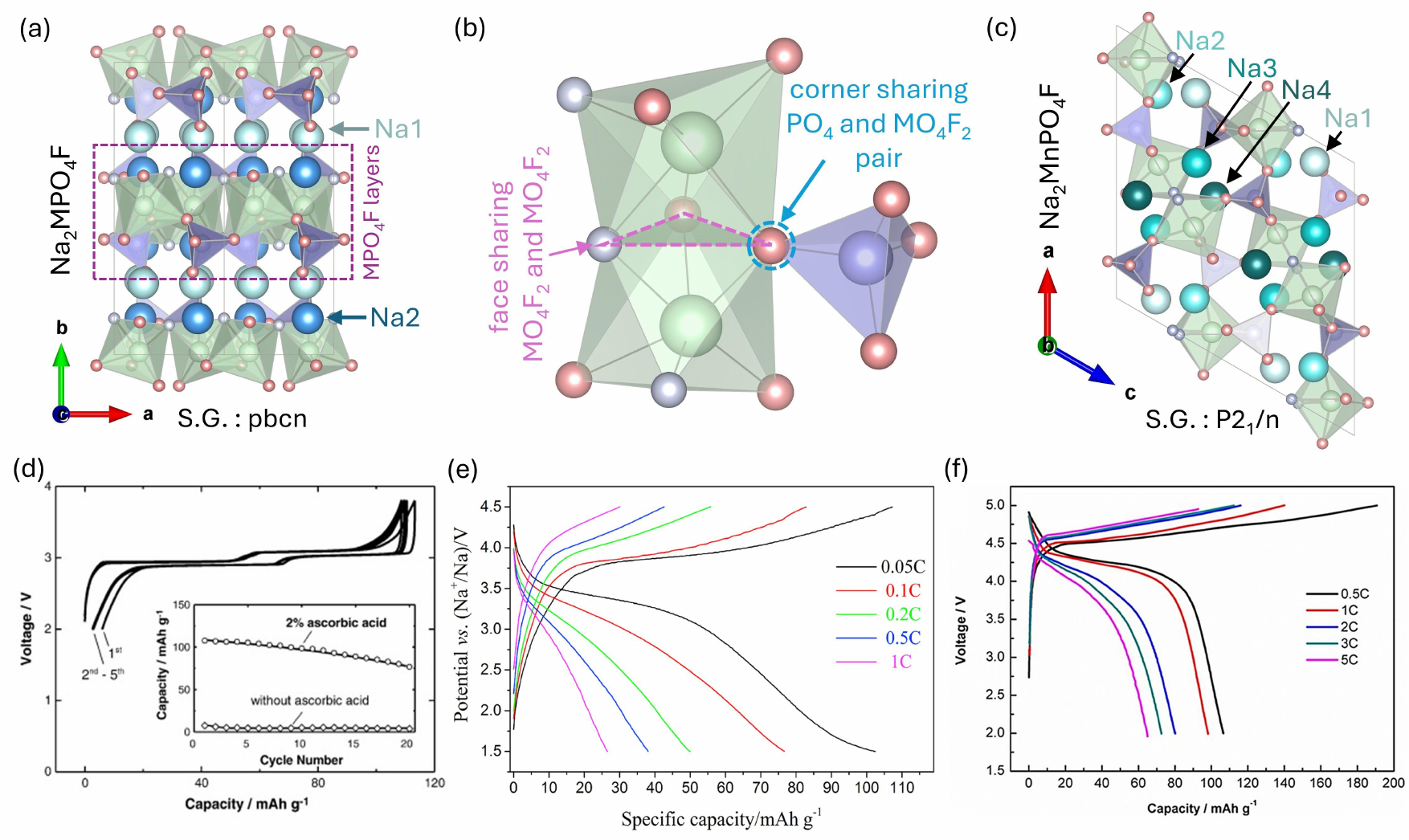}
\caption {The crystal structure of the (a) Na$_2$MPO$_4$F along the $c$ direction, (b) the polyhedra units showing the connectivity of the MO$_4$F$_2$ units and PO$_4$ units, and (c) the Na$_2$MnPO$_4$F along $b$ direction. The GCD profile of (d) the Na$_2$FePO$_4$F for five cycles and capacity retention of the samples synthesized with or without ascorbic acid is shown in the inset \cite{Kawabe_EC_11}, (e) the Na$_2$MnPO$_4$F  \cite{Wu_JPS_18} and (f) the Na$_2$CoPO$_4$F \cite{Zou_ECSEL_15} at various C rates. The CIF files to construct (a, b) are taken from (mp-1194940), and (c) from (mp-556404) \cite{Jain_APLM_13}.}
\label{Na2MPO4F}
\end{figure*}

Interestingly, by introducing fluorine (F) atoms, the charge balance and dimensionality of the crystal structure can be altered, which further enhances the inductive effect collectively with the PO$_4$ unit. The addition of F is also favorable in terms of its lower weight and elevating working voltages \cite{Ellis_NM_07}. The first in the category are the Na$_2$MPO$_4$F type cathode materials, which are expected to provide working voltage a step ahead of the NaMPO$_4$ type cathodes. The cathode materials stabilized in PO$_4$F framework are with M=Fe, Mn, Co, and Ni. Among these, the Na$_2$MPO$_4$F (M=Fe, Co, Ni) share an isostructural framework, crystallizing in the orthorhombic structure with a $Pbcn$ space group. The Na$_2$MnPO$_4$F adopts a 3D structure crystallizing in the monoclinic phase having $P2_1/n$ space group \cite{Ellis_CM_10, Wu_JMC_11}. The Na$_2$MPO$_4$F class type structures are sodium-rich, consisting of two sodium ions per formula unit, out of which one sodium ion is available for reversible storage via the M$^{2+}$/M$^{3+}$ redox [see Fig.~\ref{Na2MPO4F}(a)], as indicated by the reversible redox mechanism below: 
\begin{equation}
\small
Na_2M^{(II)}PO_4F \rightleftharpoons Na^+ + NaM^{(III)}PO_4F + e^-
\end{equation} 
The coordination environment of the transition metal involves six ligands, comprising four oxygen and two fluorine atoms (MO$_4$F$_2$), arranged in an octahedral configuration [see Fig.~\ref{Na2MPO4F}(b)]. In orthorhombic structure, these octahedra are connected through face sharing, forming M$_2$O$_6$F$_3$ bioctahedral units, which are interconnected through corner-sharing via fluorine atoms forming chains along the $a-$direction. Further, the bioctahedral chains are connected through corner-sharing PO$_4$ units along the $c-$direction to form a layered-type structure creating two sites for sodium ions between these layers, shown in Fig.~\ref{Na2MPO4F}(a). The layered structure allows the 2D sodium ion conduction connecting Na1 and Na2 sites along the $ac-$plane. The environment of these Na1 and Na2 sites is almost similar, with four oxygen and two fluorine coordination, and the angles between two of the Na1--F--M bonds are nearly 180$\degree$, while other Na1--O/F--M angles are close to 90$\degree$. The Na2 site is a little contracted with all the Na2--O/F--M bond angles around 90$\degree$ \cite{Ellis_CM_10, Li_ACIE_18}. The structural difference of Na$_2$MnPO$_4$F could be because of the bigger size of Mn$^{2+}$ ion, causing the distortion in the bioctahedral unit in which the MO$_4$F$_2$ octahedral units are connected via corner-sharing through fluorine atom instead of face sharing, forming Mn$_2$O$_8$F$_3$, which are connected with PO$_4$ tetrahedra by sharing a corner along $a$, $c$ axis to form the 3D crystal structure that stabilized in monoclinic phase, as depicted in Fig.~\ref{Na2MPO4F}(c). The distortion splits two sodium sites into four, each having a unique local environment \cite{Wu_JMC_11}. 

Addressing the electrochemical energy storage process of the orthorhombic phase of Na$_2$FePO$_4$F, the Na1 site remains inert during the charging/discharging. At the same time, the sodium ion is reversibly extracted from the Na2 site via a structural transformation with the intermediate phase Na$_{1.5}$FePO$_4$F having $P2_1/c$ symmetry and the end charged (NaFePO$_4$F)/ discharged (Na$_2$FePO$_4$F) states having the same $Pbcn$ space group \cite{Li_ACIE_18}. The participation of one sodium ion per f.u. provide the theoretical capacity of 124 mAh/g. Two distinct flat plateaus at 3.06 and 2.91 V are observed during discharging which belonged to Fe$^{2+}$/Fe$^{3+}$ redox with a capacity of 110 mAh/g at 0.05 C in a potential window of 2.0--3.8 V, also a very minimal voltage polarization between the dis/charging curves was observed, as depicted in Fig.~\ref{Na2MPO4F}(d) \cite{Kawabe_EC_11}. In the case of Na$_2$MnPO$_4$F, Ellis {\it et al.} did not find any electrochemical activity up to a potential of 5 V \cite{Ellis_CM_10}. But, later, it was observed to show an initial discharge capacity of nearly 100 mAh/g at a current density of 10 mA/g in a potential window of 1.5--4.6 V when investigated against lithium at an elevated temperature of 60\degree C \cite{Wu_JMC_11}. The poor kinetics and low intrinsic electronic conductivity were suspected to be the reasons behind the underperformance of this cathode material despite having open pathways available for sodium mobility. The morphological changes were done on this cathode material by synthesizing carbon-coated Na$_2$MnPO$_4$F hollow spheres, which delivered a specific discharge capacity of 102.3 mAh/g within a voltage platform around 3.6 V [see Fig.~\ref{Na2MPO4F}(e)] at a current rate of 0.05 C in the potential window of 1.5--4.5 V \cite{Wu_JPS_18}. However, the issues with high polarization and poor rate capability are still visible in the results. Also, the GCD profiles of isostructural Na$_2$CoPO$_4$F, in Fig.~\ref{Na2MPO4F}(f), show the plateau-like features and the redox activity of Co$^{2+}$/Co$^{3+}$ was observed at high voltage of 4.3 V accompanied with a capacity of 107 mAh/g at 0.5 C \cite{Zou_ECSEL_15}. The higher working voltage is expected from the electrochemical activity of Na$_2$NiPO$_4$F, but it may fall outside the stability window of organic electrolytes. Thus, no information about its electrochemical performance is known \cite{Ellis_CM_10}. The incorporation of F ligands can elevate the operating voltage, and the local crystal structure provides smoother charge transfer during redox processes, reducing polarization in the voltage profile. The presence of 2D ionic conduction channels, coupled with enhanced polaronic conduction along the corner-sharing M$_2$O$_6$F$_3$ chains, supports improved charge transfer in the case of Fe-based materials. However, for the Mn- and Co-based analogs, the polarization tends to increase, which may be due to the formation of less stable polarons compared to Fe \cite{Johannes_PRB_12}. 

\begin{figure*} 
\includegraphics[width=6.4in]{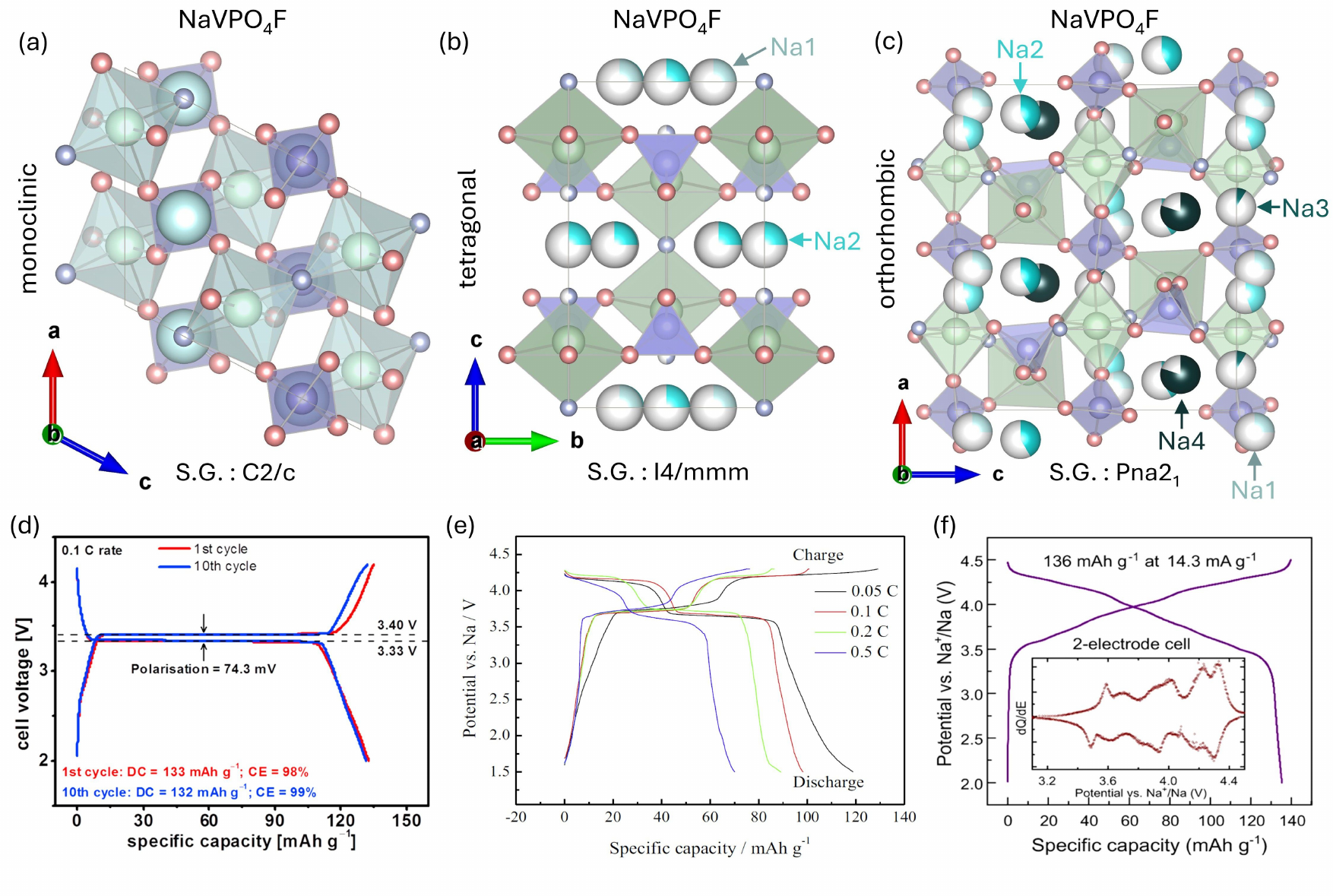}
\caption {The crystal structure of the NaVPO$_4$F in (a) monoclinic (tavorite), (b) tetragonal, and (c) orthorhombic phase, and the GCD profile in (d) monoclinic (non-tavorite) phase for 1st and 10th cycles \cite{Law_ESM_18}, (e) the tetragonal phase at various current rates \cite{Ruan_EA_15}, (f) the orthorhombic phase at a current density of 14.3 mA/g and the dQ/dE plot of the 2nd cycle is shown in the inset \cite{Shraer_NatCom_22}. The CIF file to construct (a) is taken from \cite{Boivin_JMCA_17}, (b) from \cite{Ling_ACSAMI_20}, and (c) from \cite{Shraer_NatCom_22}.}
\label{NaVPO4F}
\end{figure*}

Moreover, the cathode NaVPO$_4$F is quite intriguing and distinct from its previously discussed members. It exists in three polymorphs, namely monoclinic (m), tetragonal (t), and orthorhombic (o), and their respective crystal structures are given in Figs.~\ref{NaVPO4F}(a, b, c). An important highlight here is all of them are electrochemically active, and it is interesting to discuss how the materials having the same formula unit and varying crystal structure arrangement, in turn, modulate the electrochemical performance \cite{Boivin_JMCA_17, Law_ESM_18, Ling_ACSAMI_20, Shraer_NatCom_22}. The occurrence of these polymorphs is due to their thermal stability, where the tetragonal phase is stable below 650\degree C, which completely converts to the monoclinic phase above 750$\degree$C \cite{Ling_aem_21}. Shraer {\it et al.} successfully stabilized the NaVPO$_4$F in orthorhombic (o) phase at low sintering temperature 190\degree C by a solid-state ion exchange technique \cite{Shraer_NatCom_22}. Having shared the common formula unit and capability of contributing one sodium ion per f.u. for storage associated with the V$^{3+}$/V$^{4+}$ redox couple, they all have a theoretical capacity of around 143 mAh/g in accordance with the following reversible redox reaction: 
\begin{equation}
\small
NaV^{(III)}PO_4F \rightleftharpoons Na^+ + V^{(IV)}PO_4F + e^-
\end{equation} 
However, a significant change can be seen in the GCD profile, and the variation in the average voltage output varying from 3.33 V for the monoclinic \cite{Law_ESM_18}, 3.95 V for the tetragonal \cite{Ruan_EA_15}, and 4 V for the orthorhombic phase \cite{Shraer_NatCom_22}, as depicted in Fig. \ref{NaVPO4F}(d, e, f), the respective specific capacity values are 133 mAh/g at 0.1 C, 120.9 mAh/g at 0.05 C, and 136 mAh/g at 0.1 C. To gain insights into their individual electrochemical results, let us examine the local crystal structure where all these polymorphs share a common crystal structure units composed of corner-sharing VO$_4$F$_2$ octahedra and PO$_4$ tetrahedra linked through oxygen atoms, as shown in Fig.~\ref{NaVPO4F}(a, b, c). For clarity, the crystal structure shown in Fig. \ref{NaVPO4F}(a) is the tavorite type monoclinic structure with the same space group C2/c, as the crystal information of the non-tavorite type is not available. The electrochemical results of the monoclinic phase belong to the non-tavorite type, and the following structural information of the non-tavorite type is taken from the literature \cite{Li_AEM_18}.

On the other hand, the primary differences between these polymorphs arise from variations in their linking angles between V--P--V, and the corresponding sodium atoms arrangement categorizes each in different space groups, which in turn influences the working plateaus. For the monoclinic phase, with a V--P--V angle of 121.0\degree, the working plateau is at 3.4 V. In the tetragonal phase, which has a larger V--P--V angle of 161.9\degree, the two flat plateaus were observed, giving an average voltage of 3.95 V \cite{Ling_aem_21}. The orthorhombic phase, however, features two distinct V--P--V angles of 126.1\degree and 160.4\degree and exhibits a step-like multi-plateaus GCD profile with a working voltage of 4.0 V \cite{Shraer_NatCom_22}. This correlation suggests that the V--P--V bond angle affects the electronic structure and, consequently, the redox potential of NaVPO$_4$F. There are also claims of the non-existence of monoclinic and tetragonal phases of NaVPO$_4$F type cathodes, and depending on the synthesis temperature, exist as multiphase mixtures of Le Meins’ Na$_3$V$_2$(PO$_4$)$_2$F$_3$, unreacted VPO$_4$, and hexagonal Na$_3$V$_2$(PO$_4$)$_3$ \cite{Li_AEM_18}. The monoclinic phase is analogous to the NASICON structure, particularly Na$_3$V$_2$(PO$_4$)$_3$ with a matching flat GCD profile and the voltage level of V$^{3+}$/V$^{4+}$ redox at 3.4 V. In the tetragonal phase, the spatial configuration of VO$_4$F$_2$ and PO$_4$ corresponds with that of Na$_3$V$_2$(PO$_4$)$_2$F$_3$ and identical two sequential redox of V$^{3+}$/V$^{4+}$ at 3.6 and 4.2 V. Nonetheless, the o-NaVPO$_4$F cathode displayed 136 mAh/g capacity at a working voltage of 4.0 V, which is far ahead of other analogs of its category (133 and 120.9 mAh/g for monoclinic and tetragonal phase, respectively), its structural relative NaVOPO$_4$, and other V-based phosphates, in terms of capacity and voltage \cite{Shraer_NatCom_22}.

\subsection{~P$_2$O$_7$}

\begin{figure*} 
\includegraphics[width=6.2in]{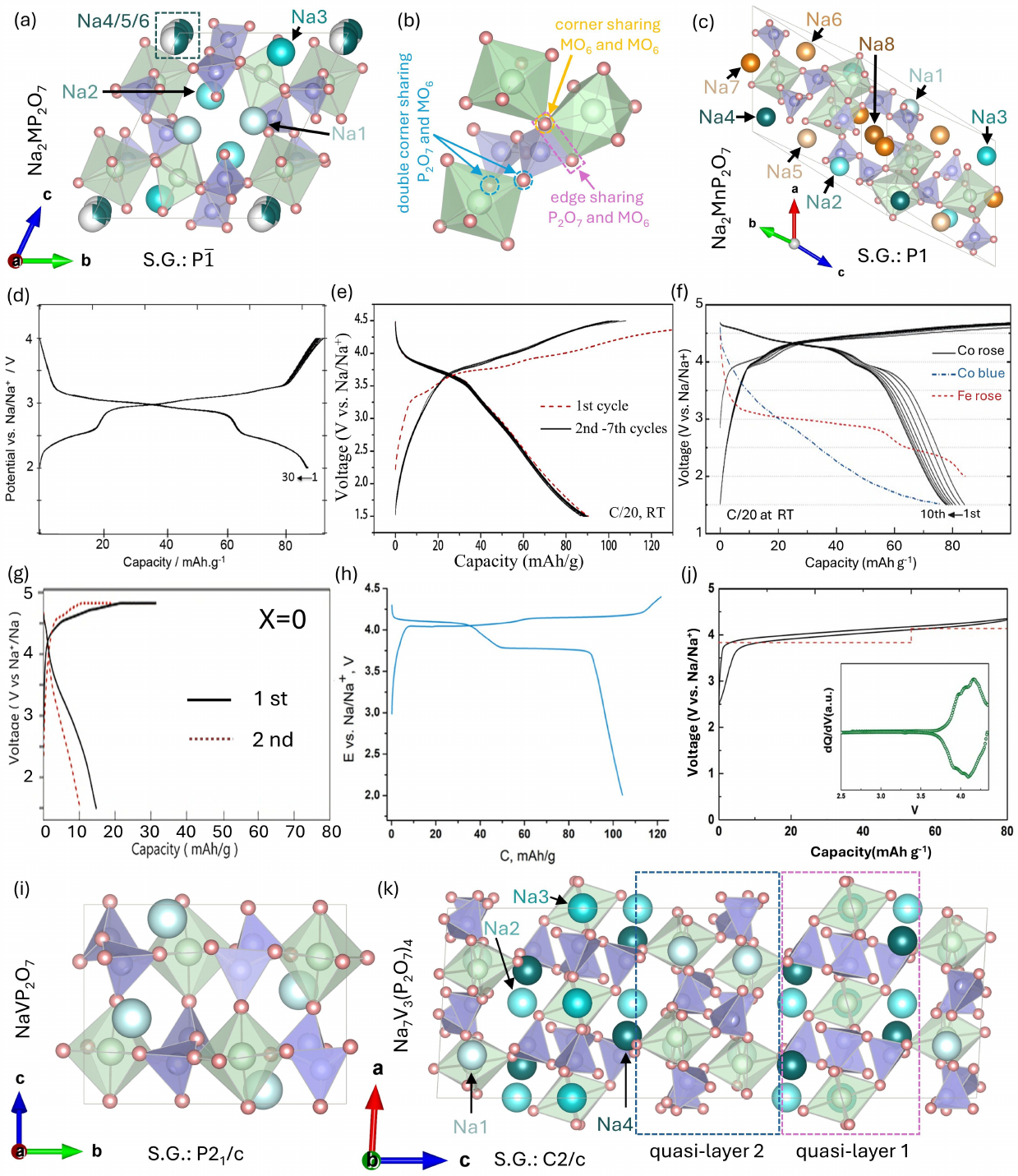}
\caption {The crystal structure of (a) the Na$_2$MP$_2$O$_7$ (space group $P\bar{1}$) along the $a$ direction showing the distinct sodium sites, (b) the connectivity among the MO$_6$ and P$_2$O$_7$ polyhedra, and (c) the Na$_2$MnP$_2$O$_7$ (space group $P1$). The GCD profile of (d) the Na$_2$FeP$_2$O$_7$ cathode for 30 cycles \cite{Barpanda_CM_13}, (e) the Na$_2$MnP$_2$O$_7$ for seven cycles \cite{Park_JACS_13}, (f) the rose-Na$_2$CoP$_2$O$_7$ for ten cycles along with the GCD profiles of Co blue, and Fe rose phases for comparison \cite{Kim_AngChem_16}, (g) the Na$_2$NiP$_2$O$_7$ for two cycles \cite{Ji_FM_20}. The GCD profile and crystal structure of (h, i) the NaVP$_2$O$_7$ \cite{Drozhzhin_CM_19}, and (j, k) the Na$_7$V$_3$(P$_2$O$_7$)$_4$ (dQ/dV plot shown in the inset) \cite{Kim_AEM_16}, respectively. The CIF files to construct (a, b) are taken from \cite{Barpanda_CM_13}, (c) from \cite{Barpanda_JMCA_13}, (i) from \cite{Drozhzhin_CM_19}, and (k) from \cite{Kim_AEM_16}.}
\label{Na2MP2O7}
\end{figure*}

The Pyrophosphates are another class of phosphate-based polyanionic materials with the general formula Na$_x$M$_y$(P$_2$O$_7$)$_z$, where $x$, $y$, and $z$ are integers, and $M$ is a transition metal. In this case, the P$_2$O$_7$ dimers provide Na diffusion paths in the three-dimensional open framework, having large channels for Na-diffusion and a robust structure that could guarantee long-term cycling performance and help buffer the volume change during cycling \cite{Clark_JMCA_14}. The Na$_2$MP$_2$O$_7$ type structure consists of MO$_6$ octahedra along with PO$_4$--PO$_4$ tetrahedra dimers and the Na atoms occupy the interstitial space, as shown in Fig.~\ref{Na2MP2O7}(a). The corner-sharing MO$_6$–MO$_6$ [M$_2$O$_{11}$] dimers form the crystal structure, which is coupled by PO$_4$--PO$_4$ [P$_2$O$_7$] units in both corner-sharing and edge-sharing fashion (see Fig.~\ref{Na2MP2O7}(b)) to form three-dimensional tortuous zigzag channels for Na$^+$-ion migration. It has been demonstrated that the triclinic ($P\bar{1}$) polymorph, which is also known as the {\textquotedblleft}rose" form due to its color, shows electrochemical activity and is potentially explored as a cathode material. Associated with the single M$^{2+}$/M$^{3+}$ redox reaction, a sodium ion per formula unit is available for the electrochemical energy storage. The Na$_2$FeP$_2$O$_7$, Na$_2$MnP$_2$O$_7$, Na$_2$CoP$_2$O$_7$, and Na$_2$NiP$_2$O$_7$ are reported to be stabilized in the isostructural triclinic $P\bar{1}$ phase, which means the charge storage process and the correlated phase changes during de/intercalation is preserved for all the transition metals \cite{Barpanda_CM_13, Huang_IC_98, Kim_AngChem_16, Ji_FM_20}. The space group $P1$ has also been used in the structural analysis for the rose phase of Na$_2$MP$_2$O$_7$, leading to ongoing confusion regarding the actual crystal structure symmetry of the Na$_2$MP$_2$O$_7$ \cite{Kim_AFM_13, Barpanda_JMCA_13, Erragh_JCSR_91}. The crystal structure of the $P1$ triclinic phase is shown in Fig.~\ref{Na2MP2O7}(c), which demonstrates that the M$_2$O$_{11}$ dimers and the P$_2$O$_7$ are arranged in a similar fashion and connectivity as in $P\bar{1}$ phase, but a significant change can be seen in the sodium ion sites. For instance, there was some discussion in the literature whether the crystal structure of Na$_2$MnP$_2$O$_7$ stabilizes in $P1$ or $P\bar{1}$ space groups. Erragh {\it et al.} showed a disorder in one of the sodium sites when $P\bar{1}$ phase was considered in the refinement, whereas, when the $P1$ phase was used, Na7 and Na8 sites were derived from the problematic sodium site \cite{Erragh_JCSR_91}. Huang {\it et al.} synthesized the Na$_2$MnP$_2$O$_7$ in $P\bar{1}$ phase, but the local structure was slightly different and described as having MnO$_6$--MnO$_6$ octahedral units forming Mn$_2$O$_{10}$ dimers via edge-sharing, which are further connected to P$_2$O$_7$ via two corner sharing \cite{Huang_IC_98}. Also, the Na$_2$MnP$_2$O$_7$ isostructural with triclinic P$\bar{1}$ Na$_2$FeP$_2$O$_7$ is reported, but the Mn$_2$O$_{10}$ dimers exist via corner-sharing of MnO$_6$ octahedra and pyramidal MnO$_5$ \cite{Park_JACS_13}. However, the nearest Mn--O distance beyond the MnO$_5$ configuration was 2.47\AA, but the average Mn--O bond length in MnO$_5$ was roughly $\sim$2.14\AA. Barpanda {\it et al.} claimed the $P1$ phase of Na$_2$MnP$_2$O$_7$, see Fig.~\ref{Na2MP2O7}(c), with corner-sharing M$_2$O$_{11}$ dimers connected by P$_2$O$_7$ units in both corner-sharing and edge-sharing fashion \cite{Barpanda_JMCA_13}. Their first principle calculations suggested the $P1$ phase to be thermodynamically most feasible in the case of Na$_2$MnP$_2$O$_7$. They also made a compelling argument, stating that the triclinic cell initially used for Na$_2$CoP$_2$O$_7$ and later applied to isostructural compounds does not represent a typical configuration, as it should have $a$ $<$ $b$, $c$, with all the angles either acute or obtuse. They also proposed that this configuration could be transformed into a more conventional form by applying the $b$+$c$, $a$+$b$+$c$, and $b$ transformations, resulting in the $P\bar{1}$ phase. 

Now, for the charge storage process, the following discussion is based on the $P\bar{1}$ phase structure, as it is characterized by inversion symmetry, exhibiting a more ordered and uniform arrangement as compared to the $P1$ structure where the inversion symmetry is absent \cite{Erragh_JCSR_91}. There are two sodium ions per formula unit distributed among six sodium ion sites in the $P\bar{1}$ phase. The sodium sites Na1, Na2, and Na3 are all fully occupied, each consisting of 0.5 sodium per f.u., while the Na4, Na5, and Na6 have almost similar atomic coordinates and are partly occupied. Here, all sites collectively consist of 0.5 sodium per f.u., so we shall collectively call the Na4/Na5/Na6 sites as Na4 for convenience \cite{Barpanda_CM_13}. By just looking at the GCD profiles shown in Fig.~\ref{Na2MP2O7}(d-f), it can be seen that the Fe, Mn, and Co analogs have almost similar features, predicting a similar process of sodium-ion de/insertion, which also suggests that they share alike local structure as well \cite{Barpanda_CM_13, Park_JACS_13, Kim_AngChem_16}. The only difference is the voltage offset and reaction kinetics relying on the transition metal ion. However, no comments can be made on the sodium storage process of Na$_2$NiP$_2$O$_7$ with the available electrochemical data [see Fig.~\ref{Na2MP2O7}(g)], despite sharing the same structure \cite{Ji_FM_20}. Figure~\ref{Na2MP2O7}(d) shows the GCD profile of Na$_2$FeP$_2$O$_7$ which indicates a stair-case type voltage profile, having a small plateau at 2.5 V followed by a large plateau at 3.0 V, and a specific capacity of 90 mAh/g at 0.1C with an average working voltage of 3.0 V. The charge-discharge profiles superimpose each other, depicting excellent capacity retention \cite{Barpanda_CM_13}. The first step near 2.5 V belongs to sodium extraction from half of the Na1 site, followed by the activation of the Na3 site at voltages above 3 V, where three voltage steps are visible. It is anticipated that the first two steps correspond to the two-step removal of complete sodium from the Na3 site, and the last step is associated with sodium removal from half of the Na4 site. In contrast, the Na2 site remains intact during the whole process \cite{Park_JACS_13}. Moreover, an average working voltage of 3.6 V was achieved for the case of Na$_2$MnP$_2$O$_7$, as shown in Fig.~\ref{Na2MP2O7}(e), which delivers a specific capacity of 90 mAh/g at 0.05 C, and maintained 96\% of the initial capacity after 30 cycles at 0.2 C \cite{Park_JACS_13}. For the Na$_2$Fe$_{0.5}$Mn$_{0.5}$P$_2$O$_7$ cathode, a binary system of Fe and Mn, a reversible capacity of 80 mAh/g at 0.05 C achieved with an average working voltage of 3.2 V, which is higher than the voltage of Na$_2$FeP$_2$O$_7$ and had better rate capability as compared to the Na$_2$MnP$_2$O$_7$ \cite{Shakoor_PCCP_16}. 

It is important to note here that the control of temperature during synthesis is crucial to achieving the electrochemical active rose phase (triclinic; $P\bar{1}$) in the case of cobalt, and if not taken care of, the reaction will result in more thermodynamic stable orthorhombic phase \cite{Erragh_JCSR_91}. The GCD profile of the Na$_2$CoP$_2$O$_7$ triclinic (rose) and orthorhombic (blue) phases are shown in Fig.~\ref{Na2MP2O7}(f) along with the discharge profile of Na$_2$FeP$_2$O$_7$ (rose) for comparison. With an average voltage of 4.3 V vs. Na/Na$^+$ along with a specific capacity of 80 mAh/g, the rose phase shows steady cycling and offers a higher energy density than its blue polymorph \cite{Kim_AngChem_16}. Due to the heavy framework of Na$_2$MP$_2$O$_7$ (M = Co, Fe, Mn, Ni, etc.), only a single redox reaction is typically feasible as per reversible redox reaction given below: 
\begin{equation}
\small
Na_2M^{(II)}P_2O_7 \rightleftharpoons Na^+ + NaM^{(III)}P_2O_7 + e^-
\end{equation}
However, this can be addressed by adopting an off-stoichiometric concentration of sodium and the transition metal, which increases the proportion of redox-active transition metal ions, enabling more redox reactions to occur. As a result, more sodium ions can be extracted, thereby enhancing the material's capacity \cite{Ha_AEM_13}. A solid solutions series Na$_{4-\alpha}$M$_{2+\alpha/2}$(P$_2$O$_7$)$_2$ (2/3$\ge$$\alpha$$\ge$7/8, M = Fe, Fe$_{0.5}$Mn$_{0.5}$, Mn) retains the same triclinic symmetry with $P\bar{1}$ space group, which offers better specific capacities, and elemental tunability.

For example, a Na$_{3.32}$Fe$_{2.34}$(P$_2$O$_7$)$_2$ cathode, crystallizing in the $P\bar{1}$ and having a theoretical capacity of 117.4 mAh/g, experimentally delivers a capacity of $\approx$100 mAh/g at 0.1 C with an average working voltage of 3.05 V and capacity retention of 89.6\% after 1100 cycles at 5 C \cite{Chen_AM_17}. The Na$_{3.12}$Mn$_{2.44}$(P$_2$O$_7$)$_2$ cathode with a theoretical capacity of 118.1 mAh/g, achieves a reversible capacity of 114 mAh/g at 0.1 C, facilitated by the reversible extraction of 2.31 sodium ions within its unique crystal structure. The high average voltage of 3.6 V was found with a 75\% capacity retention after 500 cycles at 5 C \cite{Li_ACSAMI_18}. Also, a novel cathode material, Na$_7$Fe$_{4.5}$(P$_2$O$_7$)$_4$ has been introduced with a triclinic (P$\bar{1}$) structure and a theoretical capacity of 108 mAh/g \cite{Niu_JMCA_16}. A discharge capacity of 104.8 mAh/g at a current rate of 1.5 C and 93.8\% retention after 650 cycles. If we look closely, this cathode composition seems to be a part of the solid solutions series Na$_{4-\alpha}$M$_{2+\alpha/2}$(P$_2$O$_7$)$_2$ (2/3$\ge$$\alpha$$\ge$7/8) for $\alpha$=0.5, also to mention that they show similar GCD profiles, working voltage, and the same crystal symmetry \cite{Ha_AEM_13, Niu_JMCA_16}. While the pyrophosphate-based materials with low transition metal mass ratio of 20 wt.\% may not offer high theoretical capacity, their higher operating voltage and exceptional thermal and electrochemical stability make them valuable candidates for long-life stationary energy storage applications \cite{Barpanda_CM_13}.

Moreover, the NaVP$_2$O$_7$ is a novel high-voltage cathode material, with a theoretical capacity of 108 mAh/g \cite{Kee_RSCA_15} where the flat plateaus can be seen in the GCD profile presented in Fig.~\ref{Na2MP2O7}(h). It shows the two-step redox activity of V$^{3+}$/V$^{4+}$, giving a discharge capacity of 104 mAh/g at a current density of 10 mA/g. The NaVP$_2$O$_7$ crystallizes in a monoclinic phase with $P2_1/c$, which can be described as a three-dimensional framework formed by corner sharing of VO$_6$ octahedra and P$_2$O$_7$ units as clearly shown in Fig.~\ref{Na2MP2O7}(i). During the oxidation, two plat plateaus with only a minor voltage gap took an average voltage of 4.15 V, but during reduction, it shows two distinct voltage plateaus located at 4.1 and 3.8 V. This voltage hysteresis could be related to the two-phase transition between the $P2_1/c$ and $P\bar{1}$ during cycling; however, no comments can made on their correlation \cite{Drozhzhin_CM_19}. Kim {\it et al.} introduced a novel 4 V class cathode material, Na$_7$V$_3$(P$_2$O$_7$)$_4$, having a $C2/c$ space group and capable of providing three sodium ions per formula unit on the basis of three V$^{3+}$/V$^{4+}$ redox couple with a theoretical capacity of 80 mAh/g \cite{Kim_AEM_16}, following the reversible redox reaction outlined below: 
\begin{equation}
\small
Na_7V^{(III)}_3(P_2O_7)_4 \rightleftharpoons 3Na^+ + Na_4V^{(IV)}_3(P_2O_7)_4 + 3e^-
\end{equation} 
During the electrochemical testing, the charge and discharge curves, see Fig.~\ref{Na2MP2O7}(j), are obtained at an outstanding 4.13 V with a very minimal polarization gap between them and a specific capacity of 80 mAh/g at 0.1 C. These extraordinary results are the outcome of the robust structure having the corner sharing VO$_6$ octahedra and P$_2$O$_7$ units, which form large tunnels that act as 3D pathways for sodium ion diffusion. Fig.~\ref{Na2MP2O7}(k) depicts the structure comprising a sequence of two quasi-layers and four unique sodium ion sites where the Na(2), Na(3), and Na(4) sites in quasi-layer 1 have nearly the same activation energies and are interconnected. The Na1 site in the quasi-layer 2 has higher activation energies and remains intact during sodium de/intercalation. While, among the three Na sites, the two Na ions from the Na2 site are extracted first, followed by the removal of one Na ion from the Na3 site, while the sodium extraction from the Na4 site is subjected to V$^{4+}$/V$^{5+}$ redox at voltages above 4.5 V \cite{Kim_AEM_16, Deng_ESM_16}. The V$^{4+}$/V$^{5+}$ redox couple can be accessed in tetragonal Na$_2$(VO)P$_2$O$_7$, which has a $P4bm$ symmetry and offers a theoretical capacity of 93.4 mAh/g. The structure consists of VO$_5$ square pyramids connected to P$_2$O$_7$ at their four corners, forming [VP$_2$O$_{11}$] units. These units are linked together to create infinite slabs in the $ab$ plane, which are stacked along the $c$ direction, and all the apexes of the VO$_5$ and PO$_4$ units point in the same direction. The sodium ions are stacked between these slabs, resulting in a layered framework that follows the one-dimensional Na-ion diffusion. In this case, a reversible capacity of 80 mAh/g was achieved at C/20 with an average working voltage of 3.8 V \cite{Barpanda_CEC_14}. Notably the vanadium ion in NaVP$_2$O$_7$ adopts a VO$_6$ octahedral geometry and V$^{3+}$/V$^{4+}$ redox, while in Na$_2$VOP$_2$O$_7$, it forms a VO$_5$ square pyramidal and V$^{4+}$/V$^{5+}$ redox transition. Despite the differences in oxidation states of the redox couple, the redox voltage remains constant at 3.8 V in both cases, which underscores the significant role of local coordination and CFS, which have been discussed earlier in this article.

\subsection{~(PO$_4$)P$_2$O$_7$}

\begin{table*}[]
\centering
\small
\caption{A summary of different parameters of phosphate-based polyanionic cathode materials utilizing P$_2$O$_7$ and (PO$_4$)P$_2$O$_7$ groups.}
\label{tab:Summary2}
\resizebox{\textwidth}{!}{%
\begin{tabular}{lcccccccc}
\hline
{Formula}     & 
{\begin{tabular}[c]{@{}c@{}}Space\\ Group\end{tabular}}	&
{\begin{tabular}[c]{@{}c@{}}Active\\ Redox\end{tabular}}  & 
{\begin{tabular}[c]{@{}c@{}}Working\\ Voltage\\(V vs. Na/Na$^+$)\end{tabular}}  &
{\begin{tabular}[c]{@{}c@{}}Q$_{th}$(mAh/g)\\(n Na/f.u.)\end{tabular}} & {\begin{tabular}[c]{@{}c@{}}Q$_{dis}$\\ (mAh/g)\end{tabular}} & 
{Electrolyte}                                                                        & {\begin{tabular}[c]{@{}c@{}}Voltage\\ window\\(V vs. Na/Na$^+$)\end{tabular}} & {Ref.}                   \\ \hline
\multicolumn{8}{c}{{P$_2$O$_7$}}                                                                                                                                                                                                                                                                                                                                                                                                                                                                                                                                       \\ \hline
Na$_2$FeP$_2$O$_7$            & $P\bar{1}$       & Fe$^{2+}$/Fe$^{3+}$                                                                                       & 3                                                                     & 97 (1 Na)                                                                  & 90 (0.1 C)                                                           & 1 M NaClO$_4$ in PC                                                                         & 2.0-4.0                                                               & \cite{Barpanda_CM_13}               \\ \hline
Na$_2$MnP$_2$O$_7$             & $P\bar{1}$       & Mn$^{2+}$/Mn$^{3+}$                                                                                       & 3.6                                                                   & 97.5 (1 Na)                                                                & 90 (0.05 C)                                                          & 1 M NaClO$_4$ in PC                                                                         & 1.5-4.5                                                               & \cite{Park_JACS_13}               \\ \hline
Na$_2$Fe$_{0.5}$Mn$_{0.5}$P$_2$O$_7$            & $P\bar{1}$       &   \begin{tabular}[c]{@{}c@{}} Fe$^{2+}$/Fe$^{3+}$\\ Mn$^{2+}$/Mn$^{3+}$\end{tabular}                                                                                        & 3.2                                                                   & 97.5 (1 Na)          & 80 (0.05 C)                                                         & 1 M NaClO$_4$ in PC   & 2.0-4.5  & \cite{Shakoor_PCCP_16}               \\ \hline
Na$_2$CoP$_2$O$_7$                & $P\bar{1}$     & Co$^{2+}$/Co$^{3+}$                                                                                       & 4.3                                                                   & 95 (1 Na)                                                                  & 80 (0.05 C)                                                          & \begin{tabular}[c]{@{}c@{}}1 M NaPF$_6$ in EC/DEC\\ (1:1 v/v)\end{tabular}                     & 1.5-4.7                                                               & \cite{Kim_AngChem_16}          \\ \hline
Na$_2$NiP$_2$O$_7$              & $P\bar{1}$       & Ni$^{2+}$/Ni$^{3+}$                                                                                       & 4.6                                                                   & 92.3 (1 Na)                                                                & 14.6 (0.1 C)                                                         & \begin{tabular}[c]{@{}c@{}}1 M NaPF$_6$ in EC/DEC\\ (1:1 v/v)\end{tabular}                     & 1.5-4.9                                                               & \cite{Ji_FM_20}        \\ \hline
NaVP$_2$O$_7$      	    & $P2_1/c$      & V$^{3+}$/V$^{4+}$                                                                                         & 3.9                                                                   & 108 (1 Na)                                                                 & 104 (0.1 C)                                                          & \begin{tabular}[c]{@{}c@{}}1 M NaPF$_6$ \\ (1:1 v/v, EC/PC)\end{tabular}                    & 2.0-4.4                                                               & \cite{Drozhzhin_CM_19}   \\ \hline
Na$_7$V$_3$(P$_2$O$_7$)$_4$      & $C2/c$    & V$^{3+}$/V$^{4+}$                                                                                         & 4.13                                                                  & 80 (3 Na)                                                                  & 80 (0.1 C)                                                           & \begin{tabular}[c]{@{}c@{}}1 M NaPF$_6$\\ (EC/PC, 1:1 v/v)\end{tabular}                     & 2.5–4.35                                                              & \cite{Kim_AEM_16}          \\ \hline

Na$_2$(VO)P$_2$O$_7$       & $P4bm$   & V$^{4+}$/V$^{5+}$                                                                                         & 3.8                                                                  & 93.4 (1 Na)                                                                  & 80 (0.05 C)                                                           & \begin{tabular}[c]{@{}c@{}}1 M NaClO$_4$ in EC/DEC\\with 3\% FEC\end{tabular}                     & 1.5–4.0                                                              & \cite{Barpanda_CEC_14}          \\ \hline
\multicolumn{8}{c}{{(PO$_4$)P$_2$O$_7$}}                                                                                                                                                                                                                                                                                                                                                                                                                                                                                                                               \\ \hline
Na$_4$Fe$_3$(PO$_4$)$_2$P$_2$O$_7$ 	& $Pn2_1a$    & Fe$^{2+}$/Fe$^{3+}$                                                                                       & 3.2                                                                   & 128.9 (3 Na)                                                               & 113 (0.05 C)                                                         & \begin{tabular}[c]{@{}c@{}}1M NaClO$_4$ in EC/PC\\ (v/v 1:1) with 5 vol.\% FEC\end{tabular} & 1.9-4.1                                                               & \cite{Chen_NatCom_19}      \\ \hline
Na$_4$Mn$_3$(PO$_4$)$_2$P$_2$O$_7$ 	& $Pn2_1a$		& Mn$^{2+}$/Mn$^{3+}$                                                                                       & 3.84                                                                  & 129.6 (3 Na)                                                              & 109 (C/20)                                                           & 1 M NaBF$_4$ in EC/PC                                                                       & 1.7-4.5                                                               & \cite{Kim_EES_15_Mn}              \\ \hline
Na$_4$MnFe$_2$(PO$_4$)$_2$P$_2$O$_7$            & $Pn2_1a$       &  \begin{tabular}[c]{@{}c@{}} Fe$^{2+}$/Fe$^{3+}$\\ Mn$^{2+}$/Mn$^{3+}$ \end{tabular}  & 3.4                                                                  & 129.1 (3 Na)          & 110 (0.05 C)                                                          & \begin{tabular}[c]{@{}c@{}}1M NaBF$_4$ in EC/PC\\ (v/v 1:1) \end{tabular}                                                                       & 1.7-4.5  & \cite{Kim_CM_16}               \\ \hline
Na$_4$Co$_3$(PO$_4$)$_2$P$_2$O$_7$	& $Pn2_1a$		 & Co$^{2+}$/Co$^{3+}$                                                                                       & 4.5                                                                   & 128 (3 Na)                                                               & 95 (0.2 C)                                                         & \begin{tabular}[c]{@{}c@{}}1 M NaPF$_6$ in EC/DEC \\ (1:1 v/v)\end{tabular}                    & 3.0-4.7                                                               & \cite{Nose_JPS_13}  \\ \hline
Na$_4$Ni$_3$(PO$_4$)$_2$P$_2$O$_7$ 	& $Pn2_1a$		& Ni$^{2+}$/Ni$^{3+}$                                                                                       & 4.8                                                                   & 127.2 (3 Na)                                                               & 63 (10 mA/g)                                                         & \begin{tabular}[c]{@{}c@{}}NaTFSI in Py13FSI\\ (1:9 in mole ratio)\end{tabular}             & 3.0-5.1                                                               & \cite{Zhang_NAM_17}              \\ \hline
Na$_7$V$_4$(P$_2$O$_7$)$_4$PO$_4$   	& $P$-$421c$		 & V$^{3+}$/V$^{4+}$                                                                                         & 3.85                                                                  & 92.8 (4 Na)                                                                & 92 (0.05 C)                                                          & \begin{tabular}[c]{@{}c@{}}1M NaClO$_4$ in \\ EC:DEC:DMC (v/v 1:1:1)\end{tabular}           & 2.0-4.2                                                               & \cite{Fang_RSCAdv_18}              \\ \hline
Na$_3$Fe$_2$(PO$_4$)P$_2$O$_7$ 		& $P2_12_12_1$		& Fe$^{2+}$/Fe$^{3+}$                                                                                       & 3.1                                                                   & 119 (2 Na)                                                               & 110.2 (0.1 C)                                                         & \begin{tabular}[c]{@{}c@{}}1M NaClO$_4$ in EC:DEC\\ (1:1 vol) with 5\% FEC\end{tabular}             & 1.5-3.5                                                               & \cite{Cao_ACSEL_20}              \\ \hline
Na$_3$Mn$_2$(P$_2$O$_7$)(PO$_4$) 	& $P2_12_12_1$			& Mn$^{2+}$/Mn$^{3+}$                                                                                       & 3.6                                                                   & 119 (2 Na)                                                               & 101 (0.1 C)                                                         & \begin{tabular}[c]{@{}c@{}}1M NaClO$_4$ in EC:DMC\\ (1:1 vol) with 5\% FEC\end{tabular}             & 1.8-4.3 & \cite{Li_JPS_22}              \\ \hline
Na$_3$MnFe(PO$_4$)P$_2$O$_7$            & $P2_12_12_1$       &   \begin{tabular}[c]{@{}c@{}} Fe$^{2+}$/Fe$^{3+}$\\ Mn$^{2+}$/Mn$^{3+}$ \end{tabular}                                                                                        & 3.27                                                                   & 120 (2 Na)          & 105 (0.1 C)                                                          & \begin{tabular}[c]{@{}c@{}}1M NaPF$_6$ in EC/DEC\\ (v/v 1:1) with 5\% FEC \end{tabular}                                                                       & 1.5-4.4  & \cite{Wang_CEJ_22}               \\ \hline

\end{tabular}%
}
\end{table*}

\begin{figure*} 
\includegraphics[width=6.5in]{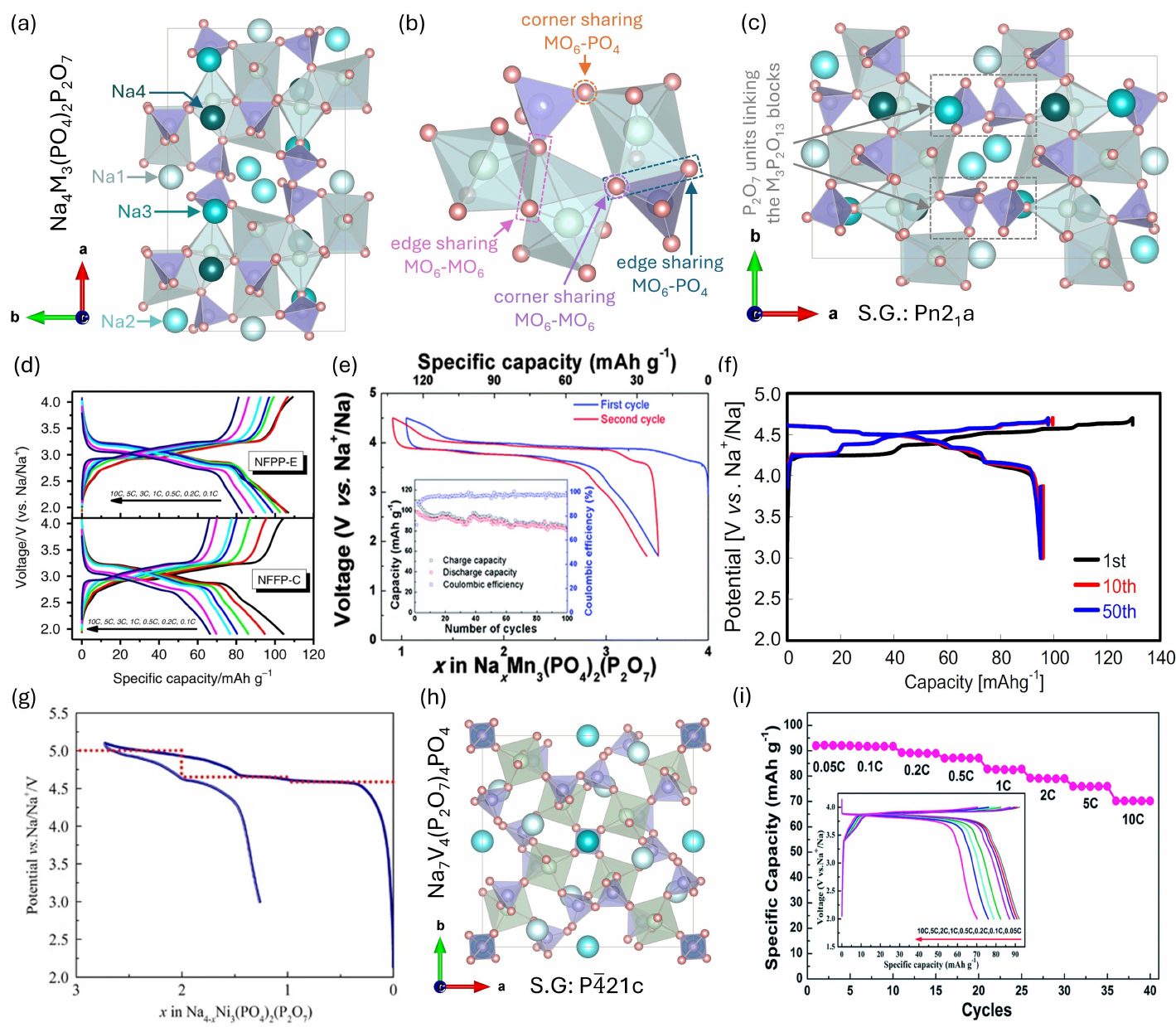}
\caption{The crystal structure of Na$_4$M$_3$(PO$_4$)$_2$P$_2$O$_7$ (a) showing the distinct sodium sites, (b) the connectivity among the MO$_6$ and PO$_4$ polyhedra in a M$_3$P$_2$O$_{13}$ block, and (c) the P$_2$O$_7$ units linking the M$_3$P$_2$O$_{13}$ blocks. The charge/discharge profiles of (d) the Na$_4$Fe$_3$(PO$_4$)$_2$P$_2$O$_7$-E and -C at various current rates between 0.1 C and 10 C \cite{Chen_NatCom_19}, (e) the Na$_4$Mn$_3$(PO$_4$)$_2$P$_2$O$_7$ for the first two cycles and the cycle stability shown in the inset \cite{Kim_EES_15_Mn}, (f) the Na$_4$Co$_3$(PO$_4$)$_2$P$_2$O$_7$ for 1st, 10th, and 50th cycles \cite{Nose_JPS_13}, and (g) the Na$_4$Ni$_3$(PO$_4$)$_2$P$_2$O$_7$ for the first cycle at a current density of 10 mA/g \cite{Zhang_NAM_17}. (h) The crystal structure of Na$_7$V$_4$(P$_2$O$_7$)$_4$PO$_4$ along the $c$ direction (sodium ion at different sites follows the same color code as given in Fig.~\ref{Na4M3(PO4)2P2O7}(a), and (i) its rate capability at various current densities, and the corresponding GCD profile at various current densities shown in the inset \cite{Fang_RSCAdv_18}. The CIF files to construct (a--c) are taken from (mp-1203835) \cite{Jain_APLM_13}, (h) from \cite{Lim_PNAS_14}.}
\label{Na4M3(PO4)2P2O7}
\end{figure*}

Now let's discuss the mix-ortho and pyrophosphate materials, which are characterized by the structural robustness of the PO$_4$ unit combined with the higher inductive effect of the P$_2$O$_7$ units. In this case, the presence of the P$_2$O$_7$ unit significantly enhances the redox potential, facilitates rapid sodium ion diffusion, and exhibits minimal volume changes during cycling. The Na$_4$M$_3$(PO$_4$)$_2$P$_2$O$_7$ type cathodes are stabilized in the crystal structure having $Pn2_1a$ space group for M= Fe, Mn, Co, Ni. The structure contains four distinct interconnected sodium sites, as shown in Fig.~\ref{Na4M3(PO4)2P2O7}(a), and each sodium site occupies one sodium ion per formula unit from which three sodium ions actively participate in the charge storage process. All four sodium sites share the common 4$a$ Wyckoff position and have six-oxygen coordination with distorted octahedral. Each sodium site exhibits distinct coordinated environments distinguished by MO$_6$ and PO$_4$ units attached, all of which collectively contribute to the complex electrochemical behavior. The structure is formed by a unique arrangement of bonding through corner-sharing and edge-sharing between MO$_6$ and PO$_4$ units, where MO$_6$ octahedra has a variety of units attached via corner/edge sharing causing inhomogeneous inductive pull from each direction, this complex arrangement visually depicted in Fig.~\ref{Na4M3(PO4)2P2O7}(b). This arrangement forms the M$_3$P$_2$O$_{13}$ blocks, which consist of three MO$_6$ octahedra and two PO$_4$ groups. These blocks are linked by the P$_2$O$_7$ along the $a$ axis, as illustrated in Fig.~\ref{Na4M3(PO4)2P2O7}(c). The strong inductive pull from the P$_2$O$_7$ units and the edge-sharing of two MO$_6$ octahedra having a greater repulsion between transition metal ions together results in elevated voltage. In addition, the structure creates extensive tunnels between the Na1 and Na2 sites in the $b$ direction, and these tunnels are linked through channels in the $a$ direction via Na3 and Na4 sites facilitating the smooth conduction of sodium ions \cite{Kim_JACS_12}.

For the electrochemical performance, a multi-step-like feature is observed in the GCD profiles of Na$_4$M$_3$(PO$_4$)$_2$P$_2$O$_7$ (M= Fe, Mn, Co, Ni) cathodes. The contribution of each sodium site towards the sodium ion de/intercalation was predicted using DFT, and it was suggested that the first sodium-ion extraction is favored from the Na2 sites via uni-phasic reaction, causing no significant change in the structural positioning. Nonetheless, the additional removal of two sodium ions and only one remaining in the structure, which is distributed as one-fourth of the sodium occupancy at the Na1 and Na3 sites and fifty percent of the sodium occupancy at the Na4 site. The unavailability of extraction of the fourth sodium ion is because, in all the cases, the M$^{3+}$/M$^{4+}$ is expected at higher voltage, which may fall beyond the stability window of the electrolyte \cite{Kim_EES_15_Mn}. Taking this into consideration, the theoretical capacity is 129 mAh/g as per three sodium reactions via M$^{2+}$/M$^{3+}$ redox by three transition metal ions in the formula unit, as shown in the sodium-ion redox process detailed below: 
\begin{equation}
\footnotesize
Na_4M^{(II)}_3(PO_4)_2P_2O_7 \rightleftharpoons 3Na^+ + NaM^{(III)}_3(PO_4)_2P_2O_7 + 3e^-
\end{equation} 
Howbeit, reversible sodium ion insertion is also explored in NCPP at lower voltages (0.01-3.0 V), giving a capacity of 250 mAh/g at 0.5 C, showcasing the potential of this category as a negative electrode \cite{Dwivedi_AAEM_21}. At first, the iron analog as a positive electrode, the GCD profile in Fig.~\ref{Na4M3(PO4)2P2O7}(d) shows the comparative results of samples prepared with ethylenediaminetetra acetic acid and citric acid, denoted as NFPP-E and NFPP-C, respectively. At a lower current rate of 0.05 C, both the samples have an average working voltage of 3.2 V, along with a decent capacity of 113 mAh/g in the voltage window of 1.9--4.1 V. The GCD profile shows a staircase-like feature due to multi-redox steps corresponding to the sequential activation of different sodium sites at a particular state of charge, as discussed above. The average working voltage is almost near to the voltage of LFP. Also, this is the highest when compared with Fe$^{2+}$/Fe$^{3+}$ redox in other phosphate polyanionic categories. Additionally, the material exhibits a volume change of only 4\%, which allows it to maintain 69.1\% capacity after 4400 cycles at a high current rate of 20 C \cite{Chen_NatCom_19}. This robust structure was also able to sustain the distortions caused by Jahn-Teller's active Mn$^{3+}$ state, which offers better reversibility of Mn$^{2+}$/Mn$^{3+}$ redox as compared to other Mn-based polyanion cathodes. For example, the Na$_4$Mn$_3$(PO$_4$)$_2$P$_2$O$_7$ cathode exhibits an initial discharge capacity of 109 mAh/g at 0.05 C with a high average working voltage of 3.84 V, see Fig.~\ref{Na4M3(PO4)2P2O7}(e), and highly reversible in subsequent cycles \cite{Kim_EES_15_Mn}. 

An interesting study on binary mix-phosphate phases of Na$_4$Mn$_x$Fe$_{3-x}$(PO$_4$)$_2$P$_2$O$_7$ ($x=$ 0, 1, 2, and 3) shows with the introduction of Mn there is a monotonous upshifting of Fe$^{2+}$/Fe$^{3+}$ redox with $x$ increasing from 0 to 2, correlated with the elevation of overall inductive pull due to longer M--O bond length with increase in Mn content \cite{Kim_CM_16}. They also emphasized the role of Mn in higher energy densities, while Fe is critical for structural stability and better kinetics. Their optimized Na$_4$MnFe$_{2}$(PO$_4$)$_2$P$_2$O$_7$ cathode exhibited exceptionally low volume change of 2.1\% only, which helps in sustaining 83\% of initial capacity after 3000 cycles at a current rate of 1 C. Likewise, theoretical calculations on binary mix-phosphate phases of Fe and Ni are expected to result in a significant increase in cell voltage. For example, the composition Na$_4$Fe$_2$Ni(PO$_4$)$_2$P$_2$O$_7$ is anticipated to have a cell voltage of 3.7 V \cite{Wood_JPCC_15}. Also, the GCD profile of Na$_4$Co$_3$(PO$_4$)$_2$P$_2$O$_7$ exhibits staircase like features associated with the multi-redox couples of Co$^{2+}$/Co$^{3+}$. The first redox couple appears at 4.26/4.18 V, while the highest redox potential is observed at 4.65/4.62 V versus Na$^+$/Na. The structure is capable of sustaining high-voltage redox reactions reversibly, achieving a capacity of 95 mAh/g at a current rate of 0.2 C, see Fig.~\ref{Na4M3(PO4)2P2O7}(f). An exceptional average voltage of 4.5 V was achieved with minimal polarization in the charge/discharge profile, which is also maintained at higher current rates \cite{Nose_JPS_13}. Following the trend of increase in average voltage, we discuss about Na$_4$Ni$_3$(PO$_4$)$_2$P$_2$O$_7$ (NNPP) cathode, which is expected to have redox activity of Ni$^{2+}$/Ni$^{3+}$ at a notably high voltage of 4.9 V \cite{Wood_JPCC_15}. The concern of possible oxidation of carbonate-based electrolytes at elevated voltages was mitigated by using an ionic liquid-based electrolyte (NaTFSI in Py13FSI, in a 1:9 mole ratio). This approach allows for a reversible discharge capacity of 63 mAh/g, which corresponds to 1.3 sodium ions being inserted at an operating potential as high as 4.8 V relative to Na$^+$/Na. \cite{Zhang_NAM_17}. The GCD profile shown in Fig.~\ref{Na4M3(PO4)2P2O7}(g) exhibits similar steplike features to those discussed above, indicating that the NNPP undergoes a similar sodium de/intercalation process. The mixed ortho- and pyrophosphate structure, derived from the combination of NaMPO$_4$ (offering high theoretical capacity) and Na$_2$MP$_2$O$_7$ (known for high voltage and stability), represents a remarkable achievement in structural engineering having a decent TM mass ratio of 27 wt.\%. This hybrid framework delivers best-in-class sodium storage capabilities where the open tunneled architecture enables smooth Na$^+$ conduction, while the extensive connectivity among MO$_6$ octahedra through edge- and corner-sharing pathways facilitates efficient polaron hopping. This complementing polaronic and ionic conduction mechanism is so effective that it enables reversible Ni$^{2+}$/Ni$^{3+}$ redox activity, a challenging feat in polyanionic cathodes due to the typically unfavorable formation of small hole polarons on the Ni \cite{Johannes_PRB_12, Wood_JPCC_15}. 

\begin{figure} [h!]
\includegraphics[width=5.8in]{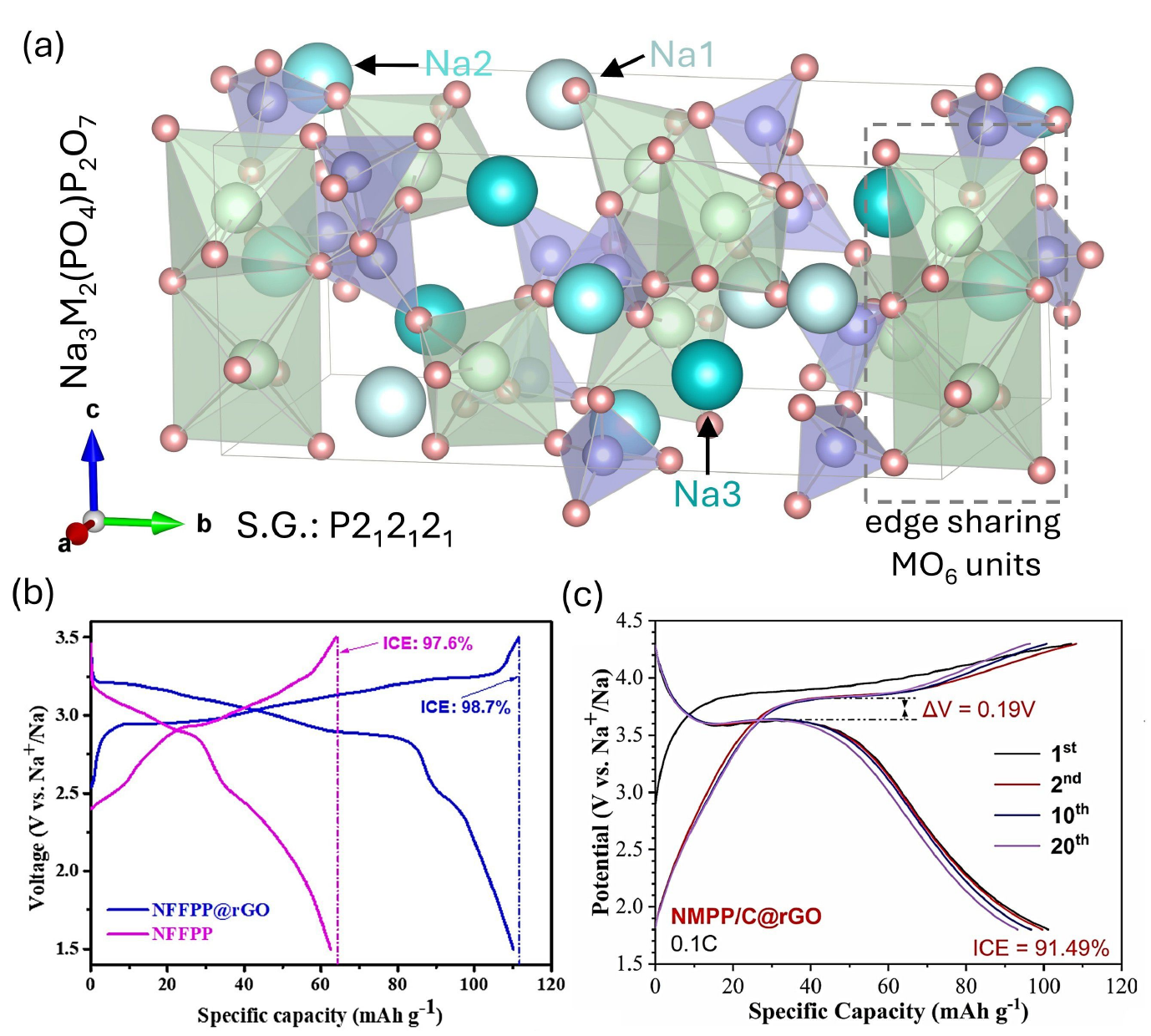}
\caption{The crystal structure of Na$_3$M$_2$(PO$_4$)P$_2$O$_7$ (a) showing the distinct sodium sites and the connectivity among MO$_6$, PO$_4$, and P$_2$O$_7$ polyhedra units. The charge/discharge profile of (b) Na$_3$Fe$_2$(PO$_4$)P$_2$O$_7$ and its rGO composite \cite{Cao_ACSEL_20}, (c) the first 20 cycles of Na$_3$Mn$_2$(PO$_4$)P$_2$O$_7$ cathode \cite{Li_JPS_22}. The CIF file for constructing (a) is taken from \cite{Li_JPS_22}.}
\label{Na3M2(PO4)P2O7}
\end{figure}

Further, the vanadium-based mixed phosphate Na$_7$V$_4$(P$_2$O$_7$)$_4$(PO$_4$), features a tetragonal structure with  $P$-$421c$ space group, composed of interconnected (VP$_2$O$_7$)$_4$PO$_4$ units, as shown in Fig.~\ref{Na4M3(PO4)2P2O7}(h). Each PO$_4$ tetrahedron shares corners with four VO$_6$ octahedra, and adjacent VO$_6$ octahedra are bridged by P$_2$O$_7$ groups, creating large tunnels for 3D sodium ion diffusion. There are three unique crystallographic sites for sodium ions, namely Na1, Na2, and Na3, where Na1 has five oxygen coordination with bipyramidal geometry, and both Na2 and Na3 have four oxygen coordination with planar and tetrahedral geometries, respectively \cite{Lim_PNAS_14}. There are four sodium ions per f.u. that are accessible for storage, and Na$_3$V$_4$(P$_2$O$_7$)$_4$(PO$_4$) is the discharged state with the existence of stable intermediate phase Na$_5$V$_4$(P$_2$O$_7$)$_4$PO$_4$. The material exhibits a distinct voltage plateau at 3.85 V and a discharge capacity of approximately 92 mAh/g (at 0.05 C), see Fig.~\ref{Na4M3(PO4)2P2O7}(i), based on the V$^{3+}$/V$^{4+}$ redox reaction with minimal polarization \cite{Fang_RSCAdv_18}. This stable structure offers the dis/charge with only a small volume change of 2.4\%, which could sustain 92.1\% capacity of the initial value after 500 cycles at 0.5 C.

We can create another category of mix-ortho and pyrophosphate materials by removing one NaMPO$_4$ unit from the Na$_4$M$_3$(PO$_4$)$_2$P$_2$O$_7$ formula, which is Na$_3$M$_2$(PO$_4$)P$_2$O$_7$. The theoretical calculation predicted the working voltage in this structure to be 3.4, 3.7, 4.7, and 5.2 V for M = Fe, Mn, Co, and Ni, respectively. These voltage values are quite decent, along with the two sodium ion reactions per formula unit, which correspond to the M$^{2+}$/M$^{3+}$ redox process outlined below \cite{Wang_CEJ_22}: 
\begin{equation}
\footnotesize
Na_3M^{(II)}_2(PO_4)P_2O_7 \rightleftharpoons 2Na^+ + NaM^{(III)}_2(PO_4)P_2O_7 + 2e^-
\end{equation} 
The structure exists in an orthorhombic system with $P2_12_12_1$ space group, consisting of MO$_6$ octahedra and PO$_4$ tetrahedra units. There are two transition metal sites based on the local environment and coordination. One MO$_6$ octahedra site has all the corners shared with PO$_4$ and P$_2$O$_7$ units, and it also isolates some of the PO$_4$ units to stop polymerizing into P$_2$O$_7$. The other MO$_6$ octahedra are in edge-sharing configuration with MO$_6$ to form edge-sharing chains along the $c$ direction, which are connected to the first MO$_6$ octahedra site via corner-sharing P$_2$O$_7$ units forming the 3D open framework as shown in Fig.~\ref{Na3M2(PO4)P2O7}(a) \cite{Chen_JALCOM_20}. Here, three sodium sites, Na1, Na2, and Na3, have the same 4$a$ Wyckoff position but are differentiated by oxygen coordination of eight, seven, and six, respectively. Two sodium ions are reversibly extracted favorably from Na1 and Na3 sites, which offer lower activation energies than Na2 sites. The GCD profile of Na$_3$Fe$_2$(PO$_4$)P$_2$O$_7$ cathode material shown in Fig.~\ref{Na3M2(PO4)P2O7}(b) displays an average voltage of 3.1 V and a reversible discharge capacity of 110.2 mAh/g at 0.1 C, which is nearly 92\% of the theoretical value of 119 mAh/g. It delivered an exceptional initial Coulombic efficiency of 98.7\%, maintains 89.7\% of the specific capacity after 6400 cycles at 20 C, and experiences only a minimal volume change of 4.95\% \cite{Cao_ACSEL_20}. Similarly, the Na$_3$Mn$_2$(PO$_4$)P$_2$O$_7$ cathode achieves a considerable discharge capacity of 101 mAh/g at 0.1 C, which is 85\% of the theoretical value, as depicted in Fig.~\ref{Na3M2(PO4)P2O7}(c). A working voltage of 3.6 V was obtained with a surprisingly minimal volume change of only 0.87\%. Through in-situ XRD measurements, the authors demonstrated that the changes in crystal structure follow the solid-solution reaction during the dis/charging process, which justifies the sturdy performance with almost no strain \cite{Li_JPS_22}. Also, a series of binary system Na$_3$Fe$_{2-x}$Mn$_x$(PO$_4$)(P$_2$O$_7$) ($x=$ 0--2) cathodes are electrochemically explored where the optimized cathode with Fe : Mn in 1 : 1 ratio delivers a specific capacity of 105 mAh/g at 0.1 C with a working voltage of 3.27 V \cite{Wang_CEJ_22}. 

\subsection{~Na$_x$MM$^\prime$(PO$_4$)$_3$}

\begin{figure*} 
\includegraphics[width=5.3in]{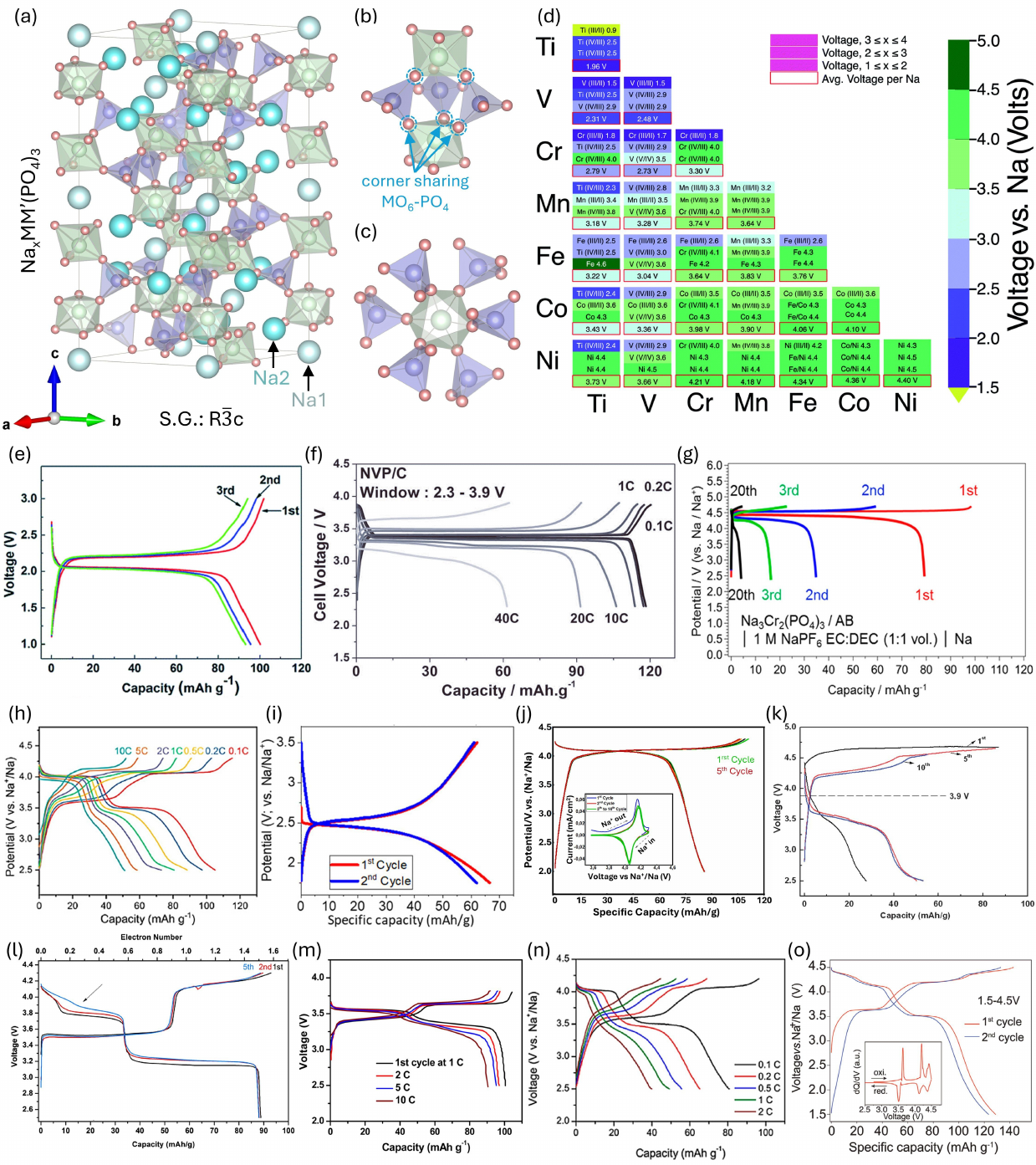}
\caption {(a) The crystal structure of the Na$_x$MM'(PO$_4$)$_3$ in NASICON phase, (b) the lantern units showing the connectivity among MO$_6$ and PO$_4$ units, and (c) the homogeneous environment of MO$_6$ provided by PO$_4$ units, (d) the voltage map showing calculated redox potentials for 3$d$ NaSICON electrode materials, Na$_x$MM'(PO$_4$)$_3$ (M, M' = Ti, V, Cr, Mn, Fe, Co, Ni). Here, each box represents a specific redox pair with color indicating the voltage vs. Na/Na$^+$ (color bar). The intensity reflects voltage magnitude (scale on the right). The red-outlined bottom box denotes the average voltage per Na (1$\le$x$\le$4) \cite{Singh_JMCA_21}. The GCD profile of (e) NaTi$_2$(PO$_4$)$_3$ for the first three cycles \cite{Pang_JMCA_14}, (f) the Na$_3$V$_2$(PO$_4$)$_3$ at various C-rates \cite{Saravanan_AEM_13}, (g) the Na$_3$Cr$_2$(PO$_4$)$_3$ for 1st, 2nd, 3rd, and 20th cycle \cite{Kawai_ACSAEM_18}, (h) the Na$_3$MnZr(PO$_4$)$_3$ at various C-rates \cite{Gao_JACS_18}, (i) the Na$_3$FeZr(PO$_4$)$_3$ for the first two cycles \cite{Paidi_IC_23}, (j) the Na$_3$CoZr(PO$_4$)$_3$ for the first 5 cycles and the CV curves for the first 10 cycles shown in the inset \cite{Chari_JPS_22}, and (k) the Na$_3$NiZr(PO$_4$)$_3$ for the 1st, 5th, and 10th cycle \cite{Pan_EES_13}, (l) the Na$_3$VCr(PO$_4$)$_3$ for 1st, 2nd, and 5th cycle \cite{Liu_ACSAMI_17}, (m) the Na$_4$MnV(PO$_4$)$_3$ for 1st cycle at 1 C, 2 C, 5 C, and 10 C each \cite{Zhou_NL_16}, (n) the Na$_3$MnTi(PO$_4$)$_3$ at various current rates from 0.1 C to 2  C \cite{Gao_CM_16}, and (o) the Na$_4$MnCr(PO$_4$)$_3$ for two cycles where the dQ/dV plot is shown in the inset \cite{Wang_AEM_20}. The CIF files to construct (a--c) are taken from \cite{Zhou_NL_16}.}
\label{NaxMM'(PO4)3}
\end{figure*}

In this section, we discuss NAtrium Super Ionic CONductors (NASICON) type cathode materials, which were first reported by J.B. Goodenough's group as sodium ion conductors and are found to stabilize in rhombohedral and monoclinic phases  \cite{Hong_MRB_76, Guin_JPS_15}. These are the most versatile and flexible cathode materials in terms of transition metals and sodium ion occupancies and also have extensive applications in all-solid-state ion batteries as solid-state electrolytes, providing better thermal stability and ionic conductivities. The majority of these active cathodes under this category are stabilized in rhombohedral phases with R$\bar{3}$c space group at room temperature. In the rhombohedral phase, there are two sodium sites, Na1 and Na2, see Fig.~\ref{NaxMM'(PO4)3}(a), categorized on the basis of oxygen coordination where Na1 has six oxygen coordinated, and Na2 has eight oxygen coordination, with Wyckoff positions 6$b$ and 18$e$, respectively. The Na1 sites can hold one sodium ion per formula unit, while the Na2 sites can accommodate three sodium ions per formula unit \cite{Jian_AFM_14}. In general, Na1 sites are more stable than Na2 and provide a pillar to the structure and hold the structural integrity during the charging/discharging. The sodium ions at Na2 sites have lesser activation energy and actively participate in the smooth de/intercalation via three-dimensional Na2-Na1-Na2 pathways, and the sodium ions at Na1 sites remain intact until all the sodium ions from the Na2 sites are extracted. The interesting thing here is the GCD profile remains flat as the sodium site energy doesn't change during dis/charging for a particular M$^{n+}$/M$^{(n+1)+}$ redox reaction.

The structure is made up of basic lantern units that are connected by sharing the corners of MO$_6$ octahedra and PO$_4$ tetrahedra, as seen in Fig.~\ref{NaxMM'(PO4)3}(b). When we look at the MO$_6$ octahedron's local environment in Fig.~\ref{NaxMM'(PO4)3}(c), we can see that it is connected to PO$_4$ at each corner, which creates a highly symmetric and homogeneous inductive pull from each corner. Moreover, the open structure creates a three-dimensional network of connected voids that allows sodium ions for ionic conduction, making NASICON materials useful in battery applications \cite{Singh_JMCA_21}. However, in orthophosphates (NaMPO$_4$), the edge-sharing of MO$_6$ octahedra and PO$_4$ tetrahedra provides more inductive pull, hence offering higher voltages than NASICON \cite{Padhi_JES_97}. In the NASICON framework, the choice of transition metal and polyanion group in the structure allows for tunable operating potential, as illustrated in Fig.~\ref{NaxMM'(PO4)3}(d), which presents the theoretically calculated voltages for combinations of 3$d$ TMs in the Na$_x$MM$^\prime$(PO$_4$)$_3$ framework, where M and M$^\prime$ are present in a 1:1 ratio \cite{Singh_JMCA_21}. The ability to substitute different metals into the structure without disrupting its framework is a key factor in its adaptability, and tunability of average voltage. Also, higher valence states of transition metals are available for the reversible redox activity in the NASICON structure \cite{Fan_EMA_24}. In NASICON-type materials, the total TM content accounts for approximately $\sim$23--24 wt.\% when considering only one sodium-ion reaction per TM. However, if multiple sodium-ion reactions per TM are enabled, the effective active mass ratio can increase to $\sim$34--36 wt.\%, which is comparable to that of LFP (35.4 wt.\%) \cite{Liu_MF_23}. The diagonal from the top left to the bottom right in Fig.~\ref{NaxMM'(PO4)3}(d) represents the single-transition-metal NASICON systems (M = M$^\prime$), where the average voltage is observed to increase monotonically with the atomic number of the transition metal, reaching the highest values for Co and Ni-based systems. In a similar systematic approach, our discussion will follow in the order of increasing the atomic number of the transition metals. Beginning with NaTi$_2$(PO$_4$)$_3$ (N1TP) having a rhombohedral structure, stabilizing Ti in a 4+ oxidation state and gives two sodium ion reactions at a stable redox voltage of 2.1 V with a theoretical capacity of 132.8 mAh/g in a potential window of 1.0--3.0 V \cite{Pang_JMCA_14}. Initially, Ti is in $d$$^0$ configuration, and no Na ions are available for extraction. During discharge, Na$_3$Ti$_2$(PO$_4$)$_3$ (N3TP) is formed, and the Ti$^{4+}$ ion reduces to Ti$^{3+}$ with the insertion of two sodium ions, and the GCD profile is shown in Fig.~\ref{NaxMM'(PO4)3}(e). Further lowering the potential from 2.0 V to 0 V and one extra sodium can be inserted at 0.4 V, causing half of the Ti$^{3+}$ reduce to Ti$^{2+}$ forming Na$_4$Ti$^{2+}$Ti$^{3+}$(PO$_4$)$_3$ (N4TP). This subsequent insertion of sodium ions causes the crystal structure to go through symmetry changes from rhombohedral (N1TP) to triclinic (N3TP) to rhombohedral again (N4TP) \cite{Senguttuvan_JACS_13}. This metal-based dense material with three sodium reactions and moderate working voltage above both the sodium plating and electrolytic reduction potential makes this a better anode material, which can be coupled with a high-voltage cathode for full-cell applications.

Note that the Na$_3$V$_2$(PO$_4$)$_3$ (NVP) cathode material needs no introduction as it has shown great potential as a positive electrode for SIB with exceptional stability, almost no polarization between GCD curves, higher rate capabilities, and 100\% Coulombic efficiencies \cite{Saravanan_AEM_13, Wang_SM_19, Sapra_ACSAMI_24}. The specific capacity of NVP is 117 mAh/g, assuming the contribution of 2 sodium ions from the V$^{3+}$/V$^{4+}$ redox at a voltage of 3.4 V, resulting in an energy density 397 Wh/kg, and follows the reversible redox reaction below: 
\begin{equation}
\small
Na_3V^{(III)}_2(PO_4)_3\rightleftharpoons 2Na^+ + NaV^{(IV)}_2(PO_4)_3 + 2e^-
\end{equation} 
These values have also been experimentally verified, see Fig.~\ref{NaxMM'(PO4)3}(f), showing excellent performance even at higher current rates \cite{Saravanan_AEM_13}. By lowering the potential window, another sodium ion can be inserted at a potential of 1.7 V, belonging to one of the two V$^{3+}$ to undergo V$^{2+}$/V$^{3+}$ with a capacity of nearly 50 mAh/g. However, the V$^{4+}$/V$^{5+}$ redox was absent in the pristine sample, which could be because of the higher activation energy of the Na1 site, giving no contribution to electrochemical processes. It has been observed that V$^{4+}$/V$^{5+}$ redox could be triggered by partial substitution of the V sites by other transition metal ions \cite{Aragon_CEC_15, Boisse_JECS_16, Zhao_AFM_24}. However, due to increasing concerns about the cost and environmental impact of vanadium, it has comparable challenges as cobalt in the context of  LIBs, and the same story follows, to opt for vanadium-free cathodes \cite{Zhao_CSC_22}. In the case of Na$_3$Cr$_2$(PO$_4$)$_3$, the chromium is in a 3+ state with the crystal field energy stabilizing the 3d$^3$ configuration (stable t$_{2g}^3$) within the octahedral environment in a rhombohedral structure having three sodium ions per f.u., out of which two are accessible for the charge storage process at a high voltage of 4.5 V corresponding to Cr$^{3+}$/Cr$^{4+}$ redox. The GCD profile shown in Fig.~\ref{NaxMM'(PO4)3}(g) features the cycling at a charging/discharging rate of 0.5 C/1 C, a charging capacity of 98 mAh/g was reached, which shows that the cathode material is reaching the capacity contributions from 2 sodium ions. The first discharge gives a 79 mAh/g capacity value but degraded significantly in subsequent cycles and is inappreciable after just 20 cycles \cite{Kawai_ACSAEM_18}. The authors suspected that the degradation is time-dependent, hampering the smooth formation of the discharged phase, ultimately leading to a decline in capacity. Interestingly, the Cr$^{3+}$/Cr$^{4+}$ redox is stabilized when half of the Cr sites are replaced with Ti, which made a huge impact on the reversibility of the 4.5 V redox \cite{Kawai_CC_19}. For the higher atomic numbers elements, the Fe analog Na$_3$Fe$_2$(PO$_4$)$_3$ stabilizes in the monoclinic phase at room temperature \cite{Liu_ACSSCE_17}, while the Na$_x$Mn$_2$(PO$_4$)$_3$, and Na$_x$Co$_2$(PO$_4$)$_3$ are theoretically found to be thermodynamically stable \cite{Singh_JMCA_21}, but no experimental evidence so far. The 2+ state of Mn, Co, and Ni can be stabilized in the NASICON structure by having another transition metal with oxidation states 3+ or above to maintain the charge neutrality and the Na ions content to $\le$4 per formula unit. To discuss the electrochemical activity of the mentioned transition metal, a binary system is considered in which half of the transition metal sites are replaced with Zr having d$^0$ configuration and no electrochemical activity within the cycled voltage window, which stabilizes the 2+ state of metal ions within the NASICON framework. 

\begin{table*}[]
\centering
\small
\caption{A summary of different parameters of phosphate-based polyanionic cathode materials utilizing (PO$_4$)$_3$ and (PO$_4$)$_2$F$_3$ groups.}
\label{tab:Summary3}
\resizebox{\textwidth}{!}{
\begin{tabular}{lcccccccc}
\hline
{Formula}     & 
{\begin{tabular}[c]{@{}c@{}}Space\\ Group\end{tabular}} &
{\begin{tabular}[c]{@{}c@{}}Active\\ Redox\end{tabular}}  & 
{\begin{tabular}[c]{@{}c@{}}Working\\ Voltage\\(V vs. Na/Na$^+$)\end{tabular}}  & 
{\begin{tabular}[c]{@{}c@{}}Q$_{th}$(mAh/g)\\(n Na/f.u.)\end{tabular}} & {\begin{tabular}[c]{@{}c@{}}Q$_{dis}$\\ (mAh/g)\end{tabular}} & 
{Electrolyte}                                                                        & 
{\begin{tabular}[c]{@{}c@{}}Voltage\\ window\\(V vs. Na/Na$^+$)\end{tabular}} & 
{Ref.}                   \\ \hline
\multicolumn{8}{c}{{(PO$_4$)$_3$}}                                                                                                                                                                                                                                                                                                                                                                                                                                                                                                                                     \\ \hline
NaTi$_2$(PO$_4$)$_3$          & $R$-$3c$       & Ti$^{3+}$/Ti$^{4+}$                                                                                       & 2.1                                                                   & 132.8 (2 Na)                                                               & 101 (0.2 C)                                                          & \begin{tabular}[c]{@{}c@{}}1M NaClO$_4$ in EC/PC\\ (1:1 in volume)\end{tabular}             & 1.0-3.0                                                               & \cite{Pang_JMCA_14}              \\ \hline
Na$_3$V$_2$(PO$_4$)$_3$        & $R$-$3c$      & V$^{3+}$/V$^{4+}$                                                                                         & 3.36                                                                  & 117.6 (2 Na)                                                               & 116 (0.1 C)                                                          & \begin{tabular}[c]{@{}c@{}}1 M NaClO$_4$ in (EC/PC)\\ (1:1 v/v)\end{tabular}                & 2.3-3.9                                                               & \cite{Saravanan_AEM_13}          \\ \hline
Na$_3$Cr$_2$(PO$_4$)$_3$        & $R$-$3c$     & Cr$^{3+}$/Cr$^{4+}$                                                                                       & 4.5                                                                   & 117 (2 Na)                                                                 & \begin{tabular}[c]{@{}c@{}}79 \\ (0.5C Chg/1C Dchg)\end{tabular}      & \begin{tabular}[c]{@{}c@{}}1 M NaPF$_6$ in EC/DEC\\ (1:1 vol)\end{tabular}                     & 2.5-4.7                                                               & \cite{Kawai_ACSAEM_18}          \\ \hline
Na$_3$MnZr(PO$_4$)$_3$        & $R$-$3c$       & \begin{tabular}[c]{@{}c@{}}Mn$^{2+}$/Mn$^{3+}$,\\ Mn$^{3+}$/Mn$^{4+}$\end{tabular}                        & 3.7                                                                   & 107 (2 Na)                                                                 & 105 (0.1 C)                                                          & \begin{tabular}[c]{@{}c@{}}1 M NaClO$_4$ in PC/FEC\\ (9:1 v/v)\end{tabular}                 & 2.5-4.3                                                               & \cite{Gao_JACS_18}            \\ \hline
Na$_2$ZrFe(PO$_4$)$_3$          & $R$-$3c$     & Fe$^{2+}$/Fe$^{3+}$                                                                                       & 2.5                                                                   & 112.2 (2 Na)                                                        & 65 (0.1C)                                                            & \begin{tabular}[c]{@{}c@{}}1M NaClO$_4$ in EC/PC\\ (v/v 1:1)\end{tabular}                   & 1.75-3.5                                                              & \cite{Paidi_IC_23}  \\ \hline
Na$_3$CoZr(PO$_4$)$_3$        & $R$-$3c$       & Co$^{2+}$/Co$^{3+}$                      & 4.05                                                                  & 106 (2 Na)                                                                & 85 (0.1 C)                                                           & \begin{tabular}[c]{@{}c@{}}1 M NaPF$_6$ in EC/DEC\\ (1:1 vol)\end{tabular}                     & 2.0-4.4                                                               & \cite{Chari_JPS_22}  \\ \hline
Na$_3$NiZr(PO$_4$)$_3$         & $R$-$3c$      & Ni$^{2+}$/Ni$^{3+}$                                                                                       & 3.9                                                                   & 106.4 (2 Na)                                                        & 50 (-)                                                               & -                                                                                           & 2.5-4.7                                                               & \cite{Pan_EES_13}              \\ \hline
Na$_3$VCr(PO$_4$)$_3$         & $R$-$3c$       & \begin{tabular}[c]{@{}c@{}}V$^{3+}$/V$^{4+}$,\\ V$^{4+}$/V$^{5+}$\end{tabular}                            & 3.7                                                                   & 117 (2 Na)                                                                                                                       & 90 (0.1 C)                                                           & \begin{tabular}[c]{@{}c@{}}1 M NaClO$_4$ in PC/FEC\\ (98:2 v/v)\end{tabular}                & 2.5-4.3                                                               & \cite{Liu_ACSAMI_17}         \\ \hline
Na$_3$FeV(PO$_4$)$_3$           & $C2/c$     & \begin{tabular}[c]{@{}c@{}}Fe$^{2+}$/Fe$^{3+}$,\\ V$^{3+}$/V$^{4+}$\end{tabular}                            & 2.9            & 116 (2 Na)      & 103 (1 C)    & \begin{tabular}[c]{@{}c@{}}1 M NaClO$_4$ in PC:FEC\\ (10:1 v/v)\end{tabular}           & 2.0-3.8                                                              & \cite{Zhou_NL_16}         \\ \hline
Na$_4$MnV(PO$_4$)$_3$         & $R$-$3c$       & \begin{tabular}[c]{@{}c@{}}V$^{3+}$/V$^{4+}$\\ Mn$^{2+}$/Mn$^{3+}$,\\ Mn$^{3+}$/Mn$^{4+}$,\end{tabular}   & 3.5 (2 Na)                                                                   & \begin{tabular}[c]{@{}c@{}}111 (2 Na)\\ 166 (3 Na)\end{tabular}                                                                                                                                                                                     & \begin{tabular}[c]{@{}c@{}}101 (1 C)\\ 91.2 (1 Na/20 h)\end{tabular}                                                          & \begin{tabular}[c]{@{}c@{}}1 M NaClO$_4$ in PC/FEC\\(10:1 v/v)\\1 M NaPF$_6$ in EC/DMC\\1:1 wt\%+3 wt\% FEC\end{tabular}               & \begin{tabular}[c]{@{}c@{}}2.5-3.8\\ 2.5-4.3\end{tabular}                                                               & \begin{tabular}[c]{@{}c@{}}\cite{Zhou_NL_16} \\ \cite{Chen_SM_19} \end{tabular}     \\ \hline
Na$_3$MnTi(PO$_4$)$_3$         & $R$-$3c$      & \begin{tabular}[c]{@{}c@{}}Ti$^{3+}$/Ti$^{4+}$\\ Mn$^{2+}$/Mn$^{3+}$,\\ Mn$^{3+}$/Mn$^{4+}$,\end{tabular} & \begin{tabular}[c]{@{}c@{}}3.6 (2 Na) \\ 3.0 (3 Na)  \end{tabular}                                                   & \begin{tabular}[c]{@{}c@{}}117 (2 Na)\\ 170 (3 Na)\end{tabular}                                                                                                                               & \begin{tabular}[c]{@{}c@{}}80 (0.1 C)\\ 160 (0.2 C)\end{tabular}                                                        &  \begin{tabular}[c]{@{}c@{}}1 M NaClO$_4$ in PC/FEC\\(9:1 v/v)\\1 M NaClO$_4$ in EC/PC\\(1:1 w/w)+5\%FEC\end{tabular}                                                                      & \begin{tabular}[c]{@{}c@{}}2.5-4.2\\ 1.5-4.2\end{tabular}                                                             & \begin{tabular}[c]{@{}c@{}}\cite{Gao_CM_16}\\\cite{Zhu_AEM_19}\end{tabular}     \\ \hline
Na$_4$MnCr(PO$_4$)$_3$       & $R$-$3c$        & \begin{tabular}[c]{@{}c@{}}Mn$^{2+}$/Mn$^{3+}$,\\ Mn$^{3+}$/Mn$^{4+}$,\\ Cr$^{3+}$/Cr$^{4+}$\end{tabular} & 3.53 (3 Na)                                                                     & \begin{tabular}[c]{@{}c@{}}110 (2 Na)\\ 165 (3 Na)\end{tabular}                                                                & \begin{tabular}[c]{@{}c@{}}114 (0.1 C)\\ 160.5 (0.05 C)\end{tabular}                                                       & \begin{tabular}[c]{@{}c@{}}1 M NaPF$_6$ in EC/DEC\\ (1:1 v/v)\end{tabular}                     & \begin{tabular}[c]{@{}c@{}}1.4-4.3\\ 1.4-4.6\end{tabular}                                                              & \cite{Zhang_AM_20}          \\ \hline
\multicolumn{8}{c}{{(PO$_4$)$_2$F$_3$}}                                                                                                                                                                                                                                                                                                                                                                                                                                                                                                                                     \\ \hline
Na$_3$V$_2$(PO$_4$)$_2$F$_3$         & $Amam$        & V$^{3+}$/V$^{4+}$                                                                              & 3.9                                                                   & 128 (2 Na)                                                               & 120 (0.1 C)                                                          & \begin{tabular}[c]{@{}c@{}}1M NaPF$_6$ in PC/EC/DMC\\ (1:1:1 in volume)\end{tabular}             & 3.0-4.4                                                               & \cite{Yan_NatCom_19}              \\ \hline
Na$_3$(VO)$_2$(PO$_4$)$_2$F           & $P4_2/mnm$      & V$^{4+}$/V$^{5+}$                                                                              & 3.77                                                                   & 130 (2 Na)                                                               & $\sim$120 (0.1 C)                                                          & \begin{tabular}[c]{@{}c@{}}1M NaBF$_4$ in EC/PC\\ (1:1 in volume)\end{tabular}             & 2.0-4.5                                                               & \cite{Park_AFM_14}              \\ \hline
Na$_3$Ti$_2$(PO$_4$)$_2$F$_3$           & $P4_2/mnm$         & Ti$^{3+}$/Ti$^{4+}$                                                                            & 2.6                                                                  & 133 (2 Na)                                                               & 57.7 (0.2 mA/g)                                                          & 1M NaClO$_4$ in PC             & 1.5-3.2                                                               & \cite{Chihara_JPS_13}              \\ \hline
Na$_3$Fe$_2$(PO$_4$)$_2$F$_3$          & $P4_2/mnm$        & Fe$^{2+}$/Fe$^{3+}$                                                                          & 2.5                                                                   & 125 (2 Na)                                                               & 23.6 (0.2 mA/g)                                                          & 1M NaClO$_4$ in PC             & 2.3-4.3                                                               & \cite{Chihara_JPS_13}              \\ \hline
Na$_3$Cr$_2$(PO$_4$)$_2$F$_3$            & $P4_2/mbc$       & Cr$^{3+}$/Cr$^{4+}$                                                                           & 4.7                                                                   & 63.8 (1 Na)                                                               & 55.1 (0.1 C)                                                          & 1M NaPF$_6$ in PC            & 2.7-5.0                                                               & \cite{Kawai_CM_21}              \\ \hline
\end{tabular}%
}
\end{table*}

The Mn holds a very important role in NASICON structure, and to explain that, we need to look into the electronic configuration of the Mn$^{2+}$/Mn$^{3+}$/Mn$^{4+}$ ions \cite{Soundharrajan_JMCA_22}. The crystal field splitting (CFS) of $d-$orbitals in an octahedron with oxygen ligands forms low energy t$_{2g}$ and high energy e$_g$ orbitals. The d$^4$ configuration of Mn$^{3+}$ is altered by the CFS, resulting in a high spin t$_{2g}^3$e$_g^1$ configuration. This places one electron in the higher-energy e$_g$ orbital, which is the primary reason why Mn$^{3+}$ is highly susceptible to Jahn-Teller distortion. The presence of this unpaired electron in the e$_g$ orbital leads to an uneven distribution of electron density, driving the system to distort in order to lower its energy \cite{Zheng_AEM_24}. Here, removing one electron from the e$_g$ orbital not only lowers the energy of the system but also stabilizes the half-filled t$_{2g}$ orbitals of Mn$^{4+}$. This accounts for the ease with which Mn$^{3+}$ can undergo Mn$^{3+}$/Mn$^{4+}$ redox. The NASICON structure, which facilitates multi-electron reactions, allows the Mn ions to further oxidize to higher oxidation states via the Mn$^{3+}$/Mn$^{4+}$ redox couple. In comparison to other transition metal systems, a voltage difference of about 1.7 V is observed between the redox couples V$^{2+}$/V$^{3+}$ and V$^{3+}$/V$^{4+}$, as well as in Ti. However, in the case of Mn, the voltage difference is much smaller, around 0.5 V, between the Mn$^{2+}$/Mn$^{3+}$ and Mn$^{3+}$/Mn$^{4+}$ redox couples. This property of Mn makes it the most favorable transition metal, giving two sodium reactions per single Mn ion, both at high voltage and within the electrolytic stability window. The GCD profile of Na$_3$MnZr(PO$_4$)$_3$ is shown in Fig.~\ref{NaxMM'(PO4)3}(h) depicting two flat regions at voltage values of 3.6 and 4.1 V corresponding to redox of Mn$^{2+}$/Mn$^{3+}$ and Mn$^{3+}$/Mn$^{4+}$, respectively \cite{Gao_JACS_18}, and exhibit relatively low polarization. 

Moving on to the discussion about NASICON type Na$_3$Fe$_2$(PO$_4$)$_3$ cathode where the Fe mostly stabilizes in a 3+ state, making the material Na-ion deficient, so the Na ions need to be introduced during the first discharge, which reduces the Fe to 2+ state. The cooperative deformation of the FeO$_6$ octahedron stabilizes the crystal structure in a monoclinic form at room temperature, adopting a $C12/c1$ configuration \cite{Liu_ACSSCE_17, Rajagopalan_AM_17}. When half of the vanadium metal in Na$_3$V$_2$(PO$_4$)$_3$ is replaced with Fe, it also causes the octahedral distortion, transforming the rhombohedral phase to monoclinic phase with no change in the Na content \cite{Zhou_NL_16}. However, replacing two PO$_4$ units with SO$_4$ units in Na$_3$Fe$_2$(PO$_4$)$_3$ reduces the Na content to 1 with the structure stabilizing in the rhombohedral phase \cite{Singh_PRB_24}. Similarly, when Zr$^{4+}$ replaces half of Fe$^{3+}$ forming Na$_2$ZrFe(PO$_4$)$_3$ with 2 Na ions also preserves the rhombohedral phase \cite{Paidi_IC_23}. Sunkyu {\it et al.} prepared Na$_4$FeV(PO$_4$)$_3$ in the rhombohedral phase and thoroughly examined the extraction/insertion mechanism, local environments using M\"ossbauer and x-ray absorption spectroscopy \cite{Park_CM_22}. It has been found that during cycling, precisely at three sodium ions in the structure, Na$_3$FeV(PO$_4$)$_3$ adopt the monoclinic phase, while the deviation of sodium ions from 3 per f.u., the crystal structure acquires a rhombohedral phase. All these results predict that the deformation of the FeO$_6$ could be related to the three sodium ions per f.u. in the NASICON structure. The electrochemical results of Na$_3$Fe$_2$(PO$_4$)$_3$, in which iron is in 3+ state with three sodium ions in the structure, show a specific capacity of 61 mAh/g, equivalent to one sodium reaction, in a potential window of 1.5--3.5 V. This means only one sodium ion is available for de/intercalation corresponding to redox at 2.5 V. Using XPS, the authors claimed partial reduction of Fe$^{3+}$ at discharge state with the formation of Na$_4$Fe$_2$(PO$_4$)$_3$ having only one sodium ion insertion \cite{Liu_ACSSCE_17}. The GCD profile of Na$_2$ZrFe(PO$_4$)$_3$ in Fig.~\ref{NaxMM'(PO4)3}(i)  shows the activity of Fe$^{2+}$/Fe$^{3+}$ redox at 2.5 V where a capacity value $\sim$60mAh/g approaching one sodium reaction in the potential window of 1.75--3.5 V because of the availability of only one Fe to be active. However, a tiny contribution from the Fe$^{3+}$/Fe$^{4+}$ redox was also seen in the broad potential window of 1.5--4.3 V \cite{Paidi_IC_23}. The carbon-coated modified Na$_3$Fe$_2$(PO$_4$)$_3$, when cycled in the potential window of 1.5--4.2 V, showed a discharge capacity of 109 mAh/g, with the activation of Fe$^{3+}$/Fe$^{4+}$ redox at higher voltages, enhancing the energy density \cite{Rajagopalan_AM_17}.

The electrochemical activity of cobalt in the NASICON structure is under explored in SIBs owing to its high cost. In one of the few reports, the cobalt-based NASICON material Na$_3$CoZr(PO$_4$)$_3$ exhibits electrochemical activity at a high voltage of 4.05 V with minimal polarization. This cathode material offers a theoretical capacity of 106 mAh/g, considering two Na-ion reactions. The discharge capacity obtained was 85 mAh/g, equivalent to 1.6 Na-ion contribution, and the GCD profile is shown in Fig.~\ref{NaxMM'(PO4)3}(j). The authors reported that high voltage can be attributed to the Co$^{2+}$/Co$^{3+}$ redox \cite{Chari_JPS_22}. As there is only one cobalt to undergo redox, it can provide a single sodium ion to participate in the reversible charge storage, which makes the 1.6 sodium ion contribution questionable. Also, the possibility of Co$^{3+}$/Co$^{4+}$ is not an option here as it is expected at a voltage of nearly 0.7 V above the Co$^{2+}$/Co$^{3+}$ redox \cite{Singh_JMCA_21}. Although this cathode material offers a high voltage with minimal volume changes during cycling, the poor average Coulombic efficiency of $\sim$83.5\% suggests further optimizations for its practical utilization. On a similar note, nickel is also less explored in this category of cathodes because of no pure phase formation or the activity of Ni$^{2+}$/Ni$^{3+}$ redox falls very near or beyond the electrolyte oxidation potential \cite{Zhou_NL_16}. The Na$_3$NiZr(PO$_4$)$_3$ gives a capacity of 50 mAh/g consistent with one sodium ion extraction, see Fig.~\ref{NaxMM'(PO4)3}(k). A large polarization of 0.8 V corresponds to poor kinetics, and an average potential of 3.9 V vs. Na$^+$/Na is associated with Ni$^{2+}$/Ni$^{3+}$ \cite{Pan_EES_13}. However, when tested as an anode in a potential window of 0.01--2.5 V, it exhibited a discharge capacity of 140 mAh/g at 25 mA/g along with good Coulombic efficiency and rate performance \cite{Tayoury_IEEE_21}. The capability of NASICON-based phosphate cathodes to provide the accessibility of multi-electron reactions and higher capacity has been explored by facilitating two or more transition metals in the structure. Fig.~\ref{NaxMM'(PO4)3}(d) presents the average voltages for 21 possible combinations of binary transition metal systems capable of multi-electron reactions. A slight variation in the redox voltage of a given M is observed with different accompanying M$^\prime$ ions, which can be attributed to changes in the local inductive effects \cite{Singh_JMCA_21}. Here, each transition metal comes with its unique properties, and there's always an optimum level of concentration for exceptional results \cite{Ma_NR_20, Li_ACSEL_23}. As discussed previously, only Mn gives subsequent redoxes with a small difference between them, which makes it a priority for mid or high-entropy NASICON cathodes \cite{Liu_ESM_23}. Nonetheless, going beyond one Mn ion per formula unit ends up with the occurrence of side impurity phases, which puts a restriction on the Mn concentration \cite{Liu_JMCA_21}. However, if the pairing is done perfectly with other transition metals, it is helpful in facilitating multielectron reactions, and exceptional results can be expected in terms of capacity, voltage, and stability. The following discussion will be on the binary system of active transition metals in the NASICON framework, which holds importance and potential for bridging the gap for the commercialization of SIBs. Many combinations are reported in the literature, but our discussion is confined to the cathodes with high-voltage outputs. 

The first in line is based on the activation of the high voltage V$^{4+}$/V$^{5+}$ redox by replacing one of the vanadium sites with chromium in Na$_3$V$_2$(PO$_4$)$_3$. Both V and Cr exist in a 3+ state in Na$_3$VCr(PO$_4$)$_3$, and the cathode material was cycled in a potential window of 2.5--4.3 V. The Cr is inactive in the given potential range, and the GCD profile in Fig.~\ref{NaxMM'(PO4)3}(l) shows that vanadium undergoes two redoxes, V$^{3+}$/V$^{4+}$ and V$^{4+}$/V$^{5+}$ at 3.4 and 4.1 V, respectively. A specific capacity of 90 mAh/g was achieved at 0.1 C, equaling 1.5 sodium ion contribution \cite{Liu_ACSAMI_17}. Further, Fe and Ti are also used for swapping one vanadium site in NVP, forming Na$_3$FeV(PO$_4$)$_3$ and Na$_2$VTi(PO$_4$)$_3$, respectively. The sodium deficiency of the pristine samples and redox activity at low voltages of 2.5 V and 2.1 V for Fe$^{2+}$/Fe$^{3+}$ and Ti$^{3+}$/Ti$^{4+}$, respectively, presents major drawbacks \cite{Zhou_NL_16, Wang_NC_17}. When combined with Mn, the sodium-rich Na$_4$MnV(PO$_4$)$_3$ NASICON structure emerges as a promising cathode material undergoing multi-redox reactions, namely V$^{3+}$/V$^{4+}$, Mn$^{2+}$/Mn$^{3+}$, and mixed V$^{4+}$/V$^{5+}$ and Mn$^{3+}$/Mn$^{4+}$ states. When the cathode material is cycled up to 3.8 V, the V$^{3+}$/V$^{4+}$, and Mn$^{2+}$/Mn$^{3+}$ redox can be reversibly accessed at a voltage of 3.4 and 3.6 V, see Fig.~\ref{NaxMM'(PO4)3}(m), respectively, and a specific capacity of 101 mAh/g was obtained at 1 C \cite{Zhou_NL_16}. When cycled above 3.8 V, a specific capacity of 156 mAh/g  was achieved with a slanting plateau during the charge at nearly 3.9 V, which is assigned to mixed V$^{4+}$/V$^{5+}$ and Mn$^{3+}$/Mn$^{4+}$ redox \cite{Chen_SM_19}. During discharge, the GCD profile did not replicate the charging profile, and no flat plateau was obtained due to the irreversible phase change in the structure, also resulting in drastic capacity decay \cite{Zakharkin_ACSAEM_18}. This cathode material could be the answer to the quest for SIBs if three sodium ion reactions reversibly accessed with structural stability.

Moreover, the Na$_3$MnTi(PO$_4$)$_3$ is a low-cost vanadium-free cathode material, capable of giving reversible three sodium reactions with a phenomenal theoretical capacity of 176 mAh/g. The pristine state can only give two sodium ions during charge, see Fig.~\ref{NaxMM'(PO4)3}(n), corresponding to two consecutive Mn$^{2+}$/Mn$^{3+}$/Mn$^{4+}$ redox at 3.6 and 4.1 V \cite{Gao_CM_16}. The third sodium ion is reversible and accessible via Ti$^{3+}$/Ti$^{4+}$ redox at 2.1 V by lowering the cut-off voltage to 1.5 V. All the three redox combined can give a three sodium reaction delivering a specific capacity of 173 mAh/g, approaching its theoretical capacity \cite{Zhu_AEM_19}. Nevertheless, there have been issues reported with the voltage hysteresis in the GCD profile as a result of Mn$^{2+}$ hopping between the sodium sites and the transition metal sites \cite{Zhang_ACSEL_21}. Recently, a study by Liu {\it et al.} introduced a minor structural change by replacing 5\% of Ti with Mo, which restricted the hopping of Mn$^{2+}$ ions and, in turn, restored the hysteresis and energy density efficiency \cite{Liu_NatE_23}. The partial substitution of Ti with electron-rich V has been reported to suppress Mn$^{2+}$ migration by increasing the energy barrier, which in turn helps reduce voltage hysteresis observed during GCD cycling \cite{Zhang_AM_24}.  With an exceptional capacity, high working voltage, and non-toxicity, this cathode material can become the perfect choice for the upscaling of SIBs-based energy storage systems \cite{Hu_ACSEL_21}. Also, the Na$_4$MnCr(PO$_4$)$_3$, a sodium-rich cathode composed of transition metals that could deliver three sodium ion reactions with high redox voltages, exemplifies an effective cathode design. The GCD profile in Fig.~\ref{NaxMM'(PO4)3}(o) shows all the active redox couples, Mn$^{2+}$/Mn$^{3+}$, Mn$^{3+}$/Mn$^{4+}$, and Cr$^{3+}$/Cr$^{4+}$ at 3.5, 4.0, and 4.35 V, respectively, which are higher than that of LFP (3.3 V) and NVP (3.4 V) \cite{Wang_AEM_20}. An extraordinary reversible discharge capacity of 160.5 mAh/g was achieved as an outcome of multi-redox reactions with an average working voltage of 3.53 V \cite{Zhang_AM_20}. However, there are still issues with the cycling stability possibly because of the structural and electrolyte degradation at higher voltage. To achieve structural improvements for reversible multi-redox reactions, researchers are opting for high-entropy cathode materials. This trend is gaining significant momentum and could lead to the development of optimal cathode material, making the dream of low-cost SIBs a reality \cite{Li_AS_22, Garcia_ESM_24, Ouyang_NC_24}.

\subsection{~Na$_3$(MO$_{1-x}$PO$_4$)$_2$F$_{1+2x}$}

\begin{figure*} 
\includegraphics[width=6.6in]{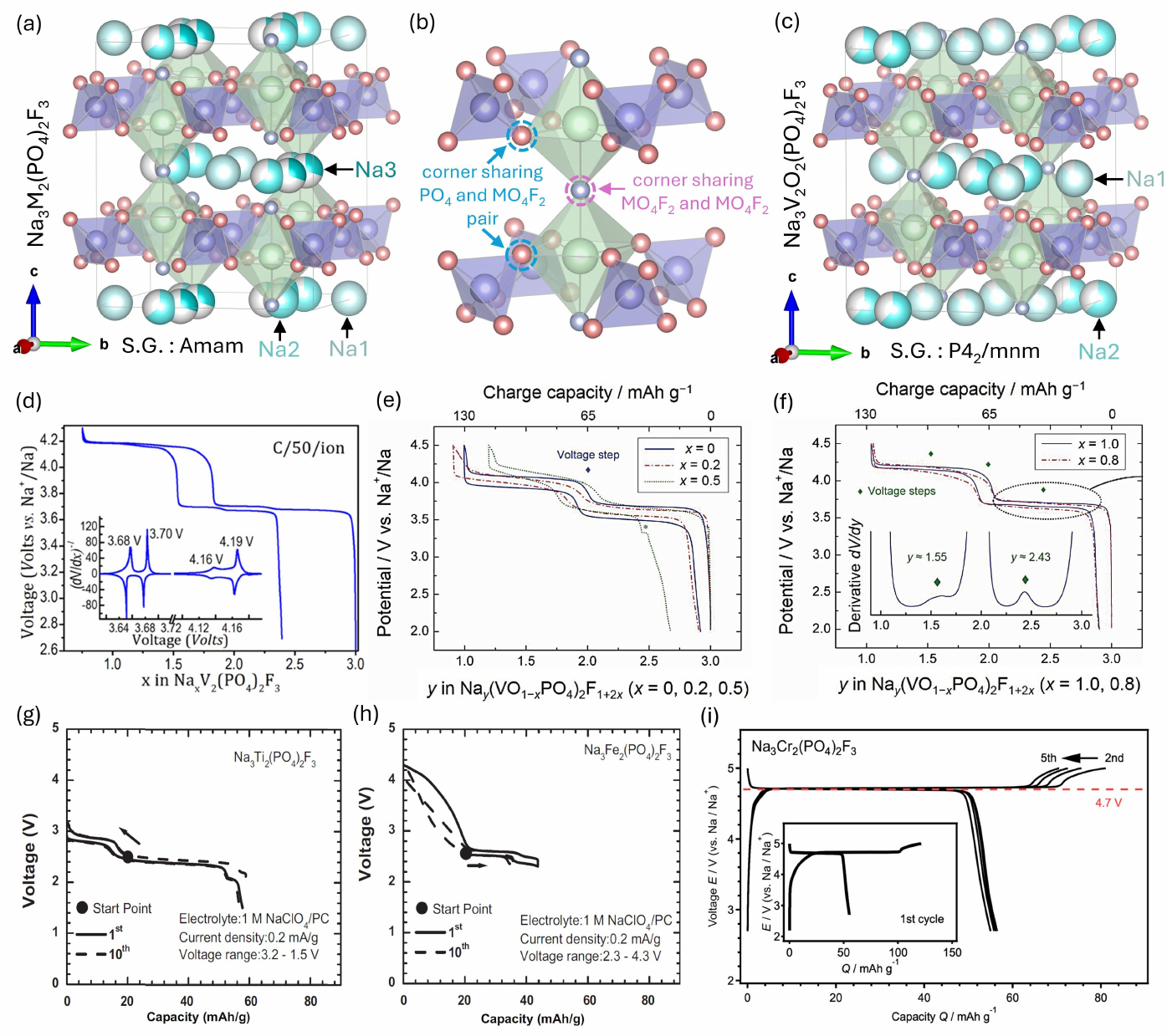}
\caption{Crystal structure of Na$_3$V$_2$(PO$_4$)$_2$F$_3$ (a) showing the distinct sodium sites, (b) the connectivity among the M$_2$O$_8$F$_3$ bioctahedra and PO$_4$ polyhedra, and (c) the crystal structure of Na$_3$(VO)$_2$(PO$_4$)$_2$F with labeled Na1 and Na2 sites. Charge/discharge profile of (d) Na$_3$V$_2$(PO$_4$)$_2$F$_3$ at a very low current rate of C/50 with dQ/dV curve shown in the inset \cite{Bianchini_CM_15}, (e,f) Na$_3$(VO$_{1-x}$PO$_4$)$_2$F$_{1+2x}$ ($x=$ 0, 0.2, 0.5, 0.8, 1.0) \cite{Park_AFM_14}, (g) Na$_3$Ti$_2$(PO$_4$)$_2$F$_3$ for 1$^{st}$ and 10$^{th}$ cycle \cite{Chihara_JPS_13}, (h) Na$_3$Fe$_2$(PO$_4$)$_2$F$_3$ for 1$^{st}$ and 10$^{th}$ cycle, and (i) Na$_3$Cr$_2$(PO$_4$)$_2$F$_3$ for the first five cycles at 0.1 C (first cycle shown in inset) \cite{Kawai_CM_21}. CIF files: (a,b) from \cite{Bianchini_CM_14}, and (c) from \cite{Park_AFM_14}.}
\label{Na3M2(PO4)2F3}
\end{figure*}

Finally, we discuss the most competitive cathode material among polyanions, which is Na$_3$V$_2$(PO$_4$)$_2$F$_3$ (NVPF). This material, having a tetragonal phase with $P4_2/mnm$ space group, features a high working voltage of 3.9 V and has a theoretical capacity of 128.2 mAh/g \cite{Meins_JSSC_99}. However, Bianchini {\it et al.} found a discrepancy in the assignment of crystal symmetry and used high angular resolution synchrotron radiation diffraction, which revealed a slight orthorhombic distortion adopting a $Amam$ space group \cite{Bianchini_CM_14}. Their inspection of the reported data of the $P4_2/mnm$ pointed out a meaningless value (8.3 \AA$^2$) of the thermal motion factor of the Na2 site. The authors found that sodium occupies three interstitial sites as shown in Fig.~\ref{Na3M2(PO4)2F3}(a): Na1 (pyramidal) is fully occupied, while Na2 (pyramidal) and Na3 (capped prism) are partially occupied and cannot be occupied simultaneously due to their smaller inter distance of 0.93 \AA~\cite{Meins_JSSC_99, Bianchini_CM_14}. Nonetheless, the reason for most of the literature adopting the $P4_2/mnm$ space group could be because of the synthesis conditions as the volatile nature of the F atom cause variations in the O/F ratio and mixed V$^{3+}$-V$^{4+}$ oxidation state, which makes the orthorhombic distortion insignificant \cite{Bianchini_CM_14, Broux_CM_16}. Regardless of the space group, the arrangement of the crystal structure units is equivalent, as demonstrated in Fig.~\ref{Na3M2(PO4)2F3}(b). The MO$_4$F$_2$ octahedra are connected via corner shared F forming M$_2$O$_8$F$_3$ bioctahedra units, which are further connected by corner sharing PO$_4$ polyhedra units via oxygen \cite{Meins_JSSC_99, Broux_CM_16, Bianchini_CM_15}. The sodium ions distribute themselves in large tunnels along the [110] and [1$\bar{1}$0] directions between the pseudo layers created by MO$_4$F$_2$ and PO$_4$ units \cite{Gu_ACIE_24}. The Na$_3$(VO)$_2$(PO$_4$)$_2$F (NVOPF) and Na$_3$V$_2$(PO$_4$)$_2$F$_3$ have similar frameworks where two F atoms (except the common F atom between the two VO$_4$F$_2$ units of NVPF) gets replaced by two oxygen atoms forming M$_2$O$_{10}$F bioctahedra unit as visible in Fig.~\ref{Na3M2(PO4)2F3}(c), and the intermediate O/F ratio between NVOPF and NVPF give rise to a family of vanadium oxyfluorophosphates Na$_3$(VO$_{1-x}$PO$_4$)$_2$F$_{1+2x}$ (0$\le$x$\le$1) existing in $P4_2/mnm$ space group with sodium ions distributed in two sodium sites \cite{Park_AFM_14}.

The electrochemical activity of NVPF was first explored by Gover {\it et al.} against lithium anode and found reversible insertion/extraction of sodium ions during initial charge/discharge followed by the dominant reversible activity of lithium after subsequent cycles \cite{Gover_SSI_06}. When tested against sodium in a potential window of 3.0--4.4 V, two sodium ions can be reversibly accessed during charging/discharging, two equal-value plateaus centered at $\sim$3.7 and $\sim$4.2 V are visible in the GCD profile shown in Fig.~\ref{Na3M2(PO4)2F3}(d) \cite{Bianchini_CM_15}. The first sodium extraction occurs at 3.7 V plateau and corresponds to a single V$^{3+}$/V$^{4+}$ redox process. The second sodium extraction at around 4.2 V activates an additional V$^{3+}$/V$^{4+}$ redox. However, during this step, V$^{4+}$ ions undergo disproportionation into equal amounts of V$^{3+}$ and V$^{5+}$, leading to a distinct charge ordering, and this process continues throughout sodium extraction and is reversed during discharge \cite{Bianchini_CM_14, Yan_NatCom_19}, following the reversible redox reaction given below: 
\begin{equation}
\footnotesize
Na_3V^{(III)}_2(PO_4)_2F_3 \rightleftharpoons 2Na^+ + NaV^{(III)}V^{(V)}(PO_4)_2F_3 + 2e^-
\end{equation} 
Upon closer inspection of the GCD profile at lower current rates, the dQ/dV curves show that each plateau splits into two, resulting in four voltage regions at 3.68 V, 3.70 V, 4.16 V, and 4.19 V during charging, while the crystal structure goes through a series of phase transitions depending on the sodium ion extracted from the structure. The pristine sample is in the $Amam$ space group, but during charge, it adopts tetragonal symmetry ($I4/mmm$) with the appearance of weak superstructural reflections. At the end of the charge, the structure transitions to the Na$_1$V$_2$(PO$_4$)$_2$F$_3$ phase with the $Cmc2_1$ space group after two sodium ions are extracted. During discharge, it returns to its original form ($Amam$), ensuring reversibility with a specific capacity of 120 mAh/g at 0.1 C \cite{Bianchini_CM_15, Yan_NatCom_19}. However, when the NVPF is cycled up to 4.8 V, a third sodium ion can also be extracted at nearly 4.7 V. The crystal structure monotonously shifts from $Cmc2_1$ to $I4/mmm$ above 4.4 V, with the extraction of sodium ions during charge. But in this case, the discharge curve does not have features identical to the charging curve and the third sodium ion is inserted at very low voltage of 1.6 V. It was also found that the structure reverts to the pristine $Amam$ space group only when the sodium ion extraction is limited to 2.5 Na per f.u., going beyond this limit resulted in tetragonal $I4/mmm$ discharged phase, and retains it even when discharged to 1.0 V \cite{Yan_NatCom_19}.

Further, in terms of valency, the vanadium present in V$^{3+}$ and V$^{4+}$ states in NVPF and NVPOF, respectively, and for intermediate samples, it exists in mixed V$^{3+/4+}$ state as per the O/F ratio. It was expected that decreasing fluorine content would decrease the average working voltage. However, the availability of V$^{4+}$/V$^{5+}$ redox compensates for the voltage difference while maintaining the charge/discharge profile and specific capacities, as visible from the comparative GCD profiles of Na$_3$(VO$_{1-x}$PO$_4$)$_2$F$_{1+2x}$ ($x=$ 0, 0.2, 0.5, 0.8, 1.0) shown in Figs.~\ref{Na3M2(PO4)2F3}(e, f). Still, it was observed that the average redox potential in NVPOF ($\sim$3.77 V) is lower than the average potential in NVPF ($\sim$3.9 V), suggesting a strong inductive effect due to fluorine \cite{Park_AFM_14}. This is also evident in the case of KTP-type NaVOPO$_4$ (V$^{4+}$/V$^{5+}$, 3.93 V) and NaVPO$_4$F (V$^{3+}$/V$^{4+}$, 4.0 V) \cite{Shraer_ESM_24, Shraer_NatCom_22}. The Na$_3$Fe$_2$(PO$_4$)$_2$F$_3$ and Na$_3$Ti$_2$(PO$_4$)$_2$F$_3$ analogs within the same framework are also electrochemically tested, see Fig.~\ref{Na3M2(PO4)2F3}(g, h), but showed extremely poor electrochemical performance \cite{Chihara_JPS_13}. Unlike the Na$_3$V$_2$(PO$_4$)$_2$F$_3$, no orthorhombic distortion was found in Na$_3$Cr$_2$(PO$_4$)$_2$F$_3$ (NCPF), adopting a $\beta$$_1$-Na$_3$Al$_2$(PO$_4$)$_2$F$_3$ type structure with $P4_2/mbc$ space group. The arrangement of M$_2$O$_8$F$_3$ bioctahedra and PO$_4$ polyhedra is identical with NVPF, while the sodium ions distribution is more complex \cite{Meins_JSSC_99, Kawai_CM_21}. The GCD cycling at 0.1 C delivers a reversible capacity of 55.1 mAh/g at a remarkable voltage of 4.7 V, as presented in Fig.~\ref{Na3M2(PO4)2F3}(i). The structures of NCPF and NVPF are identical, suggesting that further sodium extraction is possible above 5 V. We expect the experimental results to follow the same trend observed in the previously discussed cathode classes.

\begin{figure*} 
\includegraphics[width=6.4in]{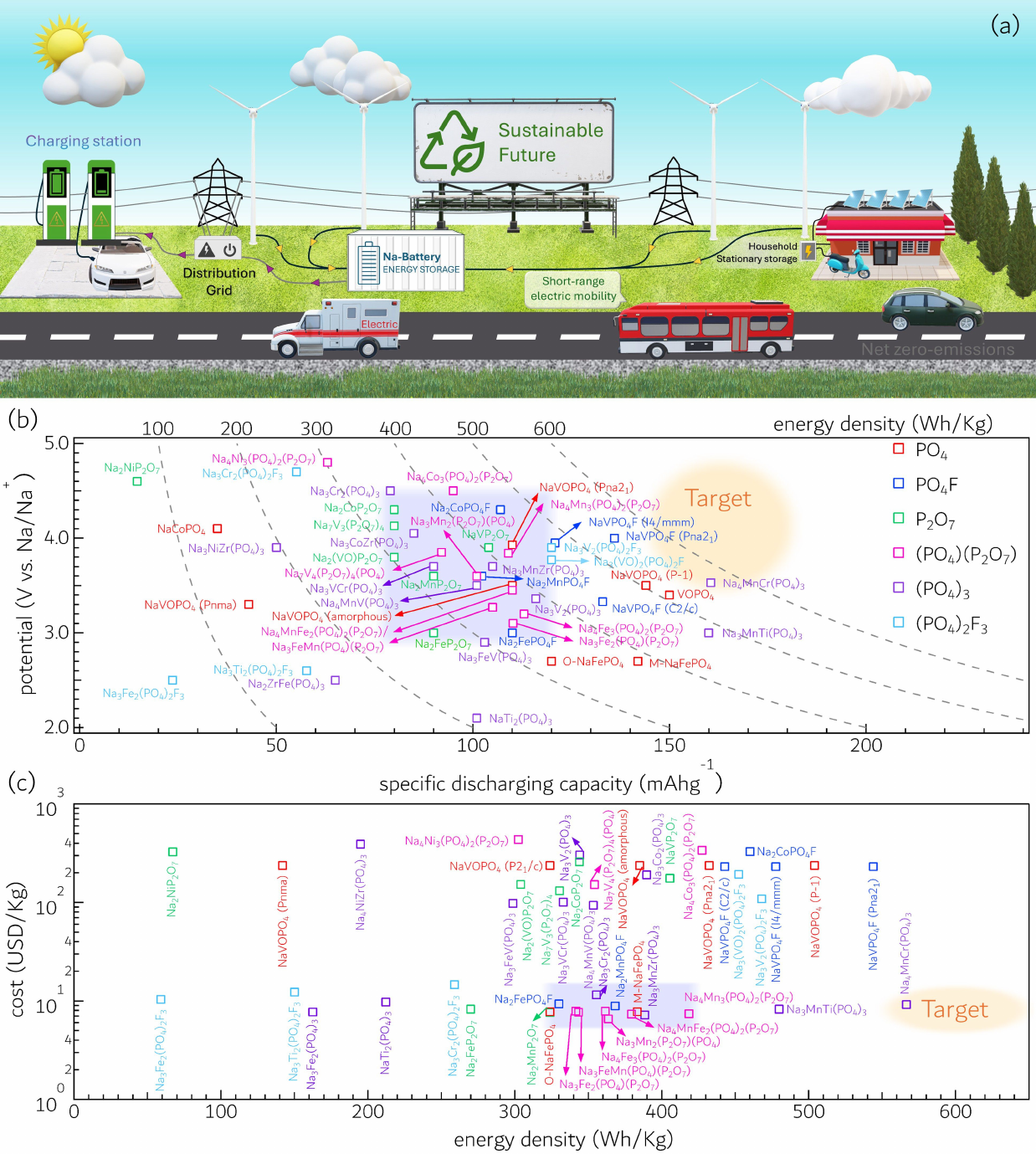}
\caption{(a) A schematic illustration of the potential applications of sodium-ion batteries as a cost-effective energy storage option. (b) The working potential vs. capacity plot for reported phosphate-based polyanionic cathodes; the square-shaded region indicates the current performance range of SIB cathodes, while the circular-shaded region represents the targeted performance for future development. (c) The estimated cost (based on raw materials only) plotted against energy density for various SIB cathodes. The rectangular shaded region denotes the current cost-effective cathodes, and the elliptical shaded region reflects the desired future target. The cost calculations consider scalable raw materials such as metal oxides, relevant to large-scale application rather than lab-scale synthesis. The prices of raw materials are sourced from \cite{TF}.}
\label{Future}
\end{figure*}

\section{\noindent ~Scale-up and performance evaluation}

To enhance the practical relevance of phosphate-based polyanionic cathode materials, this section outlines commercial-scale performance evaluations reported in the literature, with an emphasis on cost-effectiveness and sustainability. The vanadium-based polyanionic cathodes initially attracted considerable attention as potential successors to LFP due to their high operating voltage and stable cycling performance. For example, sodium-rich materials such as Na$_3$V$_2$(PO$_4$)$_3$ (NVP), Na$_3$V$_2$(PO$_4$)$_2$F$_3$ (NVPF), and Na$_3$(VO)$_2$(PO$_4$)$_2$F (NVPOF) have exhibited strong electrochemical performance and have been extensively investigated for practical sodium-ion battery applications. Notably, the combination of NVPF cathode and hard carbon anode has been commercialized by Tiamat in a cylindrical cell format. Their 2020-generation cell demonstrated a capacity of 0.61 Ah and an energy density of 68 Wh/kg at 1 C, showcasing excellent high-power capability \cite{He_JPS_23}. The NVPOF offers an additional advantage due to its room-temperature synthesis, enabling easier scalability without the need for high-temperature heat treatments \cite{Zhang_EMA_22}. However, despite their promising performance, V-based polyanionic cathodes face commercialization challenges due to the high cost and toxicity of vanadium. Moreover, a closed-loop recycling strategy proposed by Liu {\it et al.} offers a sustainable pathway, enabling the recovery of $\sim$100\% NVP and $\sim$99.1\% elemental aluminum without generating toxic waste \cite{Liu_NC_19}. Another strategy for cost reduction is by partially substituting vanadium with low-cost transition metals. For instance, Tang {\it et al.} developed a bulk synthesis method, {\textquotedblleft}suspendoid quick drying" to prepare Na$_{3.5}$V$_{1.5}$Mn$_{0.5}$(PO$_4$)$_3$, producing over 3 kg in a single batch. An 8.61 Ah pouch cell using this material delivered an energy density of 112.75 Wh/kg and retained 75.03\% of its capacity after 3000 cycles at 1 C \cite{Tang_JMCA_24}.

\begin{table*}[]
\centering
\small
\caption{A summary of long cycling performance of some of the potential phosphate polyanionic cathode materials.}
\label{tab:Retention1}
\begin{tabular}{p{4cm}p{2cm}p{3cm}p{3cm}p{3cm}p{1cm}}
\hline
Formula & 
 C-rate &
  \begin{tabular}[c]{@{}c@{}}Initial discharge\\ capacity (mAh/g)\end{tabular} &
  \begin{tabular}[c]{@{}c@{}}Retention\\ (\%)\end{tabular} &
  No. of cycles &
  Ref. \\ \hline
\multicolumn{6}{c}{PO$_4$}                                                                                                   \\ \hline
o-NaFePO$_4$                &        0.1    &   100  &  90   & 100       & \cite{Zhu_NS_13}                             \\
m-NaFePO$_4$                  &    0.05   &   142  &  95   &  200      & \cite{Kim_EES_15_Fe}                             \\
NaVOPO$_4$ ($P\bar{1}$)     &  0.5    &  112   & 67    &  1000    & \cite{Fang_C_18}     \\
NaVOPO$_4$ ($Pna2_1$)      &    2     &     92   & 87   & 1000     & \cite{Shraer_ESM_24}                   \\
NaVOPO$_4$ (amorphous)   &     5    &   50     & 96   & 2000     & \cite{Fang_CCSC_21}           \\ \hline

\multicolumn{6}{c}{PO$_4$F}                                                                                                  \\ \hline
Na$_2$FePO$_4$F                 &    5     &  56.4   & 97   &  700      &  \cite{Cao_S_22}         \\
Na$_2$MnPO$_4$F              &     0.1   &   113  & 77   &  200      & \cite{Ling_CI_19} \\
NaVPO$_4$F ($C2/c$)          &       1    &  121  & 83   &  2500     & \cite{Law_ESM_18}                     \\
NaVPO$_4$F ($I4/mmm$)    &     0.05 &  120.9    & 97.7 & 50     & \cite{Ruan_EA_15}                  \\
NaVPO$_4$F ($Pna2_1$)      &       5    & $\sim$120   & 82   &  800     & \cite{Shraer_NatCom_22}                      \\ \hline

\multicolumn{6}{c}{P$_2$O$_7$}                                                                                               \\ \hline
Na$_2$FeP$_2$O$_7$           &       60   &  55  & 84   &  10000    & \cite{Song_JMCA_17}             \\
Na$_2$MnP$_2$O$_7$           &       2    &   63  & 83   &  600      & \cite{Li_ESM_19}     \\
Na$_2$Fe$_{0.5}$Mn$_{0.5}$P$_2$O$_7$&  0.05 &  80  & 84   & 90     & \cite{Shakoor_PCCP_16}                            \\
NaVP$_2$O$_7$                    &     20   &  90  & 99   &  100         & \cite{Drozhzhin_CM_19}                  \\
Na$_7$V$_3$(P$_2$O$_7$)$_4$  &       1    &  $\sim$70   & 75   &  600     & \cite{Kim_AEM_16}                         \\ \hline
\end{tabular}%
\end{table*}

Further, Cao {\it et al.} evaluated the cost of raw materials for Na$_2$FePO$_4$F (NFPF) and found it to be significantly lower than that of LiFePO$_4$ (LFP); 1389.6 USD/ton for NFPF compared to 13,487 USD/ton for LFP (cost conversion 11-Apr-2025, 1 USD = 7.31 CNY) \cite{Cao_S_22}. Motivated by this cost advantage, they scaled up the synthesis of NFPF using a spray-drying method, successfully producing pure-phase material with promising electrochemical performance. An 186 mAh pouch cell assembled with a hard carbon‖NFPF configuration delivered a high energy density of 210 Wh/kg and retained 92\% of its capacity after 200 cycles at 0.3 C \cite{Cao_S_22}. The potential of mixed ortho- and pyrophosphate frameworks has gained significant attention due to their favorable electrochemical performance and cost-effectiveness. Very recently, Liu {\it et al.} explored cathodes such as Na$_3$Fe$_2$(PO$_4$)P$_2$O$_7$ and Na$_4$Fe$_3$(PO$_4$)P$_2$O$_7$ for their scalability and practicality \cite{Liu_AFM_25}. Lu {\it et al.} reported a green and efficient synthesis route for Na$_3$Fe$_2$(PO$_4$)P$_2$O$_7$ using a sand mill assisted spray drying process with dual iron sources (iron phosphate and ferrous oxalate), avoiding nitrate-based precursors. A 4.5 Ah pouch cell assembled with the synthesized Na$_3$Fe$_2$(PO$_4$)P$_2$O$_7$ cathode and a hard carbon anode demonstrated excellent cycling stability, retaining 82.7\% of its capacity after 3500 cycles at 1 C, with a minimal capacity decay rate of 0.00495\% per cycle \cite{Lu_JES_25}. Furthermore, a 20 kg batch of Na$_{3.6}$Fe$_{2.6}$(PO$_4$)P$_2$O$_7$ was successfully synthesized using a sand mill-assisted spray drying process. When tested in a 12 Ah pouch cell configuration with a hard carbon anode, the cell retained 82.4\% of its capacity after 2000 cycles at 1 C \cite{Dai_JPS_24}. In another study, a 0.35 Ah pouch cell fabricated using an iron-deficient Na$_4$Fe$_{2.91}$(PO$_4$)P$_2$O$_7$ cathode achieved 87.4\% capacity retention over 1000 cycles at 1 C \cite{Zhao_NE_22}. The Na$_4$Fe$_3$(PO$_4$)P$_2$O$_7$ cathode has also demonstrated promising performance in an anode-less pouch cell configuration, utilizing hard-carbon-derived interphase on an aluminum current collector, which significantly reduces the weight of the fabricated pouch cell. The 2.28 Ah cell delivered an impressive energy density of 203 Wh/kg, comparable to that of LFP systems, and exhibited excellent cycling stability, retaining 95.2\% capacity after 180 cycles at 0.2 C and 80\% after 770 cycles \cite{Ruan_NS_25}.

As we know, currently the LFP dominates the battery market, but the growing accumulation of spent LFP batteries raises serious recycling concerns \cite{Mrozik_EES_21}. The recovered olivine FePO$_4$ from LFP recycling presents a promising raw material for Fe-based phosphate cathodes, offering a sustainable and cost-effective solution \cite{Gan_S_24}. Additionally, battery-grade FePO$_4$.2H$_2$O can be prepared from hazardous industrial steel swarf (69\% Fe), preventing thousands of tons of waste from ending up in landfills and promoting circular resource utilization \cite{Ottink_RCR_22, Xu_SPT_24}. Together, these advancements not only highlight the viability of low-cost, scalable iron-based phosphate chemistries, but also pave the way for a sustainable, circular economy for next-generation SIBs.

\begin{table*}[]
\centering
\small
\caption{A summary of long cycling performance of some of the potential phosphate polyanionic cathode materials.}
\label{tab:Retention2}
\begin{tabular}{p{4cm}p{2cm}p{3cm}p{3cm}p{3cm}p{1cm}}
\hline
Formula & 
 C-rate &
  \begin{tabular}[c]{@{}c@{}}Initial discharge\\ capacity (mAh/g)\end{tabular} &
  \begin{tabular}[c]{@{}c@{}}Retention\\ (\%)\end{tabular} &
  No. of cycles &
  Ref. \\ \hline
\multicolumn{6}{c}{(PO$_4$)(P$_2$O$_7$)}                                                                                     \\ \hline
Na$_4$Fe$_3$(PO$_4$)$_2$(P$_2$O$_7$)   &   20   & 80.3 &  69.1 &  4400    	& \cite{Chen_NatCom_19}                     \\
Na$_4$Mn$_3$(PO$_4$)$_2$(P$_2$O$_7$) &   0.2  & $\sim$85 & 86   & 100    	 & \cite{Kim_EES_15_Mn}                             \\
Na$_4$MnFe$_2$(PO$_4$)$_2$(P$_2$O$_7$)& 1    & $\sim$80 & 83   &  3000    	& \cite{Kim_CM_16}                  \\
Na$_7$V$_4$(P$_2$O$_7$)$_4$PO$_4$    &      0.5  & 92 & 92.1 &  300    	 & \cite{Fang_RSCAdv_18}                             \\
Na$_3$Fe$_2$(PO$_4$)(P$_2$O$_7$)      &        20   & 61.6 & 89.7 &  6400   	 & \cite{Cao_ACSEL_20}                  \\
Na$_3$Mn$_2$(PO$_4$)(P$_2$O$_7$)    &         5    & 32 & 73   &  500     	&  \cite{Li_JPS_22}                   \\
Na$_3$MnFe(PO$_4$)(P$_2$O$_7$)      &           5    & 60 & 65.5 &  1000   	 & \cite{Wang_CEJ_22}                     \\ \hline

\multicolumn{6}{c}{(PO$_4$)$_3$}                                                                                             \\ \hline
NaTi$_2$(PO$_4$)$_3$               &   0.5  & 76 & 88.8 &  800                         & \cite{Pang_JMCA_14}                             \\
Na$_3$V$_2$(PO$_4$)$_3$       &          40   &62 & 50   & 30000 		   & \cite{Saravanan_AEM_13}                         \\
Na$_3$MnZr(PO$_4$)$_3$         &      0.5  & $\sim$90 & 91   &  500      			& \cite{Gao_JACS_18}                           \\
Na$_4$MnV(PO$_4$)$_3$ (2 Na)   &        1   &   $\sim$100   & 97   &  1000     		 &  \cite{Zhou_NL_16}                 \\
Na$_3$MnTi(PO$_4$)$_3$ (3 Na)   &        2    & 129   & 92   &  500      		&  \cite{Zhu_AEM_19}                           \\
Na$_4$MnCr(PO$_4$)$_3$ (2 Na)     &       5   &  60.5    & 86.5   &  600  		   & \cite{Zhang_AM_20}         \\ \hline

\multicolumn{6}{c}{(PO$_4$)$_2$F$_3$}                                                                                        \\ \hline
Na$_3$V$_2$(PO$_4$)$_2$F$_3$   &      30   &  91  & -  &  10000           & \cite{Subramanian_CEJ_21}      \\
Na$_3$(VO)$_2$(PO$_4$)$_2$F      &      45   &  70  & 81   &  2000         & \cite{Deng_ESM_16}     \\ \hline
\end{tabular}%
\end{table*}

\section{\noindent ~Future perspective and strategies}

The future outlook for sodium-ion batteries remains aligned with the established goal: emerging as a cost-effective alternative to lithium-ion batteries. The SIBs are still in the early stages of commercialization, presenting significant opportunities for improvement. The target working voltage levels should be around 4.0 V or higher, with a practical energy density range of 160-220 Wh/kg in pouch or cylindrical cell configurations. Ideally, the cost per kWh of SIBs should be 20 to 40\% lower compared to LIBs. Nevertheless, it is also important to emphasize cycle life, as the lower cost per cycle should be considered alongside energy density. This could only be possible by a breakthrough cathode material's electrochemical performance based on sodium chemistry. The marvelous flexibility of polyanionic cathodes may deliver the promise of a low-cost energy storage system that can be utilized in the stationary applications, short-range electric mobility, and potential avenues as visualized in Fig.~\ref{Future}(a). The phosphate-based polyanionic cathodes are also well-known for their robust structural stability, enabling long-term cycling performance even at high current rates, as summarized in Tables~\ref{tab:Retention1} and \ref{tab:Retention2}. The detailed discussion on the relationship between voltage and specific capacities concerning the host structure reveals that the specific capacity is largely influenced by the molecular weight of the host material, of which the atomic weight of the sodium is only a fractional part. We also elucidated the voltage perspective of the cathodes and stated that the correct choice of host structure could elevate the redox potentials, which in turn make SIBs capable of competing with voltage levels of LIBs and making the pessimistic term {\textquotedblleft}sodium reduction potential less negative than that of lithium" insignificant. The working potential versus specific capacity plot [see Fig.~\ref{Future}(b)] illustrates that most of the currently reported cathode materials lie within an energy density range of 300--450 Wh/kg, based on the active material weight. The square highlighted region indicates the target energy density range for future cathodes, which can only be achieved by enabling multi-electron reactions per transition metal in combination with higher operating voltages. Therefore, a perfect blend of ideal transition metal fitted in a host structure could keep in check the intrinsic ionic and electronic conductivities, thereby enabling multi-electron reactions at elevated voltages. As an illustration, the orthophosphate suffers from the thermal stability of the electrochemical active Olivine phase but offers superior theoretical capacity. On the other hand, pyrophosphates, with impressive thermal durability, have a bulky structure, which has an adverse effect on the specific capacities. The mix-pyro and orthophosphate host structures, with an improved ratio of active sodium ions per formula unit weight, have demonstrated a complementary combination of inductive effects, along with structural and cycling stability. This combination results in an iron-based 3.2 V air-moisture stable cathode, which is also capable of withstanding the Jahn-Teller distortion associated with the Mn$^{3+}$ state, as well as exhibits exceptionally high redox activity of Ni. Such strategies lay the foundation for the future development of promising host structures of SIBs. The cost analysis presented in Fig.~\ref{Future}(c) supports the same observation: the rectangular shaded region, representing cost-effective cathodes with moderate energy density, is primarily populated by Fe, Mn, and Fe+Mn-based mix-pyro and orthophosphate structures. However, to reach the target performance region, future efforts should focus on designing cathodes capable of multi-electron reactions per transition metal at higher voltages, enabling higher energy densities. An example is Na$_4$MnCr(PO$_4$)$_3$ (NMCP), which lies near the edge of the proposed target region [see Fig.~\ref{Future}(b, c)]. Unfortunately, the NMCP currently suffers from rapid capacity fading, electrolyte instability, and other challenges that must be addressed before it can be considered for commercialization. Therefore, it is critical to extend the electrochemical stability window of electrolytes beyond 5 V, as it would open the doors for numerous unexplored electrochemically active cathodes that could potentially exhibit significant performance improvements if provided with the appropriate voltage window. 

\section{\noindent ~Conclusions}

In summary, the electrochemical performance is governed by both the host structure and the transition metal in the discussed Phosphate-Based Polyanionic Cathode Materials for Sodium-Ion Batteries. The charge storage process and associated phase transformations are intrinsic properties of the host structure, and at the same time, the reaction kinetics are majorly influenced by the transition metal. The working voltage, however, is a combined effect of both. Regarding transition metals, the Fe-based cathode materials exhibit promising structural stability and cost-effectiveness, although their working voltage is constrained to the 2.5--3.2 V range within phosphate-based frameworks. Conversely, Mn, being another economical option, suffers from structural instability due to its large ionic size, significant polarization effects, and considerable volume changes induced by Jahn-Teller distortion. The binary transition metal system of Mn and Fe, however, proves to be a highly efficient approach, where the stability provided by Fe and the high voltage characteristics of Mn complement each other. The cathodes with an optimized Fe-Mn ratio show great potential for the development of low-cost, high-performance materials, positioning them as a {\textquotedblleft}hidden gem" in the search for affordable and effective cathode solutions. At the same time, the Co-based chemistries offer higher working voltage and stable cycling. However, their toxicity and cost have prompted researchers to pursue Co-free alternatives. On the other hand, some Ni-based phosphate materials have shown no electrochemical activity up to 5 V. While in the mix-pyrophosphate framework, Ni demonstrates high-voltage electrochemical activity, albeit with capacities well below theoretical values. The vanadium-based polyanionic cathode materials within phosphate frameworks, due to their exceptional structural stability, minimal polarization effects, consistent electrochemical performance, and superior cycling stability, have emerged as the optimal choices. However, given the concerns about the rising costs and toxicity of vanadium, alternatives like Fe and Mn, as well as high entropy materials, offer promising solutions for sustainable, low-cost, and eco-friendly energy options. Looking ahead, the future research on phosphate-based polyanionic cathodes should focus on optimizing the balance between energy density and structural stability, exploring high entropy, facilitating multi-electron reactions to unlock synergistic effects, and advancing scalable, green synthesis methods. We hope the detailed discussion on the relationship between electrochemical activity and structural integrity will inspire new researchers to explore novel high-voltage materials and chemistries for SIB cathodes.

\section{\noindent ~Acknowledgments}

MS and RG thank CSIR-HRDG and MHRD for the fellowship support, respectively, through IIT Delhi. MS and RG thank Pooja Sindhu for useful discussions. RSD acknowledges the DST for financially supporting the research facilities for sodium-ion batteries through {\textquotedblleft}DST-IIT Delhi Energy Storage Platform on Batteries" (project no. DST/TMD/MECSP/2K17/07) and from SERB-DST (now ANRF) through a core research grant (file no.: CRG/2020/003436).

\end{document}